\documentclass[runningheads]{llncs}
\usepackage{amsmath}
\usepackage{amsfonts}
\usepackage{graphicx}
\usepackage{framed}
\usepackage[colorlinks, anchorcolor=blue, linkcolor=blue, citecolor=blue, urlcolor=cyan]{hyperref}
\usepackage{tikz}
\usepackage{lscape}
\usepackage{hyperref}

\usepackage{breakcites}
\usepackage{colortbl}
\usepackage{amssymb}

\hypersetup{
    colorlinks=true,    
    urlcolor=blue       
}

\newcommand{\Cuckoo}{\mathsf{Cuckoo}}
\newcommand{\Simple}{\mathsf{Simple}}
\newcommand{\PRF}{\mathsf{PRF}}
\newcommand{\FuncROT}{\mathcal{F}_\mathsf{rot}}

\newcommand{\FuncbOPPRF}{\mathcal{F}_\mathsf{bOPPRF}}
\newcommand{\FuncbssPMT}{\mathcal{F}_\mathsf{bssPMT}}
\newcommand{\FuncmssROT}{\mathcal{F}_\mathsf{mss\text{-}rot}}

\newcommand{\FuncMS}{\mathcal{F}_\mathsf{ms}}

\newcommand{\FuncssPEQT}{\mathcal{F}_\mathsf{ssPEQT}}
\newcommand{\FuncMPSU}{\mathcal{F}_\mathsf{mpsu}}

\newcommand{\FuncMPID}{\mathcal{F}_\mathsf{MPID}}

\newcommand{\ProtoMPSUTwo}{\Pi_\mathsf{SK\text{-}MPSU}}
\newcommand{\ProtoMPSUFour}{\Pi_\mathsf{PK\text{-}MPSU}}
\newcommand{\ProtomssROT}{\Pi_\mathsf{mss\text{-}rot}}

\newcommand{\ProtobssPMT}{\Pi_\mathsf{bssPMT}}

\newcommand{\ProtoMPID}{\Pi_\mathsf{MPID}}
\newcommand{\Encode}{\mathsf{Encode_H}}
\newcommand{\Decode}{\mathsf{Decode_H}}
\newcommand{\Gen}{\mathsf{Gen}}
\newcommand{\Enc}{\mathsf{Enc}}
\newcommand{\Dec}{\mathsf{Dec}}

\newcommand{\ParDec}{\mathsf{ParDec}}

\newcommand{\ReRand}{\mathsf{ReRand}}
\newcommand{\Hash}{\mathsf{H}}
\newcommand{\Honest}{\mathcal{H}}
\newcommand{\Corr}{\mathcal{C} orr}
\newcommand{\Elem}{\mathsf{Elem}}
\newcommand{\ct}{\mathsf{ct}}
\newcommand{\Sd}{\mathcal{S}}
\newcommand{\Rcv}{\mathcal{R}}

\usepackage{pifont}
\newcommand{\cmark}{\text{\ding{51}}}
\newcommand{\xmark}{\text{\ding{55}}}

\usepackage{lineno}

\usepackage{multirow}

\begin{document}

\title{Breaking Free: Efficient Multi-Party Private Set Union Without Non-Collusion Assumptions}
\titlerunning{Efficient Multi-Party Private Set Union Without Non-Collusion Assumptions}

\author{Minglang Dong\inst{1,2,3}\orcidID{0009-0002-5323-7119} \and
Yu Chen\inst{1,2,3}\orcidID{0000-0003-2553-1281} \and
Cong Zhang\inst{4}\orcidID{0009-0000-5403-2866} \and 
Yujie Bai\inst{1,2,3}\orcidID{0009-0006-7500-3531}}

\authorrunning{M. Dong et al.}

\institute{School of Cyber Science and Technology, Shandong University, \\Qingdao 266237, China
\and
State Key Laboratory of Cryptology, P.O. Box 5159, Beijing 100878, China
\and
Key Laboratory of Cryptologic Technology and Information Security, Ministry of Education, 
Shandong University, Qingdao 266237, China \email{minglang\_dong@mail.sdu.edu.cn, yuchen@sdu.edu.cn, baiyujie@mail.sdu.edu.cn}
\and
Institute for Advanced Study, BNRist, Tsinghua University, Beijing, China \\ \email{zhangcong@mail.tsinghua.edu.cn}
}

\maketitle            

\begin{abstract}
Multi-party private set union (MPSU) protocol enables $m$ $(m > 2)$ parties, each holding a set, to collectively compute the union of their sets without revealing any additional information to other parties. There are two main categories of multi-party private set union (MPSU) protocols: The first category builds on public-key techniques, where existing works require a super-linear number of public-key operations, resulting in their poor practical efficiency. The second category builds on oblivious transfer and symmetric-key techniques. The only work in this category, proposed by Liu and Gao (ASIACRYPT 2023), features the best concrete performance among all existing protocols, but still has super-linear computation and communication. Moreover, it does not achieve the standard semi-honest security, as it inherently relies on a non-collusion assumption, which is unlikely to hold in practice. 

There remain two significant open problems so far: no MPSU protocol achieves semi-honest security based on oblivious transfer and symmetric-key techniques, and no MPSU protocol achieves both linear computation and linear communication complexity. In this work, we resolve both of them. 

\begin{itemize}
\item We propose the first MPSU protocol based on oblivious transfer and symmetric-key techniques in the standard semi-honest model. This protocol is $3.9-10.0 \times$ faster than Liu and Gao in the LAN setting. Concretely, our protocol requires only $4.4$ seconds in online phase for 3 parties with sets of $2^{20}$ items each.

\item We propose the first MPSU protocol achieving both linear computation and linear communication complexity, based on public-key operations. This protocol has the lowest overall communication costs and shows a factor of $3.0-36.5\times$ improvement in terms of overall communication compared to Liu and Gao.
\end{itemize} 

We implement our protocols and conduct an extensive experiment to compare the performance of our protocols and the state-of-the-art. To the best of our knowledge, our code is the first correct and secure implementation of MPSU that reports on large-size experiments. 

\end{abstract}

\section{Introduction}

Over the last decade, there has been growing interest in private set operation (PSO), which consists of private set intersection (PSI), private set union (PSU), and private computing on set intersection (PCSI), etc. 
Among these functionalities, PSI, especially two-party PSI~\cite{PSZ-USENIX-2014,KKRT-CCS-2016,PRTY-CRYPTO-2019,CM-CRYPTO-2020,PRTY20,RS-EUROCRYPT-2021,RR-CCS-2022}, has made tremendous progress and become highly practical with extremely fast and cryptographically secure implementations. Meanwhile, multi-party PSI~\cite{KMPRT-CCS-2017,NTY-CCS-2021,CDGOSS-CCS-2021,BNOP-AsiaCCS-2022} is also well-studied. In contrast, the advancement of PSU has been sluggish until recently, several works proposed efficient two-party PSU protocols~\cite{KRTW-ASIACRYPT-2019,GMRSS-PKC-2021,Jia-USENIX-2022,ZCLZL-USENIX-2023,ZCZDL-2024,Jia0ZG24}. However, multi-party PSU has still not been extensively studied. In this work, we focus on PSU in the multi-party setting.

Multi-party private set union (MPSU) enables $m$ $(m > 2)$ mutually untrusted parties, each holding a private set of elements, to compute the union of their sets without revealing any additional information. 
MPSU and its variants have numerous applications, such as information security risk assessment~\cite{LV04}, IP blacklist and vulnerability data aggregation~\cite{HLSSSYY16}, joint graph computation~\cite{BllS05}, distributed network monitoring~\cite{KS-CRYPTO-2005}, building block for private DB supporting full join~\cite{KRTW-ASIACRYPT-2019}, private ID~\cite{GMRSS-PKC-2021}, etc.

According to the underlying techniques, existing MPSU protocols\footnote{We only consider the special-purpose solutions for MPSU, excluding those based on circuit-based generic techniques of secure computation, due to their unacceptable performance.} can be divided into two categories: 
  \begin{itemize}
    \item \textbf{PK-MPSU}: This category is primarily based on public-key techniques, and has been explored in a series of works~\cite{KS-CRYPTO-2005,Frikken-ACNS-2007,VCE22,GNT-eprint-2023}. A common drawback of these works is that each party has to perform a substantial number of public-key operations, leading to unsatisfactory practical efficiency.
    \item \textbf{SK-MPSU}: This category is primarily based on symmetric-key techniques, and includes only one existing work~\cite{LG-ASIACRYPT-2023} to date. This work exhibits much better performance than all prior works but fails to achieve standard semi-honest security due to its inherent reliance on a non-collusion assumption, assuming the party who obtains the union (we call it leader hereafter) not to collude with other parties.
  \end{itemize}
Both categories share one common limitation: neither of them includes a protocol achieving linear computation and communication complexity\footnote{In the context of MPSU, linear complexity means that the complexity per party scales linearly with the total size of all parties' sets. In this paper, we consider the balanced setting where each party holds sets of equal size, thus linear complexity denotes the complexity per party to scale linearly with both the number of parties $m$ and the set size $n$. Meanwhile, following current conventions, linear complexity only considers the online phase.}. Motivated by the context, we raise the following two questions:

\begin{center}
\emph{Can we construct an MPSU protocol based on oblivious transfer and symmetric-key operations, without any non-collusion assumptions? 
Can we construct an MPSU protocol with both linear computation and linear communication complexity?}
\end{center}

\subsection{Our Contribution}

In this work, we answer the above two questions affirmatively.
Our contributions are summarized as follows:

\begin{trivlist}
\item \textbf{Efficient batch ssPMT.} We present a new primitive called batch secret-shared private membership test (batch ssPMT).
Compared to multi-query secret-shared private membership test (mq-ssPMT), which is the technical core of the state-of-the-art MPSU protocol~\cite{LG-ASIACRYPT-2023} (hereafter referred to as LG), batch ssPMT admits a much more efficient construction. Combined with hashing to bins, batch ssPMT can replace mq-ssPMT in the context of MPSU. 
Looking ahead, our batch ssPMT serves as a core building block in our two MPSU protocols, and significantly contributes to our speedup to LG. 

\item \textbf{SK-MPSU in standard semi-honest model.} We generalize random oblivious transfer (ROT) into multi-party setting and present a new primitive called multi-party secret-shared random oblivious transfer (mss-ROT).
Based on batch ssPMT and mss-ROT, we propose the first SK-MPSU in the standard semi-honest model. In addition to enhanced security, our SK-MPSU has superior online / total performance with a $3.9-10.0 \times$ / $1.2-7.8 \times$ improvement in the LAN setting.

\item \textbf{PK-MPSU with linear complexity.} Based on batch ssPMT and multi-key rerandomizable public-key encryption (MKR-PKE)~\cite{GNT-eprint-2023}, we propose the first MPSU protocol with both linear computation and communication. Our PK-MPSU has the lowest overall communication costs with a factor of $3.0-36.5\times$ improvement compared to LG. 
Along the way, we find that the MPSU protocol of Gao et al.~\cite{GNT-eprint-2023} is insecure against arbitrary collusion and give a practical attack to demonstrate that it necessitates non-collusion assumption as well.
\footnote{After we pointed out the security flaw of the GNT protocol~\cite{GNT-eprint-2023} with a concrete attack, 
the authors of~\cite{GNT-eprint-2023} contacted us and confirmed our attack. 
Subsequently, they updated their paper and revised their original protocol (c.f. Appendix B in the new version of~\cite{GNT-eprint-2023}) to a new one, which is similar to our PK-MPSU. The main difference lies in that they implement our batch ssPMT by invoking multiple instances of ssPMT separately, which renders their new protocol still has superlinear complexities. 
See also the summary of the relationship in their updated paper.}
\end{trivlist}

\begin{figure}[!hbtp]
\begin{framed}
\begin{tikzpicture}
\node (v6) at (-2.5,1.5) {Our SK-MPSU};
\node (v9) at (2,1.5) {Our PK-MPSU};

\node at (0,0.15) {batch ssPMT};
\draw  (-1.15,0.5) rectangle (1.15,-0.2);
\node (v8) at (0,0.45){};
\draw [->] (v8) edge (v6);
\draw [->] (v8) edge (v9);

\node (v2) at (0,-0.15) {};

\node(v1) at (-1.5,-1.25) {batch OPPRF};
\draw [dashed] (-2.75,-0.9) rectangle (-0.25,-1.6);
\node (v13) at (-1.5,-0.95) {};
\draw [->] (v13) edge (v2);

\node (v3) at (1.5,-1.25) {ssPEQT};
\draw [dashed] (0.7,-0.9) rectangle (2.25,-1.6);
\node (v14) at (1.5,-0.95) {};
\draw [->] (v14) edge (v2);

\node at (-2.5,0.15) {mss-ROT};
\draw [->] (-3.4,0.5) rectangle (-1.6,-0.2);
\node (v7) at (-2.5,0.4){};

\node (v5) at (-5.85,0.15) {multi-party secret-shared};
\node (v5) at (-5.85,-0.25) {shuffle};
\draw [dashed] (-7.85,0.5) rectangle (-3.85,-0.6);
\node (v12) at (-5.85,0.45){};

\node (v10) at (2.6,0.15) {MKR-PKE};
\draw [dashed] (1.6,0.5) rectangle (3.6,-0.2);
\node (v11) at (2.6,0.4){};

\draw [->] (v7) edge (v6);

\draw [->] (v11) edge (v9);

\draw [->] (v12) edge (v6);

\end{tikzpicture}
\end{framed}
\caption{Technical overview of our MPSU framework. The newly introduced primitives are marked with solid boxes. The existing primitives are marked with dashed boxes.}\label{fig:relation}
\end{figure}
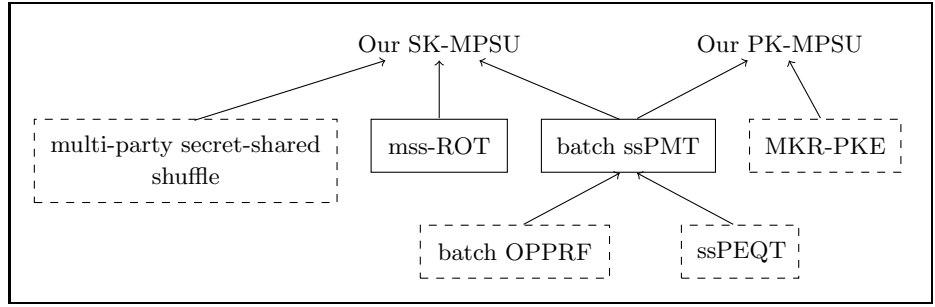

Figure~\ref{fig:relation} depicts the technical overview of our new MPSU framework. We will elaborate the details in Section~\ref{technical}.

\subsection{Related Works}
We review the existing semi-honest MPSU protocols below. 

\begin{trivlist}
\item \textbf{PK-MPSU.} Kisser and Song~\cite{KS-CRYPTO-2005} introduced the first MPSU protocol, based on polynomial representations and additively homomorphic encryption (AHE). This protocol requires a substantial number of AHE operations and high-degree polynomial calculations, so it is completely impractical.

Frikken~\cite{Frikken-ACNS-2007} improved~\cite{KS-CRYPTO-2005} by decreasing the polynomial degree. However, the number of AHE operations remains quadratic in the set size due to the necessity of performing multi-point evaluations on the encrypted polynomials.

Vos et al.~\cite{VCE22} proposed an MPSU protocol based on the bit-vector representations, where the parties collectively compute the union by performing the private OR operations on the bit-vectors, using ElGamal encryption. It shows poor concrete efficiency reported by~\cite{LG-ASIACRYPT-2023} and the leader requires quadratic computation and communication in the number of parties.

Recently, Gao et al.~\cite{GNT-eprint-2023} proposed an MPSU protocol,  
representing the most advanced MPSU in terms of asymptotic complexity. Unfortunately, their protocol turns out to be insecure against arbitrary collusion. We propose a practical attack to show that it requires the same non-collusion assumption as LG (see Appendix~\ref{sec:attack} for details).

\item \textbf{SK-MPSU.} Recently, Liu and Gao~\cite{LG-ASIACRYPT-2023} proposed a practical MPSU protocol based on oblivious transfer and symmetric-key operations. This protocol is several orders of magnitude faster than the prior works. For instance, when computing on datasets of $2^{10}$ element, it is $109 \times$ faster than~\cite{VCE22}. However, their protocol is not secure in the standard semi-honest model.

\item \textbf{Other Related Works.} Blanton et al.~\cite{BA-ASIACCS-2012} proposed an MPSU protocol based on oblivious sorting and generic multi-party computation (MPC). 
The heavy dependency on general MPC leads to inefficiency.
\end{trivlist}

Table \ref{tab:comparisons} provides a comprehensive theoretical comparison between existing MPSU protocols and our proposed protocols. Leader denotes the participant who obtains the union result. Client refers to the remaining participants. 

\begin{table*}
\centering
\resizebox{\textwidth}{!}{%
  \begin{tabular}{|c|c|c|c|c|c|c|}
        \hline 
        \multirow{2}{*}{\textbf{Protocol}} & \multicolumn{2}{c|}{\textbf{Computation}} & \multicolumn{2}{c|}{\textbf{Communication}} & \multirow{2}{*}{\textbf{Round}} & \multirow{2}{*}{\textbf{Security}}\\ 
        \cline{2-5}
        & Leader & Client & Leader & Client & & \\
        \hline
      ~\cite{KS-CRYPTO-2005} & $m^2 n^3$ pub & $m^2 n^3$ pub & $\lambda m^3 n^2$ &$ \lambda m^3 n^2$ & $m$ &$\cmark$\\
        \hline
      ~\cite{Frikken-ACNS-2007} & $m n^2$ pub & $m n^2$ pub & $\lambda m n$ & $\lambda m n$ & $m$ &$\cmark$\\
        \hline
      ~\cite{VCE22} & $l m^2 n$ pub & $l m n$ pub & $\lambda l m^2 n$ & $\lambda l m n$ & $l$ &$\cmark$\\
        \hline
      ~\cite{BA-ASIACCS-2012} & \multicolumn{2}{c|}{$\sigma m n \log n + m^2$ sym} & \multicolumn{2}{c|}{$\sigma^2 m n \log n + \sigma m^2$} & $\log m$ &$\cmark$\\
        \hline
        \cite{GNT-eprint-2023} & \multicolumn{2}{c|}{$m n (\log n / \log \log n)$ pub} & \multicolumn{2}{c|}{$(\gamma + \lambda) m n (\log n / \log \log n)$} & $\log \gamma + m$ &$\xmark$\\
        \hline
        \cite{LG-ASIACRYPT-2023} & $(T + l + m) m n$ sym & $(T + l) m n$ sym & $(T + l) m n + l m^2 n$ & $(T + l) m n$ & $\log (l - \log n) + m$ &$\xmark$\\
        \hline
        Our SK-MPSU& $m^2 n$ sym & $m^2 n$ sym & $\gamma m n + l m^2 n$ & $(\gamma + l+ m) m n$ & $\log \gamma + m$ &$\cmark$\\
        \hline
        Our PK-MPSU& $m n$ pub & $m n$ pub & $(\gamma + \lambda) m n$ & $(\gamma + \lambda) m n$ & $\log \gamma + m$&$\cmark$\\
        \hline
  \end{tabular}}
  \caption{Asymptotic communication (bits) and computation costs of MPSU protocols in the semi-honest setting. For the sake of comparison, we omit the Big $O$ notations and simplify the complexity by retaining only the dominant terms. We use $\cmark$ to denote protocols in the standard semi-honest model and $\xmark$ to denote protocols requiring non-collusion assumption. pub: public-key operations; sym: symmetric cryptographic operations.
  $n$ is the set size. $m$ is the number of parties. $\lambda$ and $\sigma$ are computational and statistical security parameter. $T$ is the number of AND gate in a SKE decryption circuit in \cite{LG-ASIACRYPT-2023}. $l$ is the bit length of input elements. $\gamma$ is the output length of OPPRF. In the typical setting, $n \le 2^{24}$, $m \le 32$, $\lambda = 128$, $\sigma = 40$, $T \approx 600$, $l \le 128$, $\gamma \le 64$. }
  \label{tab:comparisons}
\end{table*}

\section{Technical Overview}\label{technical} 

\subsection{LG Revisit}

We start by abstracting the high-level idea of LG. For the sake of simplicity, we focus here on the case of three parties $P_1, P_2, P_3$, with respective inputs $X_1, X_2, X_3$. We designate $P_1$ as the leader. Since $P_1$ already holds $X_1$, it needs to obtain the set difference $Y_1 = (X_1 \cup X_2 \cup X_3) \setminus X_1$ from $P_2$ and $P_3$. 

LG can be summarized as a secret-sharing based MPSU framework with two phases: The first phase involves two secret-sharing processes, where $P_2$ somehow secret-shares $Y_2 = X_2 \setminus X_1$ and $P_3$ somehow secret-shares $Y_3 = X_3 \setminus (X_1 \cup X_2)$ among all parties. Since $\{Y_2,Y_3\}$ is a partition of $Y_1$, as a result, the parties secret-share $Y_1$ in the order of $Y_2, Y_3$. The second phase is to reconstruct $Y_1$ to $P_1$. In this case, a straightforward approach is insufficient because $P_1$ can identify which party each element $x \in Y_1$ originates from. The solution is to let the parties invoke multi-party secret-shared shuffle to randomly permute and re-share all shares. This shuffle ensures that any two parties have no knowledge of the permutation, thereby concealing the correspondence between the secret shares and individual difference sets $Y_2,Y_3$. Afterwards, $P_2$ and $P_3$ can send their shuffled shares to $P_1$, who then reconstructs $Y_1$.

LG utilizes two ingredients to realize the secret-sharing processes of the framework: (1) The secret-shared private membership test (ssPMT) \cite{CO-SCN-2018,ZMSJYX-WASA-2021}, where the sender $\Sd$ inputs a set $X$, and the receiver $\Rcv$ inputs an element $y$. If $y \in X$, $\Sd$ and $\Rcv$ receive secret shares of 1, otherwise secret shares of 0. Liu and Gao proposed a multi-query ssPMT (mq-ssPMT), which supports the receiver querying multiple elements' memberships of the sender's set simultaneously. Namely, $\Sd$ inputs $X$, and $\Rcv$ inputs $y_1,\cdots,y_n$. $\Sd$ and $\Rcv$ receive secret shares of a bit vector of size $n$, where if $y_i \in X$, the $i$th bit is 1, otherwise 0. (2) A two-choice-bit version of random oblivious transfer (ROT), where the sender $\Sd$ and the receiver $\Rcv$ each holds a choice bit $e_0, e_1$. $\Sd$ receives two random messages $r_0, r_1$. If $e_0 \oplus e_1 = 0$, $\Rcv$ receives $r_0$, otherwise $r_1$\footnote{This special ROT is identical to the standard 1-out-of-2 ROT, where $e_0$ is determined by $\Sd$ to indicate whether to swap the order of $r_0$ and $r_1$.}.
The following is to elaborate the concrete constructions of the two processes.

The process for $P_2$ to secret-share $Y_2$: $P_2$ acts as $\Rcv$ and executes the mq-ssPMT with $P_1$. For each item $x \in X_2$, $P_2$ and $P_1$ receive shares $e_{2,1}$ and $e_{1,2}$. If $x \in X_1$, $e_{2,1} \oplus e_{1,2} = 1$, otherwise $e_{2,1} \oplus e_{1,2} = 0$. Then $P_2$ acts as $\Sd$ and executes the two-choice-bit ROT with $P_1$. $P_2$ and $P_1$ each inputs $e_{2,1}, e_{1,2}$. $P_2$ receives $r_{2,1}^0, r_{2,1}^1$. If $e_{2,1} \oplus e_{1,2} = 0$, $P_1$ receives $r_{2,1}^0$, otherwise $P_1$ receives $r_{2,1}^1$. $P_1$ sets the output as its share $s_{2,1}$. $P_2$ sets its share $s_{2,2}$ to be $r_{2,1}^0 \oplus x \Vert \Hash(x)$\footnote{Suppose a lack of specific structure for set elements, then $P_1$ cannot distinguish a set element $x$ and a random value $r$ in the reconstruction. To address this, the parties append the hash value when secret sharing an element.
}. $P_3$ sets $s_{2,3}$ to 0. If $x \notin X_1$, $e_{2,1} \oplus e_{1,2} = 0$, $s_{2,1} \oplus s_{2,2} \oplus s_{2,3} = x \Vert \Hash(x)$. Otherwise, $e_{2,1} \oplus e_{1,2} = 1$, $s_{2,1} \oplus s_{2,2} \oplus s_{2,3}= r_{2,1}^1 \oplus r_{2,1}^0 \oplus x \Vert \Hash(x)$ is uniformly random. That is, $Y_2$ is secret-shared among all parties\footnote{After the re-sharing of multi-party secret-shared shuffle.}, and the elements in $X_1 \cap X_2$ are masked by random values. 

The process for $P_3$ to secret-share $Y_3$: $P_3$ acts as $\Rcv$ and executes the mq-ssPMT with $P_1$ and $P_2$ separately. For each $x \in X_3$, $P_3$ and $P_1$ receive shares $e_{3,1}$ and $e_{1,3}$, while $P_3$ and $P_2$ receive shares $e_{3,2}$ and $e_{2,3}$. Then $P_3$ acts as $\Sd$ and executes the two-choice-bit ROT with $P_1$ and $P_2$. In the ROT between $P_3$ and $P_1$, $P_3$ inputs $e_{3,1}$ and $P_1$ inputs $e_{1,3}$. $P_3$ receives $r_{3,1}^0, r_{3,1}^1$. If $e_{3,1} \oplus e_{1,3} = 0$, $P_1$ receives $r_{3,1}^0$, otherwise $P_1$ receives $r_{3,1}^1$. $P_1$ sets the output as its share $s_{3,1}$. In the ROT between $P_3$ and $P_2$, $P_3$ inputs $e_{3,2}$ and $P_2$ inputs $e_{2,3}$. $P_3$ receives $r_{3,2}^0, r_{3,2}^1$. If $e_{3,2} \oplus e_{2,3} = 0$, $P_2$ receives $r_{3,2}^0$, otherwise $P_2$ receives $r_{3,2}^1$. $P_2$ sets the output as its share $s_{3,2}$. $P_3$ sets its share $s_{3,3}$ to be $r_{3,1}^0 \oplus r_{3,2}^0 \oplus x \Vert \Hash(x)$. If $x \notin X_1$ and $x \notin X_2$, $e_{3,1} \oplus e_{1,3} = 0$ and $e_{3,2} \oplus e_{2,3} = 0$, $s_{3,1} \oplus s_{3,2} \oplus s_{3,3} =  x \Vert \Hash(x)$. 
Otherwise, there is at least one random value ($r_{3,1}^1$ or $r_{3,2}^1$) cannot be canceled out from the summation that makes it random. 

The above outlines how LG works. The protocol has two main drawbacks: First, the core component, mq-ssPMT heavily relies on expensive general MPC machinery, which becomes the bottleneck of the entire protocol. Second, LG fails to achieve security against arbitrary collusion. The next two sections are devoted to addressing these two drawbacks.

\subsection{Efficient Batch ssPMT}

To improve the efficiency of mq-ssPMT, we introduce a new functionality called batch ssPMT: The sender $\Sd$ inputs $n$ disjoint sets $X_1,\cdots,X_n$, and the receiver $\Rcv$ inputs $n$ elements $y_1,\cdots,y_n$. $\Sd$ and $\Rcv$ receive secret shares of a bit vector of size $n$, where the $i$th bit is 1 if $y_i \in X_i$, and 0 otherwise. 

Compared to mq-ssPMT, batch ssPMT enables the testing of element membership across multiple distinct sets rather than within a single common set. Their relationship is analogous to that between batched oblivious pseudorandom function (batch OPRF) \cite{KKRT-CCS-2016} and multi-point oblivious pseudorandom function (multi-point OPRF) \cite{PRTY-CRYPTO-2019,CM-CRYPTO-2020}. Thanks to its batching nature, batch ssPMT admits a much more efficient construction, 
lending it a superior alternative to mq-ssPMT in the context of MPSU, by coupling with hashing to bins.

The alternative to mq-ssPMT works as follows: First, $\Sd$ and $\Rcv$ preprocess inputs through hashing to bins technique. $\Rcv$ uses hash functions $h_1, h_2, h_3$ to assign $y_1, \cdots, y_n$ to $B$ bins via Cuckoo hashing~\cite{PR04}, ensuring that each bin accommodates at most one element. At the same time, $\Sd$ assigns each $x \in X$ to all bins determined by $h_1(x), h_2(x), h_3(x)$. Then they invoke $B$ instances of ssPMT, where in the $j$th instance, $\Sd$ inputs the subset $X_j \in X$ containing all elements mapped into its bin $j$, while $\Rcv$ inputs the sole element mapped into its bin $j$. If some $y_i$ is mapped to bin $j$ and $y_i \in X$, then $\Sd$ certainly maps $y_i$ to bin $j$ (and other bins) as well, ensuring $y_{i} \in X_j$. Therefore, for each bin $j$ of $\Rcv$, if the inside element $y_{i}$ is in $X$, $y_{i} \in X_j$, $\Sd$ and $\Rcv$ receive shares of 1 from the $i$th instance of batch ssPMT. Otherwise, $y_{i} \notin X_j$, $\Sd$ and $\Rcv$ receive shares of 0.

The functionality realized by the above construction differs slightly from mq-ssPMT, as the query sequence of $\Rcv$ is determined by the Cuckoo hash positions of its input elements. Since the Cuckoo hash positions depends on the entire input set of $\Rcv$, and all shares is arranged according to the parties' Cuckoo hash positions, a straightforward reconstruction may leak information about the parties' input sets to $P_1$. Therefore, we have to eliminate the dependence of shares' order on the parties' Cuckoo hash positions during the reconstruction phase. Our crucial insight is that the multi-party secret-shared shuffle used in the reconstruction phase eliminates not only the correspondence between secret shares and difference sets, but also this dependence on the hash positions. As a result, our construction can plug in the secret-sharing based framework seamlessly, without introducing any information leakage or additional overhead.

Unlike the mq-ssPMT, which heavily relies on general 2PC, our batch ssPMT is built on two highly efficient primitives, batched oblivious programmable pseudorandom function (batch OPPRF)~\cite{KMPRT-CCS-2017,PSTY-EUROCRYPT-2019} and secret-shared private equality test (ssPEQT)~\cite{PSTY-EUROCRYPT-2019,CGS22}, where the former has an extremely fast specialized instantiation through vector oblivious linear evaluation (VOLE), while the later is composed of a small circuit, allowing for efficient implementation with general 2PC. Therefore, our batch ssPMT greatly reduces dependency on general 2PC, leading to a substantial performance improvement. For more details, refer to Appendix \ref{compare:lg}.

\subsection{SK-MPSU from Batch ssPMT and mss-ROT}
By replacing mq-ssPMT with the construction in the previous section, we improve LG's performance significantly. However, the resulting protocol still relies on the non-collusion assumption. We proceed to show how to eliminate it. 

First, we explain why LG requires the non-collusion assumption using the three-party example: After the secret-sharing process of $P_3$, all parties hold the secret shares of
\begin{equation*}
    \begin{cases}
      x \Vert \Hash(x) & x \in X_3 \setminus (X_1 \cup X_2)\\
      r_{3,1}^0 \oplus r_{3,1}^1 \oplus x \Vert \Hash(x) & x \in (X_2 \cup X_3) \setminus X_1\\
      r_{3,2}^0 \oplus r_{3,2}^1 \oplus x \Vert \Hash(x) & x \in (X_1 \cup X_3) \setminus X_2\\
      r_{3,1}^1 \oplus r_{3,2}^1 \oplus r_{3,1}^0 \oplus r_{3,2}^0 \oplus x \Vert \Hash(x) & x \in X_1 \cap X_2 \cap X_3
    \end{cases}       
\end{equation*}

If $x \in Y_3$, $P_1$ reconstructs the secret $x \Vert \Hash(x)$. If $x \notin Y_3$, the secret is random so that $P_1$ gains no information about $x$. Nevertheless, this is only guaranteed when $P_1$ does not collude with $P_3$. Recall that $P_3$ receives $r_{3,1}^0, r_{3,1}^1$ and $r_{3,2}^0, r_{3,2}^1$ from the ROT execution with $P_1$ and $P_2$ respectively, so the coalition of $P_1$ and $P_3$ can easily check the secret, and distinguish the distinct cases, which reveals information of $P_2$'s inputs.

To fix this security issue, we generalize ROT into multi-party setting, which we call multi-party secret-shared random oblivious transfer (mss-ROT). Consider $d$ involved parties, with two parties, denoted as $P_{\text{ch}_0}$ and $P_{\text{ch}_1}$, possessing choice bits $b_0$ and $b_1$ respectively. There is a subset $J$ of $ \{1, \cdots, d\}$ so that each party $P_j$ ($j \in J$) holds $\Delta_j$ as input. The mss-ROT functionality outputs a random $r_i$ to each party $P_i$, s.t. if $b_0 \oplus b_1 = 0$, $\bigoplus_{i=1}^d r_i = 0$. Otherwise, $\bigoplus_{i=1}^d r_i = \bigoplus_{j \in J} \Delta_j$. 

Equipped with this new primitive, the secret-sharing process of $P_3$ is revised as follows: After $P_3$ and $P_1$ receiving shares $e_{3,1}$ and $e_{1,3}$, $\{P_1,P_2,P_3\}$ invoke mss-ROT, where $P_3$ and $P_1$ act as $P_{\text{ch}_0}$ and $P_{\text{ch}_1}$ holding $e_{3,1}$ and $e_{1,3}$, while $P_2$ and $P_3$ samples $\Delta_{2,31},\Delta_{3,31}$ uniformly as inputs. $P_1,P_2,P_3$ receives $r_{1,31}, r_{2,31}, r_{3,31}$ separately. If $e_{3,1} \oplus e_{1,3} = 0$, $r_{1,31} \oplus r_{2,31} \oplus r_{3,31} = 0$, otherwise $r_{1,31} \oplus r_{2,31} \oplus r_{3,31} = \Delta_{2,31} \oplus \Delta_{3,31}$. After $P_3$ and $P_2$ receiving shares $e_{3,2}$ and $e_{2,3}$, $\{P_2,P_3\}$ invoke mss-ROT, where $P_3$ and $P_2$ hold choice bits $e_{3,2}$ and $e_{2,3}$, while $P_2$ and $P_3$ samples $\Delta_{2,32},\Delta_{3,32}$ uniformly as inputs. $P_2$ and $P_3$ receives $r_{2,32}$ and $r_{3,32}$ separately. If $e_{3,2} \oplus e_{2,3} = 0$, $r_{2,32} \oplus r_{3,32} = 0$, otherwise $r_{2,32} \oplus r_{3,32} = \Delta_{2,32} \oplus \Delta_{3,32}$. Finally, $P_1$ sets $s_{3,1} = r_{1,31}$. $P_2$ sets $s_{3,2} = r_{2,31} \oplus r_{2,32}$. $P_3$ sets $s_{3,3} = r_{3,31} \oplus r_{3,32} \oplus x \Vert \Hash(x)$. 

We conclude that all parties hold the secret shares of 
\begin{equation*}
    \begin{cases}
      x \Vert \Hash(x) & x \in X_3 \setminus (X_1 \cup X_2)\\
      \Delta_{2,31} \oplus \Delta_{3,31} \oplus x \Vert \Hash(x) & x \in (X_2 \cup X_3) \setminus X_1\\
      \Delta_{2,32} \oplus \Delta_{3,32} \oplus x \Vert \Hash(x) & x \in (X_1 \cup X_3) \setminus X_2\\
      \Delta_{2,31} \oplus \Delta_{3,31} \oplus \Delta_{2,32} \oplus \Delta_{3,32} \oplus x \Vert \Hash(x) & x \in X_1 \cap X_2 \cap X_3
    \end{cases}       
\end{equation*}
It is immediate that if $x \notin Y_3$, the secret is random in the view of the coalition of $\{P_1,P_2\}$ or $\{P_1,P_3\}$. Since the coalition of $\{P_2,P_3\}$ cannot know the secret, the protocol achieves security under arbitrary collusion.

\subsection{PK-MPSU from Batch ssPMT and MKR-PKE}
There are no existing MPSU protocols with linear computation and communication complexity. Notably, it is impossible to achieve this goal within the secret-sharing based framework, given that $P_1$ must reconstruct $(m-1)n$ secrets with each secret comprising $m$ shares. Therefore, we have to seek for another technical route.
 
Our start point is the existing work~\cite{GNT-eprint-2023} with the best asymptotic complexity $m n (\log n / \log \log n)$. We distill their protocol into a framework with two phases: The first phase allows $P_1$ to somehow obtain encrypted $X_i \setminus (X_1 \cup \cdots \cup X_{i-1})$ for $2 \le i \le m$, interspersed with encrypted dummy messages filling the positions of duplicate elements. However, if $P_1$ decrypts ciphertexts by itself, then it could associate each element in $X_1 \cup \cdots \cup X_m \setminus X_1$ with the specific party it originates from. This problem is addressed by the second phase, which is the collaborative decryption and shuffle procedure, making use of a PKE variant MKR-PKE: $P_1$ sends the ciphertexts to $P_2$ after permuting them using a random permutation $\pi_1$. $P_2$ performs a partial decryption on the received ciphertexts, rerandomizes them, permutes them with a random permutation $\pi_2$, and forwards the ciphertexts to $P_3$. This iterative process continues until the last party, $P_m$, returns its permuted partially decrypted ciphertexts to $P_1$. Finally, $P_1$ fully decrypts the ciphertexts, filters out the dummy elements, retains the set $X_1 \cup \cdots \cup X_m \setminus X_1$, and appends $X_1$ to compute the union.

We identify that both the non-linear complexities and insecurity against arbitrary collusion of~\cite{GNT-eprint-2023} arise from their construction in the first phase. Therefore, the task of building a linear-complexity MPSU reduces to devising a linear-complexity and secure construction for this phase.

A rough idea is as follows: Let $P_i$ hold a set of encrypted elements in $X_i$. For $2 \le i \le m$, each $P_i$ engage in a procedure with each $P_j$ ($1 < j < i$) such that $P_j$ can ``obliviously'' replace the encrypted elements in $X_i \cap X_j$ with encrypted dummy messages. After the procedure, $P_i$ holds a set of encrypted elements in $X_i \setminus (X_2 \cup \cdots \cup X_{i-1})$. The replacement of encrypted elements in $X_1$ is handled at the end.

The replacing procedure between each $P_i$ and $P_j$: (1) $P_i$ acts as $\Rcv$ and executes batch ssPMT (after preprocessing their inputs using hashing to bins, which we omit here for simplicity) with $P_j$. For each $x \in X_i$, $P_i$ and $P_j$ receive shares $e_{i,j}$ and $e_{j,i}$. (2) If $j =2$, $P_i$ initializes $m_0$ as the encrypted $x$ and $m_1$ as an encrypted dummy message. (3) $P_i$ acts as $\Sd$ and executes two-choice-bit OT\footnote{The difference from two-choice-bit ROT is that $\Sd$ inputs two specific messages $m_0,m_1$ instead of obtaining two uniform messages.} with $P_j$, where $P_i$ and $P_j$ input $e_{i,j}$ and $e_{j,i}$ as choice bits. $P_i$ inputs $m_0, m_1$ and $P_j$ receives $\ct = m_{e_{i,j} \oplus e_{j,i}}$. If $x \notin X_j$, $\ct$ is the ciphertext of $x$; otherwise, the ciphertext of dummy message. (4) $P_j$ rerandomizes $\ct$ to $\ct'$ and returns $\ct'$ to $P_i$. (5) $P_i$ rerandomizes $\ct'$, updates $m_0$ to $\ct'$, and rerandomizes $m_1$ before repeating the procedure with the next party $P_{j+1}$. 
Note that only if all conditions of $x \notin X_j$ are satisfied, the final $m_0$ is an encrypted message of $x$.

For $2 \le i \le m$, each $P_i$ and $P_1$ execute batch ssPMT and receive secret shares as choice bits for the subsequent execution of two-choice-bit OT. For each $x \in X_i$, $P_i$ acts as $\Sd$ and inputs the final $m_0$ of the above procedure and an encrypted dummy message. As a result, if $x \notin X_1$, $P_1$ receives the final $m_0$; otherwise, it receives the encrypted dummy message. This excludes all elements in $X_1$ from the encrypted set $X_i \setminus (X_2 \cup \cdots \cup X_{i-1})$ and lets $P_1$ obtain the results.

The linear complexities of our construction are attributed to two key facts: First, this framework ensures that each interaction between $P_i$ and $P_j$ only involves the input sets of themselves. Second, our batch ssPMT protocol has linear computation and communication complexity (cf. Section~\ref{bssPMT}). 

\section{Preliminaries}\label{sec:preliminaries}

\subsection{Notation}

Let $m$ denote the number of participants. We write $[m]$ to denote a set $\{1,\cdots,m\}$. We use $P_i$ ($i \in [m]$) to denote participants, $X_i$ to represent the sets they hold, where each set has $n$ $l$-bit elements. $x \Vert y$ denotes the concatenation of two strings. We use $\lambda, \sigma$ as the computational and statistical security parameters respectively, and use $\overset{s}{\approx}$ (resp.$\overset{c}{\approx}$) to denote that two distributions are statistically (resp. computationally) indistinguishable. We denote vectors by letters with a right arrow above and $a_i$ denotes the $i$th component of $\vec{a}$. $\vec{a} \oplus \vec{b} = (a_1 \oplus b_1, \cdots, a_n \oplus b_n)$. $\pi(\vec{a}) = (a_{\pi(1)}, \cdots, a_{\pi(n)})$, where $\pi$ is a permutation over $n$ items. $\pi = \pi_1 \circ \cdots \circ \pi_n$ represents applying the permutations $\pi_1, \cdots, \pi_n$ in sequence. $x[i]$ denotes the $i$th bit of element $x$, and $X(i)$ denotes the $i$th element of set $X$. 

\subsection{Multi-party Private Set Union}

The ideal functionality of MPSU is formalized in Figure~\ref{fig:func-mpsu}.

\begin{figure}[!hbtp]
\begin{framed}
\begin{minipage}[center]{\textwidth}
\begin{trivlist}
\item \textbf{Parameters.} $m$ parties $P_1, \cdots, P_m$, where $P_1$ is the leader. Size $n$ of input sets. The bit length $l$ of set elements.
\item \textbf{Functionality.} On input $X_i = \{x_i^1,\cdots, x_i^n\} \subseteq \{0,1\}^l$ from $P_i$ ($i \in [m]$), output $\bigcup_{i=1}^m X_i$ to $P_1$.
\end{trivlist}
\end{minipage}
\end{framed}
\caption{Multi-party Private Set Union Functionality $\FuncMPSU$}\label{fig:func-mpsu}
\end{figure}

\subsection{Batch Oblivious Programmable Pseudorandom Function}\label{sec:bopprf}


Oblivious pseudorandom function (OPRF)~\cite{FIPR-TCC-2005} is a central primitive in the area of PSO. Kolesnikov et al.~\cite{KKRT-CCS-2016} introduced batched OPRF, which provides a batch of OPRF instances. In the $i$th instance, the sender $\Sd$ learns a PRF key $k_i$, while the receiver $\Rcv$ inputs $x_i$ and learns $\PRF(k_i, x_i)$. 

Oblivious programmable pseudorandom function (OPPRF)~\cite{KMPRT-CCS-2017,PSTY-EUROCRYPT-2019,CGS22,RS-EUROCRYPT-2021,RR-CCS-2022} is an extension of OPRF, which lets $\Sd$ program a PRF $F$ so that it has specific uniform outputs for some specific inputs and pseudorandom outputs for all other inputs. This kind of PRF with the additional property that the function outputs programmed values on a certain programmed set of inputs is called programmable PRF (PPRF)~\cite{PSTY-EUROCRYPT-2019}. $\Rcv$ evaluates the OPPRF with no knowledge of whether it learns a programmed output of $F$ or just a pseudorandom value. The batch OPPRF functionality is given in Figure~\ref{fig:func-bopprf}.

\begin{figure}[!hbtp]
\begin{framed}
\begin{minipage}[center]{\textwidth}
\begin{trivlist}
\item \textbf{Parameters.} Sender $\Sd$. Receiver $\Rcv$. Batch size $B$. The bit length $l$ of keys. The bit length $\gamma$ of values.
\item \textbf{Sender's inputs.} $\Sd$ inputs $B$ sets of key-value pairs including:
\begin{itemize}
\item Disjoint key sets $K_1,\cdots,K_B$.
\item The value sets $V_1,\cdots,V_B$, where $\lvert K_i \rvert = \lvert V_i \rvert$, $i \in [B]$.
\end{itemize}
\item \textbf{Receiver's inputs.} $\Rcv$ inputs $B$ queries $\vec{x} \subseteq (\{0,1\}^l)^B$.
\item \textbf{Functionality:} On input $(K_1, \cdots, K_B)$ and $(V_1,\cdots,V_B)$ from $\Sd$ and $\vec{x} \subseteq (\{0,1\}^l)^B$ from $\Rcv$,
\begin{itemize}
\item Choose a uniform PPRF key $k_i$, for $i \in [B]$;
\item Sample a PPRF $F: \{0,1\}^* \times \{0,1\}^l \to \{0,1\}^{\gamma}$ such that $F(k_i,K_i(j)) = V_i(j)$ for $i \in [B], 1 \le j \le \lvert K_i \rvert$;
\item Define $f_i = F(k_i,x_i)$, for $i \in [B]$;
\item Give vector $\vec{f} = (f_1, \cdots, f_B)$ to $\Rcv$.
\end{itemize}
\end{trivlist}
\end{minipage}
\end{framed}
\caption{Batch OPPRF Functionality $\FuncbOPPRF$}\label{fig:func-bopprf}
\end{figure}

\subsection{Hashing to Bins}
The hashing to bins technique was introduced by Pinkas et al.~\cite{PSZ-USENIX-2014,PSZ-USENIX-2015} to construct two-party PSI. At a high level, the receiver $\Rcv$ uses hash functions $h_1, h_2, h_3$ to assign its items to $B$ bins via Cuckoo hashing~\cite{PR04}, so that each bin has at most one item. The Cuckoo hashing process uses eviction and the choice of bins for each item depends on the entire set. On the other hand, the sender $\Sd$ assigns each of its items $x$ to all bins $h_1(x), h_2(x), h_3(x)$ via simple hashing. This guarantees that for each item $x$ of $\Rcv$, if $x$ is mapped into the $b$th bin of Cuckoo hash table ($b \in \{h_1(x), h_2(x), h_3(x)\}$), and $x$ is in $\Sd$'s set, then the $b$th of simple hash table certainly contains $x$.

We denote simple hashing with the following notation:
\begin{gather*}
\mathcal{T}^1, \cdots, \mathcal{T}^B \gets \Simple_{h_1,h_2,h_3}^B(X)
\end{gather*}
This expression represents hashing the items of $X$ into $B$ bins using simple hashing with hash functions $h_1, h_2, h_3: \{0, 1\}^* \to [B]$. The output is a simple hash table denoted by $\mathcal{T}^1, \cdots, \mathcal{T}^B$, where for each $x \in X$ we have $\mathcal{T}^{h_i(x)} \supseteq \{x \Vert i \vert i= 1, 2, 3\}$.\footnote{Appending the index of the hash function is helpful for dealing with edge cases like $h_1(x) = h_2(x) = i$, which happen with non-negligible probability.}

We denote Cuckoo hashing with the following notation:
\begin{gather*}
\mathcal{C}^1, \cdots, \mathcal{C}^B \gets \Cuckoo_{h_1,h_2,h_3}^B(X)
\end{gather*}
This expression represents hashing the items of $X$ into $B$ bins using Cuckoo hashing with hash functions $h_1, h_2, h_3: \{0, 1\}^* \to [B]$. The output is a Cuckoo hash table denoted by $\mathcal{C}^1, \cdots, \mathcal{C}^B$, where for each $x \in X$ there is some $i \in \{1, 2, 3\}$ such that $\mathcal{C}^{h_i(x)} = \{x \Vert i\}$. Some Cuckoo hash positions are irrelevant, corresponding to empty bins. We use these symbols throughout the subsequent sections. 

\subsection{Secret-Shared Private Equality Test}\label{sec:sspeqt}
Secret-shared private equality test protocol (ssPEQT)~\cite{PSTY-EUROCRYPT-2019,CGS22} can be viewed as an extreme case of ssPMT when the sender $\Sd$'s input set size is 1. Figure~\ref{fig:func-sspeqt} defines its functionality.

\begin{figure}[!hbtp]
\begin{framed}
\begin{minipage}[center]{\textwidth}
\begin{trivlist}
\item \textbf{Parameters.} Two parties $P_1, P_2$. The input bit length $\gamma$.
\item \textbf{Functionality.} On input $x$ from $P_1$ and input $y$ from $P_2$, sample two random bits $a, b$ 
s.t. if $x = y$, $a \oplus b = 1$. Otherwise $a \oplus b = 0$. Give $a$ to $P_1$ and $b$ to $P_2$.
\end{trivlist}
\end{minipage}
\end{framed}
\caption{Secret-Shared Private Equality Test Functionality $\FuncssPEQT$}\label{fig:func-sspeqt}
\end{figure}

\subsection{Random Oblivious Transfer}
Oblivious transfer (OT)~\cite{Rabin05} is a foundational primitive in MPC, the functionality of 1-out-of-2 random OT (ROT) is given in Figure~\ref{fig:func-rot}.

\begin{figure}[!hbtp]
\begin{framed}
\begin{minipage}[center]{\textwidth}
\begin{trivlist}
\item \textbf{Parameters.} Sender $\Sd$, Receiver $\Rcv$. The message length $l$.
\item \textbf{Functionality.} On input $b \in \{0,1\}$ from $\Rcv$, sample $r_0, r_1 \gets \{0, 1\}^l$. Give $(r_0, r_1)$ to $\Sd$ and give $r_b$ to $\Rcv$.
\end{trivlist}
\end{minipage}
\end{framed}
\caption{1-out-of-2 Random OT Functionality $\FuncROT$}\label{fig:func-rot}
\end{figure}

\subsection{Multi-Party Secret-Shared Shuffle}\label{sec:ms}

Multi-party secret-shared shuffle functionality works by randomly permuting the share vectors of all parties and then refreshing all shares, ensuring that the permutation remains unknown to any coalition of $m-1$ parties. The formal functionality is given in Figure \ref{fig:func-ms}. 

Eskandarian et al.~\cite{EB22} proposed an online-efficient protocol by extending~\cite{CGP-ASIACRYPT-2020} to the multi-party setting. In the offline phase, each party generates a random permutation and a set of correlated vectors
called share correlations. In the online phase, each party efficiently permutes and refreshes the share vectors using these share correlations. 

\begin{figure}[!hbtp]
\begin{framed}
\begin{minipage}[center]{\textwidth}
\begin{trivlist}
\item \textbf{Parameters.} $m$ parties $P_1, \cdots P_m$. The dimension of vector $n$. The item length $l$.
\item \textbf{Functionality.} On input $\vec{x}_i = {(x_i^1, \cdots, x_i^n)}$ from each $P_i$, sample a random permutation $\pi : [n] \to [n]$. For $1 \le i \le m$, sample $\vec{{x}_{i}'} \gets (\{0,1\}^l)^n$ satisfying $\bigoplus_{i=1}^m \vec{{x}_{i}'} = \pi(\bigoplus_{i=1}^m \vec{{x}_i})$. Give $\vec{{x}_{i}'}$ to $P_i$.
\end{trivlist}
\end{minipage}
\end{framed}
\caption{Multi-Party Secret-Shared Shuffle Functionality $\FuncMS$}\label{fig:func-ms}
\end{figure}

Gao et al.~\cite{GNT-eprint-2023} introduced the concept of multi-key rerandomizable public-key encryption (MKR-PKE), a variant of PKE with several additional properties. Let $\mathcal{SK}$ denote the space of secret keys, which forms an abelian group under the operation $+$, and $\mathcal{PK}$ denote the space of public keys, which forms an abelian group under the operation $\cdot$. $\mathcal{M}$ denotes the space of plaintexts, and $\mathcal{C}$ denotes the space of ciphertexts. MKR-PKE is a tuple of PPT algorithms $(\Gen, \Enc, \ParDec, \Dec, \ReRand)$ such that:
\begin{itemize}
\item The key-generation algorithm $\Gen$ takes as input a security parameter $1^\lambda$ and outputs a pair of keys $(pk, sk) \in \mathcal{PK} \times \mathcal{SK}$. 
\item The encryption algorithm $\Enc$ takes as input a public key $pk \in \mathcal{PK}$ and a plaintext message $x \in \mathcal{M}$, and outputs a ciphertext $\ct \in \mathcal{C}$. 
\item The partial decryption algorithm $\ParDec$ takes as input a secret key share $sk \in \mathcal{SK}$ and a ciphertext $\ct \in \mathcal{C}$, and outputs another ciphertext $\ct' \in \mathcal{C}$. 
\item The decryption algorithm $\Dec$ takes as input a secret key $sk \in \mathcal{SK}$ and a ciphertext $\ct \in \mathcal{C}$, outputs a message $x \in \mathcal{M}$ or an error symbol $\perp$. 
\item The rerandomization algorithm $\ReRand$ takes as input a public key $pk \in \mathcal{PK}$ and a ciphertext $\ct \in \mathcal{C}$, outputs another ciphertext $\ct' \in \mathcal{C}$.
\end{itemize}

MKR-PKE is an IND-CPA secure PKE scheme that requires the following additional properties: 

\begin{trivlist}
\item \textbf{Partially Decryptable.} For any two pairs of keys $(sk_1, pk_1) \gets \Gen(1^\lambda), (sk_2, pk_2) \gets \Gen(1^\lambda)$ and any $x \in \mathcal{M}$, 
\begin{gather*}
\ParDec(sk_1, \Enc(pk_1 \cdot pk_2,x)) = \Enc(pk_2,x)
\end{gather*}

\item \textbf{Rerandomizable.} For any $pk \in \mathcal{PK}$ and any $x \in \mathcal{M}$, $$\ReRand(pk,\Enc(pk,x)) \overset{s}{\approx} \Enc(pk,x)$$

Gao et al. \cite{GNT-eprint-2023} make use of elliptic curve (EC) based ElGamal encryption~\cite{ElGamal-IEEE-IT-1985} to instantiate MKR-PKE. 
The concrete EC MKR-PKE is described in Appendix~\ref{sec:elgamal}. 
\begin{remark}
To the best of our knowledge, there is few elliptic curves that support efficient encoding and decoding between bit-strings and EC points.
Therefore, the plaintext space of EC MKR-PKE is generally restricted to EC points to guarantee rerandomizable property. 
\end{remark}
\end{trivlist}

\section{Batch Secret-Shared Private Membership Test}\label{bssPMT}

The batch secret-shared private membership test (batch ssPMT) is a central building block in our SK-MPSU and PK-MPSU protocols. It is a two-party protocol that implements multiple instances of ssPMT between a sender $\Sd$ and a receiver $\Rcv$. Given a batch size of $B$, $\Sd$ inputs $B$ sets $X_1,\cdots,X_B$, while $\Rcv$ inputs $B$ elements $x_1,\cdots,x_B$. As a result, $\Sd$ and $\Rcv$ receive secret shares of a bit vector of size $B$, where the $i$th bit is 1 if $x_i \in X_i$, 0 otherwise. The batch ssPMT functionality is presented in Figure~\ref{fig:func-bssPMT} and the construction is given in Figure~\ref{fig:proto-bssPMT}, built with batch OPPRF and ssPEQT.

\begin{figure}[!hbtp]
\begin{framed}
\begin{minipage}[center]{\textwidth}
\begin{trivlist}
\item \textbf{Parameters.} Sender $\Sd$. Receiver $\Rcv$. Batch size $B$. The bit length $l$ of set elements. The output length $\gamma$ of OPPRF.
\item \textbf{Inputs.} $\Sd$ inputs $B$ disjoint sets $X_1, \cdots, X_B$ and $\Rcv$ inputs $\vec{x} \subseteq (\{0,1\}^l)^B$.
\item \textbf{Functionality.} On inputs $X_1, \cdots, X_B$ from $\Sd$ and input $\vec{x}$ from $\Rcv$, for each $i \in [B]$, sample two random bits $e_S^i, e_R^i$ s.t. if $x_i \in X_i,e_S^i \oplus e_R^i = 1$, otherwise $e_S^i \oplus e_R^i = 0$. Give $\vec{e}_S = (e_S^1, \cdots, e_S^B)$ to $\Sd$ and $\vec{e}_R = (e_R^1, \cdots, e_R^B)$ to $\Rcv$.
\end{trivlist}
\end{minipage}
\end{framed}
\caption{Batch ssPMT Functionality $\FuncbssPMT$}\label{fig:func-bssPMT}
\end{figure}

\begin{figure}[!hbtp]
\begin{framed}
\begin{minipage}[center]{\textwidth}
\begin{trivlist}
\item \textbf{Parameters.} Sender $\Sd$. Receiver $\Rcv$. Batch size $B$. The bit length $l$ of set elements. The output length $\gamma$ of OPPRF.
\item \textbf{Inputs.} $\Sd$ inputs $B$ disjoint sets $X_1, \cdots, X_B$ and $\Rcv$ inputs $\vec{x} \subseteq (\{0,1\}^l)^B$.
\item \textbf{Protocol.}
\begin{enumerate}
\item For each $i \in [B]$, $\Sd$ chooses random $s_i \gets \{0,1\}^\gamma$ and computes a multiset $S_i$ comprising $\lvert X_i \rvert$ repeated elements that all equal to $s_i$.
\item The parties invoke $\FuncbOPPRF$ of batch size $B$. $\Sd$ acts as sender and inputs $X_1, \cdots, X_B$ as key sets and $S_1, \cdots, S_B$ as value sets. $\Rcv$ acts as receiver with input $\vec{x}$, and receives a vector $\vec{t} = (t_1,\cdots, t_B)$.
\item The parties invoke $B$ instances of $\FuncssPEQT$, where in the $i$th instance $\Sd$ inputs $s_i$ and $\Rcv$ inputs $t_i$. In the end, $\Sd$ and $\Rcv$ receives $e_S^i, e_R^i \in \{0,1\}$ respectively.
\end{enumerate}
\end{trivlist}
\end{minipage}
\end{framed}
\caption{Batch ssPMT $\ProtobssPMT$}\label{fig:proto-bssPMT}
\end{figure}

\begin{theorem} \label{theorem:bssPMT}
Protocol $\ProtobssPMT$ securely realizes $\FuncbssPMT$ in the $(\FuncbOPPRF,\FuncssPEQT)$-hybrid model.
\end{theorem}  

\begin{trivlist}
\item \textbf{Correctness.} According to the batch OPPRF functionality, if $x_i \in X_i$, $\Rcv$ receives $t_i = s_i$. Then in the $i$th instance of ssPEQT, since $s_i = t_i$, $e_S^i \oplus e_R^i = 1$. Conversely, if $x_i \notin X_i$, $\Rcv$ receives a pseudorandom value $t_i$. The probability that any $t_i = s_i$ for $i \in [B]$ is $B \cdot 2^{-\gamma}$. By setting $\gamma \ge \sigma + \log B$, we have $B \cdot 2^{-\gamma} \le 2^{-\sigma}$, meaning the probability of any such collision is negligible\footnote{In MPSU, the batch OPPRF is invoked multiple times, and the lower bound of $\gamma$ is relevant to the total number of batch OPPRF invocations. Refer to ~\ref{sec:analysis-bsspmt} for more details.}. After the invocation of ssPEQT, we have that if $x_i \notin X_i$, $e_S^i \oplus e_R^i = 0$ with overwhelming probability.

\item \textbf{Security.} The security of the protocol follows immediately from the security of batch OPPRF and ssPEQT functionalities.
\end{trivlist}

In our MPSU protocols, the batch ssPMT is always combined with the hashing to bins technique. The combined construction has good efficiency, together with linear computation and communication in term of $n$, which mainly benefits from the following technical advances: First, we follow the paradigm in~\cite{PSTY-EUROCRYPT-2019} to construct batch OPPRF from batch OPRF and oblivious key-value store (OKVS)~\cite{PRTY20,GPRTY-CRYPTO-2021,RR-CCS-2022,BPSY-USENIX-2023}. By leveraging the technique to amortize communication, the
total communication of computing $B = O(n)$ instances of OPPRF is equal to the total number of items $3n$. Second, we utilize subfield vector oblivious linear evaluation (subfield-VOLE)~\cite{BCGIKRS19,BCGIKS19,RRT23} to instantiate batch OPRF and the construction in~\cite{RR-CCS-2022} to instantiate OKVS. This ensures the computation complexity of batch OPPRF of size $O(n)$ to scale linearly with $n$.
A comprehensive analysis is in Appendix~\ref{sec:analysis-bsspmt}. 




\section{MPSU from Symmetric-Key Techniques}\label{sec:protocol2}

In this section, we introduce a new primitive called multi-party secret-shared random oblivious transfer (mss-ROT), then we utilize it to build SK-MPSU based on oblivious transfer and symmetric-key operations in the standard semi-honest model. 

\subsection{Multi-Party Secret-Shared Random Oblivious Transfer}\label{sec:mss-rot}


In Figure~\ref{fig:func-rot}, the ROT functionality can be interpreted as $r_0 \oplus r_b = b \cdot \Delta$, where $\Delta = r_0 \oplus r_1$. As the aforementioned two-choice-bit ROT, it can be consider to secret-share the choice bit $b$ between two parties and interpreted as $r_0 \oplus r_b = (b_0 \oplus b_1) \cdot \Delta$. We further extend this secret-sharing idea to multiple parties, allowing not only any non-empty subset of the involved parties to secret-share $b$, but also any non-empty subset of the involved parties to secret-share $\Delta$.

To build our SK-MPSU, we only need two-choice-bit version of mss-ROT. For simplicity, we will refer to this as mss-ROT hereafter. The formal functionality\footnote{We choose to let the parties input the shares of $\Delta$ instead of outputting them, as the former functionality simply implies the latter.} is given in Figure~\ref{fig:func-mssrot} and the detailed construction is given in Figure~\ref{fig:proto-mssrot}.

\begin{theorem}\label{theorem:mssrot}
Protocol $\ProtomssROT$ securely implements $\FuncmssROT$ in the presence of any semi-honest adversary corrupting $t < m$ parties in the $\FuncROT$-hybrid model. 
\end{theorem}

It is easy to see that our construction essentially boils down to performing ROT pairwise. As one of the benefits, we can utilize the derandomization technique \cite{Beaver-CRYPTO-1991} to bring most tasks forward to the offline phase. And the correctness and security of the mss-ROT protocol stems from the correctness and security of ROT. For the complete proof, refer to Appendix~\ref{proof:mssrot}.

\subsection{Construction of Our SK-MPSU}

We now turn our attention to construct a SK-MPSU. The construction follows the high-level ideas we introduced in the technical overview and is formally presented in Figure~\ref{fig:proto-mpsu2}.

\begin{theorem} \label{theorem:ske}
Protocol $\ProtoMPSUTwo$ securely implements $\FuncMPSU$ against any semi-honest adversary corrupting $t < m$ parties in the $(\FuncbssPMT,\FuncmssROT, \FuncMS)$-hybrid model.
\end{theorem} 

The proof of Theorem \ref{theorem:ske} is in Appendix~\ref{proof:ske}. We provide a comprehensive complexity analysis for our SK-MPSU / LG and an exhaustive comparison between them in Appendix~\ref{sec:analysis}. 





\begin{figure}[!hbtp]
\begin{framed}
\begin{minipage}[center]{\textwidth}
\begin{trivlist}
\item \textbf{Parameters.} $m$ parties $P_1, \cdots, P_m$, where $P_{\text{ch}_0}$ and $P_{\text{ch}_1}$ provide inputs as shares of the choice bit, $\text{ch}_0, \text{ch}_1 \in [m]$. We use $J = \{j_1, \cdots, j_d\}$ to denote the set of indices for the parties who provide inputs as shares of $\Delta$, where $d \le m$. The message length $l$.
\item \textbf{Functionality.} On input $b_0 \in \{0, 1\}$ from $P_{\text{ch}_0}$, $b_1 \in \{0, 1\}$ from $P_{\text{ch}_1}$, and $\Delta_j$ from each $P_j$ where $j \in J$,
\begin{itemize} 
\item Sample $r_i \gets \{0, 1\}^l$ and give $r_i$ to $P_i$ for $2 \le i \le m$.
\item If $b_0 \oplus b_1 = 0$, compute $r_1 = \bigoplus_{i=2}^m r_i$, else compute $r_1 = \bigoplus_{i=2}^m r_i \oplus (\bigoplus_{j \in J} \Delta_j)$. Give $r_1$ to $P_1$.
\end{itemize}
\end{trivlist}
\end{minipage}
\end{framed}
\caption{Multi-Party Secret-Shared Random OT Functionality $\FuncmssROT$}\label{fig:func-mssrot}
\end{figure}

\begin{figure}[!hbtp]
\begin{framed}
\begin{minipage}[center]{\textwidth}
\begin{trivlist}
\item \textbf{Parameters.} $m$ parties $P_1, \cdots, P_m$. where $P_{\text{ch}_0}$ and $P_{\text{ch}_1}$ provide inputs as shares of the choice bit, $\text{ch}_0, \text{ch}_1 \in [m]$. We use $J = \{j_1, \cdots, j_d\}$ to denote the set of indices for the parties who provide inputs as shares of $\Delta$, where $d \le m$. The message length $l$.
\item \textbf{Inputs.} $P_{\text{ch}_0}$ has input $b_0 \in \{0, 1\}$ and $P_{\text{ch}_1}$ has input $b_1 \in \{0, 1\}$. Each $P_j$ for $j \in J$ has input $\Delta_j$,
\item \textbf{Protocol.}
\begin{enumerate}
\item For $1 \le i \le m$, $P_i$ initializes $r_i = 0$. If $ch_{j= 0,1} \in J$, $P_{ch_j}$ sets $r_{ch_j} = b_j \cdot \Delta_{ch_j}$ ($\cdot$ denotes bitwise-AND).
\item For $j \in J$: $P_{\text{ch}_0}$ and $P_j$ invoke $\FuncROT$ where $P_{\text{ch}_0}$ acts as receiver with input $b_0$ and $P_j$ as sender. $P_{\text{ch}_0}$ receives $u_{j,{\text{ch}_0}}^{b_0} \in \{0,1\}^l$. $P_j$ receives $u_{j,{\text{ch}_0}}^0, u_{j,{\text{ch}_0}}^1 \in \{0,1\}^l$.
$P_j$ updates $r_j = r_j \oplus u_{j,{\text{ch}_0}}^0$ and sends $\Delta'_{j,{\text{ch}_0}} = u_{j,{\text{ch}_0}}^0 \oplus u_{j,{\text{ch}_0}}^1 \oplus \Delta_j$ to $P_{\text{ch}_0}$. $P_{\text{ch}_0}$ updates $r_{\text{ch}_0} = r_{\text{ch}_0} \oplus u_{j,{\text{ch}_0}}^{b_0} \oplus b_0 \cdot \Delta'_{j,{\text{ch}_0}}$.
\item For $j \in J$: $P_{\text{ch}_1}$ and $P_j$ invoke $\FuncROT$ where $P_{\text{ch}_1}$ acts as receiver with input $b_1$ and $P_j$ as sender. $P_{\text{ch}_1}$ receives $u_{j,{\text{ch}_1}}^{b_1} \in \{0,1\}^l$. $P_j$ receives $u_{j,{\text{ch}_1}}^0, u_{j,{\text{ch}_1}}^1 \in \{0,1\}^l$.
$P_j$ updates $r_j = r_j \oplus u_{j,{\text{ch}_1}}^0$ and sends $\Delta'_{j,{\text{ch}_1}} = u_{j,{\text{ch}_1}}^0 \oplus u_{j,{\text{ch}_1}}^1 \oplus \Delta_j$ to $P_{\text{ch}_1}$. $P_{\text{ch}_1}$ updates $r_{\text{ch}_1} = r_{\text{ch}_1} \oplus u_{j,{\text{ch}_1}}^{b_1} \oplus b_1 \cdot \Delta'_{j,{\text{ch}_1}}$.
\item If $\lvert I = \{ch_0,ch_1\} \cup J \rvert < m$, each $P_i$ ($i \in I$) samples $r'_{i,i'}$ and sends $r'_{i,i'}$ to each $P_{i'}$ ($i' \in [m] \setminus I$) and updates $r_i = r_i \oplus r'_{i,i'}$. $P_{i'}$ updates $r_{i'} = r_{i'} \oplus r'_{i,i'}$. 
\item For $1 \le i \le m$, $P_i$ outputs the final value of $r_i$.
\end{enumerate}
\end{trivlist}
\end{minipage}
\end{framed}
\caption{Multi-Party Secret-Shared Random OT $\ProtomssROT$}\label{fig:proto-mssrot}
\end{figure}

\begin{figure}[!hbtp]
\begin{framed}
\begin{minipage}[center]{\textwidth}
\begin{trivlist}
\item \textbf{Parameters.} $m$ parties $P_1, \cdots, P_m$. Size $n$ of input sets. The bit length $l$ of set elements. Cuckoo hashing parameters: hash functions $h_1, h_2, h_3$ and number of bins $B$. A collision-resisitant hash function $\Hash(x): \{0,1\}^l \to \{0,1\}^{\kappa}$. 
\item \textbf{Inputs.} Each party $P_i$ inputs $X_i = \{x_i^1,\cdots, x_i^n\} \subseteq \{0,1\}^l$.
\item \textbf{Protocol.}
\begin{enumerate}
\item \textbf{Hashing to bins.} $P_1$ does $\mathcal{T}_1^1, \cdots, \mathcal{T}_1^B \gets \Simple_{h_1,h_2,h_3}^B(X_1)$. For $1 < j \le m $, $P_j$ does $\mathcal{C}_j^1, \cdots, \mathcal{C}_j^B \gets \Cuckoo_{h_1,h_2,h_3}^B(X_j)$ and $\mathcal{T}_j^1, \cdots, \mathcal{T}_j^B \gets \Simple_{h_1,h_2,h_3}^B(X_j)$.
\item \textbf{Batch secret-shared private membership test.} For $1 \le i < j \le m$: $P_i$ and $P_j$ invoke $\FuncbssPMT$ of batch size $B$, where $P_i$ acts as sender with inputs $\mathcal{T}_i^1, \cdots, \mathcal{T}_i^B$ and $P_j$ acts as receiver with inputs $\mathcal{C}_j^1, \cdots, \mathcal{C}_j^B$. For the instance $b \in [B]$, $P_i$ receives $e_{i,j}^b \in \{0,1\}$, and $P_j$ receives $e_{j,i}^b \in \{0,1\}$.
\item \textbf{Multi-party secret-shared random oblivious transfers.} For $1 \le i < j \le m, 1 \le b \le B$: $P_{min(2,i)}, \cdots, P_j$ invoke $\FuncmssROT$ where $P_i$ acts as $P_{\text{ch}_0}$ with input $e_{i,j}^b$ and $P_j$ acts as $P_{\text{ch}_1}$ with input $e_{j,i}^b$. For $1 < j' \le j$, $P_{j'}$ samples $\Delta_{j',ji}^b \gets \{0,1\}^ {l+\kappa}$ and inputs $\Delta_{j',ji}^b$ as shares of $\Delta$. For $min(2,i) \le d < j$, $P_d$ receives $r_{d,ji}^b \in \{0,1\}^ {l+\kappa}$ and computes $u_{d,j}^b = \bigoplus_{i=1}^{j-1} r_{d,ji}^b$. $P_j$ receives $r_{j,ji}^b \in \{0,1\}^ {l+\kappa}$ and computes $u_{j,j}^b = \bigoplus_{i=1}^{j-1} r_{j,ji}^b \oplus (\Elem(\mathcal{C}_j^b) \Vert \Hash(\Elem(\mathcal{C}_j^b)))$ if $\mathcal{C}_j^b$ is not corresponding to an empty bin, otherwise chooses $u_{j,j}^b$ at random. $\Elem(\mathcal{C}_j^b)$ denotes the element in $\mathcal{C}_j^b$.
\item \textbf{Multi-party secret-shared shuffle.} 
\begin{enumerate}
\item For $1 \le i \le m$, each party $P_i$ computes $\vec{sh}_i \in (\{0,1\}^{l+\kappa})^{(m-1)B}$ as follows: For $max(2,i) \le j \le m, 1 \le b \le B$, $sh_{i,(j-2)B+b} = u_{i,j}^b$. Set all other positions to $0$.
\item For $1 \le i \le m$, all parties $P_i$ invoke $\FuncMS$ with input $\vec{sh}_i$. $P_i$ receives $\vec{sh}_{i}'$.
\end{enumerate}
\item \textbf{Output reconstruction.} For $2 \le j \le m$, $P_j$ sends $\vec{sh}_{j}'$ to $P_1$. $P_1$ recovers $\vec{v} = \bigoplus_{i=1}^m \vec{sh}_{i}'$ and sets $Y = \emptyset$. For $1 \le i \le (m-1)B$, if $v'_i = x \Vert \Hash(x)$ holds for some $x \in \{0,1\}^l$, adds $x$ to $Y$. Outputs $X_1 \cup Y$.
\end{enumerate}
\end{trivlist}
\end{minipage}
\end{framed}
\caption{Our SK-MPSU $\ProtoMPSUTwo$}\label{fig:proto-mpsu2}
\end{figure}

\section{MPSU from Public-Key Techniques}\label{sec:protocol4}

In this section, we describe how to construct a PK-MPSU achieving linear computation and linear communication complexity.
The construction is formally presented in Figure~\ref{fig:proto-mpsu4}.

\begin{theorem}\label{theorem:pke}
Protocol $\ProtoMPSUFour$ securely implements $\FuncMPSU$ against any semi-honest adversary corrupting $t < m$ parties in the $(\FuncbssPMT,\FuncROT)$-hybrid model, assuming the rerandomizable property and indistinguishable multiple encryptions (cf. Appendix~\ref{sec:pke}) of MKR-PKE scheme.
\end{theorem}   

For the complete proof, refer to Appendix~\ref{proof:pke}. A comprehensive complexity analysis and an exhaustive comparison with~\cite{GNT-eprint-2023} can be found in Appendix~\ref{sec:analysis}. 



\begin{remark}
When instantiating our PK-MPSU framework with EC MKR-PKE, the element space is set as EC points accordingly, 
which might limit its usage in applications. We also identified the same issue in \cite{GNT-eprint-2023}. However, we argue that this limitation is not inherent in our PK-MPSU (nor to the protocol in\cite{GNT-eprint-2023}), as it can be addressed by finding an elliptic curve with efficient encoding and decoding algorithms or by exploring alternative instantiations of MKR-PKE.

Despite the limitation, our EC-element-space PK-MPSU remains useful in many scenarios. For instance, we present the first multi-party private ID protocol in Appendix~\ref{sec:mpid}, where our EC-element-space PK-MPSU MPSU serves as a core subprotocol. Our multi-party private-ID takes advantages of our PK-MPSU and achieves linear computation and communication complexity and low communication costs as well.
\end{remark}

\begin{figure*}[!hbtp]
\begin{framed}
\begin{minipage}[center]{\textwidth}
\begin{trivlist}
\item \textbf{Parameters.} $m$ parties $P_1, \cdots, P_m$. Size $n$ of input sets. The bit length $l$ of set elements. Cuckoo hashing parameters: hash functions $h_1, h_2, h_3$ and number of bins $B$. A MKR-PKE scheme $\mathcal{E} = (\Gen, \Enc, \ParDec, \Dec, \ReRand)$.
\item \textbf{Inputs.} Each party $P_i$ has input $X_i = \{x_i^1,\cdots, x_i^n\} \subseteq \{0,1\}^l$.
\item \textbf{Protocol.}
Each party $P_i$ runs $(pk_i, sk_i) \gets \Gen(1^\lambda)$, and distributes its public key $pk_i$ to other parties. Define $sk = sk_1 + \cdots + sk_m$ and each party can compute its associated public key $pk= \prod_{i =1}^m pk_i$.
\begin{enumerate}
\item \textbf{Hashing to bins.} $P_1$ does $\mathcal{T}_1^1, \cdots, \mathcal{T}_1^B \gets \Simple_{h_1,h_2,h_3}^B(X_1)$. For $1 < j \le m $, $P_j$ does $\mathcal{C}_j^1, \cdots, \mathcal{C}_j^B \gets \Cuckoo_{h_1,h_2,h_3}^B(X_j)$ and $\mathcal{T}_j^1, \cdots, \mathcal{T}_j^B \gets \Simple_{h_1,h_2,h_3}^B(X_j)$.
\item \textbf{Batch secret-shared private membership test.} For $1 \le i < j \le m$: $P_i$ and $P_j$ invoke $\FuncbssPMT$ of batch size $B$, where $P_i$ acts as sender with inputs $\mathcal{T}_i^1, \cdots, \mathcal{T}_i^B$ and $P_j$ acts as receiver with inputs $\mathcal{C}_j^1, \cdots, \mathcal{C}_j^B$. For the instance $b \in [B]$, $P_i$ receives $e_{i,j}^b \in \{0,1\}$, and $P_j$ receives $e_{j,i}^b \in \{0,1\}$.
\item \textbf{Random oblivious transfers and messages rerandomization.}  
\begin{enumerate} 
\item[(a)] For $2 \le j \le m, 1 \le b \le B$: $P_j$ defines $\vec{c}_j$ and sets $c_j^b = \Enc(pk, \Elem(\mathcal{C}_j^b))$. $\Elem(\mathcal{C}_j^b)$ denotes the element in $\mathcal{C}_j^b$.
\end{enumerate}
    \begin{itemize} 
    \item[-] For $2 \le i < j$: $P_i$ and $P_j$ invoke $\FuncROT$ where $P_i$ acts as receiver with input $e_{i,j}^b$ and $P_j$ acts as sender. $P_i$ receives $r_{i,j}^b = r_{j,i,e_{i,j}^b}^b \in \{0,1\}^ \lambda$. $P_j$ receives $r_{j,i,0}^b, r_{j,i,1}^b \in \{0,1\}^ \lambda$. $P_j$ computes $u_{j,i,e_{j,i}^b}^b = r_{j,i,e_{j,i}^b}^b \oplus c_j^b$, $u_{j,i,e_{j,i}^b \oplus 1}^b = r_{j,i,e_{j,i}^b \oplus 1}^b \oplus \Enc(pk, \perp)$, then sends $u_{j,i,0}^b, u_{j,i,1}^b$ to $P_i$. 

    \item[-] $P_i$ defines $v_{j,i}^b = u_{j,i,e_{i,j}^b}^b \oplus r_{i,j}^b$ and sends ${v}_{j,i}'^b = \ReRand(pk,v_{j,i}^b)$ to $P_j$. $P_j$ updates $c_j^b = \ReRand(pk,{v}_{j,i}'^b)$.
    \end{itemize}
\begin{enumerate} 
\item[(b)] For $2 \le j \le m, 1 \le b \le B$: 
\end{enumerate}
\begin{itemize}
\item[-] $P_1$ and $P_j$ invoke $\FuncROT$ where $P_1$ acts as receiver with input $e_{1,j}^b$ and $P_j$ acts as sender. $P_1$ receives $r_{1,j}^b = r_{j,1,e_{1,j}^b}^b \in \{0,1\}^ \lambda$. $P_j$ receives $r_{j,1,0}^b, r_{j,1,1}^b \in \{0,1\}^ \lambda$. $P_j$ computes $u_{j,1,e_{j,1}^b}^b = r_{j,1,e_{j,1}^b}^b \oplus c_j^b$, $u_{j,1,e_{j,1}^b \oplus 1}^b = r_{j,1,e_{j,1}^b \oplus 1}^b \oplus \Enc(pk, \perp)$, then sends $u_{j,1,0}^b, u_{j,1,1}^b$ to $P_1$. 
\item[-] $P_1$ defines $\vec{\ct}_1' \in (\{0,1\}^{\lambda})^{(m-1)B}$, and sets ${\ct}_1'^{(j-2)B+b} = \ReRand(pk,u_{j,1,e_{1,j}^b}^b \oplus r_{1,j}^b)$.
\end{itemize}
\item \textbf{Messages decryptions and shufflings.} 
\begin{enumerate}
    \item $P_1$ samples $\pi_1:[(m-1)B] \to [(m-1)B]$ and computes $\vec{\ct}_1'' = \pi_1(\vec{{\ct}_1'})$. $P_1$ sends $\vec{\ct}_1''$ to $P_{2}$.
    \item For $2 \le j \le m, 1 \le i \le (m-1)B$: $P_j$ computes ${\ct}_{j}^i = \ParDec(sk_j,{\ct}_{j-1}''^i)$, $pk_{A_j} = pk_1 \cdot \prod_{d = j+1}^m pk_d$, and ${\ct}_{j}'^i = \ReRand(pk_{A_j},{\ct}_{j}^i)$. Then it samples $\pi_j:[(m-1)B] \to [(m-1)B]$ and computes $\vec{\ct}_{j}'' = {\pi_j}(\vec{\ct}_{j}')$. If $j \ne m$, $P_j$ sends $\vec{\ct}_{j}''$ to $P_{j+1}$; else, $P_m$ sends $\vec{\ct}_{m}''$ to $P_{1}$.
    \item For $1 \le i \le (m-1)B$: $P_1$ computes $\mathsf{pt}_i = \Dec(sk_1,{\ct}_{m}''^i)$. 
\end{enumerate}
\item \textbf{Output reconstruction.} $P_{1}$ sets $Y = \emptyset$. For $1 \le i \le (m-1)B$, if $\mathsf{pt}_i \ne \perp$, it updates $Y = Y \cup \{\mathsf{pt}_i\}$. $P_{1}$ outputs $Y$.
\end{enumerate}
\end{trivlist}
\end{minipage}
\end{framed}
\caption{Our PK-MPSU $\ProtoMPSUFour$}\label{fig:proto-mpsu4}
\end{figure*}

\section{Performance Evaluation}

In this section, we provide the implementation details and experimental results for our works. For most previous works~\cite{KS-CRYPTO-2005,Frikken-ACNS-2007,BA-ASIACCS-2012,GNT-eprint-2023} do not provide open-source code, and the protocol in~\cite{VCE22} shows fairly inferior performance in comparison with the state-of-the-art LG, here we only compare our protocols with LG, whose implementation is available on \href{https://github.com/lx-1234/MPSU}{https://github.com/lx-1234/MPSU}. 

During our experiments, we identify several issues with the LG implementation: 
\begin{enumerate}
  \item The code is neither a complete nor a secure implementation for MPSU, as it lacks the offline share correlation generation of the multi-party secret-shared shuffle protocol~\cite{EB22}. Instead of adhering to the protocol specifications, the implementation designates one party to generate and store some ``fake'' share correlations as local files during the offline phase, while the other parties read these files and consume the share correlations in the online phase\footnote{Refer to the function \texttt{ShareCorrelation::generate()} in ShareCorrelationGen.cpp}. This results in two serious consequences: 
  \begin{itemize}
  \item The distributed execution is not supported. 
  \item The security of multi-party secret-shared shuffle is compromised, leading to information leakage.
  \end{itemize} 
  To conduct a fair and complete comparison, we integrated our correct version of the share correlation generation implementation into their code. 
  \item The code is not a correct implementation, as it produces incorrect results when the set size $n$ increases beyond a certain threshold. We figure out that this issue arises from the use of a fixed parameter\footnote{The variable \texttt{batchsize} in line 30 of circuit/TripleGen.cpp} for generating Beaver Triples in the offline phase, which leads to an insufficient number of Beaver Triples for the online phase when $n$ is large. We modified their code to ensure correct execution and mark the cases where the original code fails with $*$.
\end{enumerate}

\subsection{Experimental Setup}

We conduct our experiments on an Alibaba Cloud virtual machine with Intel(R) Xeon(R) 2.70GHz CPU (32 physical cores) and 128 GB RAM.
We emulate the two network connections using Linux \texttt{tc} command.
In the LAN setting, the bandwidth is set to be 10 Gbps with 0.1 ms RTT latency.
In the WAN setting, the bandwidth is set to be 400 Mbps with 80 ms RTT latency. 
We measure the running time as the maximal time from protocol begin to end, including messages transmission time, and the communication costs as the total data that leader sent and received.
For a fair comparison, we stick to the following settings for all protocols: 
\begin{itemize}
\item We set the computational security parameter $\kappa = 128$ and the statistical security parameter $\lambda = 40$. 

\item We test the balanced scenario by setting all $m$ input sets to be of equal size. In LG and our SK-MPSU, each party holds $n$ 64-bit strings. 
In our PK-MPSU, each party holds $n$ elements encoded as EC points in compressed form.


\item Each party uses $m - 1$ threads to interact simultaneously with all other parties and 4 threads to perform share correlation generation (in LG and our SK-MPSU), Beaver Triple generation (in all three), parallel SKE encryption (in LG), ciphertext rerandomization and partial decryption (in our PK-MPSU).
\end{itemize}

\subsection{Implementation Details}
Our protocols are written in C++, and  we use the following libraries in our implementation.

\begin{itemize}
\item VOLE: We use VOLE implemented in \textsf{libOTe}\footnote{\href{https://github.com/osu-crypto/libOTe.git}{https://github.com/osu-crypto/libOTe.git}}, instantiating the code family with Expand-Convolute codes~\cite{RRT23}.
\item OKVS and GMW: We use the optimized OKVS in \cite{RR-CCS-2022} as our OKVS
 instantiation\footnote{Since the existence of suitable parameters for the new OKVS construction of the recent work \cite{BPSY-USENIX-2023} is unclear when the set size is less than $2^{10}$, we choose to use the OKVS construction of \cite{RR-CCS-2022}. }, and re-use the implementation of OKVS and GMW by the authors of in \cite{RR-CCS-2022}\footnote{\href{https://github.com/Visa-Research/volepsi.git}{https://github.com/Visa-Research/volepsi.git}}.
\item ROT: We use SoftSpokenOT \cite{Roy22} implemented in \textsf{libOTe}, and set field bits to 5 to balance computation and communication costs. 
\item Share Correlation: We re-use the implementation of Permute+Share~\cite{MS13,CGP-ASIACRYPT-2020} by the authors of \cite{Jia-USENIX-2022}\footnote{\href{https://github.com/dujiajun/PSU.git}{https://github.com/dujiajun/PSU.git}} to build share correlation generation for our SK-MPSU and LG.
\item MKR-PKE: We implement MKR-PKE on top of the curve \texttt{NIST P-256} (also known as \texttt{secp256r1} and \texttt{prime256v1}) implementation from \textsf{openssl}\footnote{\href{https://github.com/openssl/openssl.git}{https://github.com/openssl/openssl.git}}.
\item Additionally, we use the \textsf{cryptoTools}\footnote{\href{https://github.com/ladnir/cryptoTools.git}{https://github.com/ladnir/cryptoTools.git}} library to compute hash functions and PRNG calls, and we adopt \textsf{Coproto}\footnote{\href{https://github.com/Visa-Research/coproto.git}{https://github.com/Visa-Research/coproto.git}} to realize network communication.
\end{itemize}

\subsection{Experimental Results}

We conduct extensive experiments across various numbers of parties $\{3,4,5,7,9,10\}$ and a wide range of set sizes $\{2^6,2^8,2^{10},2^{12},2^{14},2^{16},2^{18},2^{20}\}$ in the LAN and WAN settings. The performance of protocols is evaluated from four dimensions: online and total running time, and online and total communication costs. The results for running time / communication costs are depicted in Table \ref{tab:time}. As we can see in the table, our protocols outperform LG in all the case studies, even with enhanced security.

\begin{trivlist}

\item \textbf{Online computation improvement.} Our SK-MPSU achieves a $3.9-10.0 \times$ speedup compared to LG in the LAN setting, and a $1.1-2.1 \times$ speedup compared to LG in the WAN setting. For example, in the online phase to compute the union of $2^{20}$-size sets among 3 parties, our SK-MPSU runs in 4.4 seconds (LAN) and 33.1 seconds (WAN), which is $5.8 \times$ and $1.64 \times$ faster than LG that takes 25.2 seconds (LAN) and 54.1 seconds (WAN), respectively.

\item \textbf{Total computation improvement.} Our SK-MPSU achieves a $1.2-7.8 \times$ speedup compared to LG in the LAN setting. Our PK-MPSU achieves a $1.2-4.0 \times$ speedup compared to LG in the WAN setting. For example, to compute the union of $2^{20}$-size sets among 3 parties, our SK-MPSU runs in 96.2 seconds overall (LAN), which is $6.4 \times$ faster than LG that takes 610.8 seconds. To compute the union of $2^{18}$-size sets among 9 parties, our PK-MPSU runs in 2304 seconds overall (WAN), which is $2.8 \times$ faster than LG that takes 6381 seconds.

\item \textbf{Online communication improvement.} The online communication costs of our SK-MPSU is $1.2-4.9 \times$ smaller than LG. For example, in the online phase to compute the union of $2^{20}$-size sets among 3 parties, our SK-MPSU requires 455.6 MB communication, which is a $2.0 \times$ improvement of LG that requires 917.0 MB communication.

\item \textbf{Total communication improvement.} The total communication costs of our PK-MPSU is $3.0-36.5\times$
 smaller than LG. For example, to compute the union of $2^{18}$-size sets among 9 parties, our PK-MPSU requires 960.1 MB communication, a $34.8 \times$ improvement of LG that requires 33360 MB.
\end{trivlist}

In conclusion, our protocols exhibit distinct advantages and are suitable for different scenarios:
\begin{itemize}
  \item Our SK-MPSU: This protocol excels in online performance. Additionally, it also outperforms other protocols in nearly all cases in the LAN setting. Notably, the online phase takes only $4.4$ seconds for 3 parties with sets of $2^{20}$ items each, and $7$ seconds for 5 parties with sets of $2^{20}$ items each in the LAN setting. Therefore, our SK-MPSU is the optimal choice when online performance is the primary concern, or in high-bandwidth networks.
  \item Our PK-MPSU: This protocol has the best total performance in the WAN setting, so it is preferred when the total cost is crucial (e.g. in the small-sets setting), at the same time in low-bandwidth networks. Note that our PK-MPSU is the only protocol with the capability of running among 10 parties with sets of $2^{18}$ items each in our experiments, which sufficiently indicates the superiority of its linear complexities.
\end{itemize}


\begin{table*}[]
\resizebox{\textwidth}{!}{%
\begin{tabular}{|c|c|c|cccccccccccccccc|}
\hline
\multirow{3}{*}{Sett.} & \multirow{3}{*}{\begin{tabular}[c]{@{}c@{}} $m$\end{tabular}} & \multirow{3}{*}{Protocol} & \multicolumn{16}{c|}{Set size $n$}                                                                                                                                                                                                                                                                                                                                                                                                                                                                                                                                                                          \\ \cline{4-19} 
                       &                                                                           &                            & \multicolumn{8}{c|}{Online}                                                                                                                                                                                                                                                                              & \multicolumn{8}{c|}{Total = Offline + Online}                                                                                                                                                                                                                                                                         \\ \cline{4-19} 
                       &                     &                             & \multicolumn{1}{c|}{$2^6$}          & \multicolumn{1}{c|}{$2^8$}          & \multicolumn{1}{c|}{$2^{10}$}       & \multicolumn{1}{c|}{$2^{12}$}       & \multicolumn{1}{c|}{$2^{14}$}       & \multicolumn{1}{c|}{$2^{16}$}       & \multicolumn{1}{c|}{$2^{18}$}       & \multicolumn{1}{c|}{$2^{20}$}       & \multicolumn{1}{c|}{$2^6$}          & \multicolumn{1}{c|}{$2^8$}          & \multicolumn{1}{c|}{$2^{10}$}       & \multicolumn{1}{c|}{$2^{12}$}       & \multicolumn{1}{c|}{$2^{14}$}       & \multicolumn{1}{c|}{$2^{16}$}       & \multicolumn{1}{c|}{$2^{18}$}       & \multicolumn{1}{c|}{$2^{20}$} \\ \hline

\multirow{24}{*}{}  & \multirow{3}{*}{3}    & LG & \multicolumn{1}{c|}{0.050} & \multicolumn{1}{c|}{0.058} & \multicolumn{1}{c|}{0.069} & \multicolumn{1}{c|}{0.143} & \multicolumn{1}{c|}{0.425} & \multicolumn{1}{c|}{1.582} & \multicolumn{1}{c|}{6.219} & \multicolumn{1}{c|}{25.16$^*$} & \multicolumn{1}{c|}{0.982} & \multicolumn{1}{c|}{1.016} & \multicolumn{1}{c|}{1.098} & \multicolumn{1}{c|}{1.767} & \multicolumn{1}{c|}{5.650} & \multicolumn{1}{c|}{32.42} & \multicolumn{1}{c|}{153.8} & 610.8$^*$ \\ \cline{3-19} 
 &  & Our SK & \multicolumn{1}{c|}{\cellcolor{blue!30}0.005} & \multicolumn{1}{c|}{\cellcolor{blue!30}0.007} & \multicolumn{1}{c|}{\cellcolor{blue!30}0.009} & \multicolumn{1}{c|}{\cellcolor{blue!30}0.017} & \multicolumn{1}{c|}{\cellcolor{blue!30}0.050} & \multicolumn{1}{c|}{\cellcolor{blue!30}0.213} & \multicolumn{1}{c|}{\cellcolor{blue!30}1.005} & \multicolumn{1}{c|}{\cellcolor{blue!30}4.352} & \multicolumn{1}{c|}{0.361} & \multicolumn{1}{c|}{0.386} & \multicolumn{1}{c|}{\cellcolor{blue!30}0.395} & \multicolumn{1}{c|}{\cellcolor{blue!30}0.489} & \multicolumn{1}{c|}{\cellcolor{blue!30}1.170} & \multicolumn{1}{c|}{\cellcolor{blue!30}4.534} & \multicolumn{1}{c|}{\cellcolor{blue!30}19.77} & \cellcolor{blue!30}96.23 \\ \cline{3-19} 
 & \multirow{-3}{*}{3} & Our PK & \multicolumn{1}{c|}{0.092} & \multicolumn{1}{c|}{0.237} & \multicolumn{1}{c|}{0.822} & \multicolumn{1}{c|}{3.176} & \multicolumn{1}{c|}{12.69} & \multicolumn{1}{c|}{51.17} & \multicolumn{1}{c|}{205.7} & \multicolumn{1}{c|}{829.1} & \multicolumn{1}{c|}{\cellcolor{blue!30}0.117} & \multicolumn{1}{c|}{\cellcolor{blue!30}0.273} & \multicolumn{1}{c|}{0.868} & \multicolumn{1}{c|}{3.270} & \multicolumn{1}{c|}{13.08} & \multicolumn{1}{c|}{53.05} & \multicolumn{1}{c|}{215.0} & 872.6 \\ \cline{2-19} 

& \multirow{3}{*}{4}  & LG & \multicolumn{1}{c|}{0.069} & \multicolumn{1}{c|}{0.076} & \multicolumn{1}{c|}{0.093} & \multicolumn{1}{c|}{0.162} & \multicolumn{1}{c|}{0.478} & \multicolumn{1}{c|}{1.637} & \multicolumn{1}{c|}{6.375} & \multicolumn{1}{c|}{25.79$^*$} & \multicolumn{1}{c|}{1.292} & \multicolumn{1}{c|}{1.358} & \multicolumn{1}{c|}{1.514} & \multicolumn{1}{c|}{2.432} & \multicolumn{1}{c|}{8.716} & \multicolumn{1}{c|}{45.37} & \multicolumn{1}{c|}{197.2} & 762.9$^*$ \\ \cline{3-19} 
 &  & Our SK & \multicolumn{1}{c|}{\cellcolor{blue!30}0.008} & \multicolumn{1}{c|}{\cellcolor{blue!30}0.010} & \multicolumn{1}{c|}{\cellcolor{blue!30}0.013} & \multicolumn{1}{c|}{\cellcolor{blue!30}0.023} & \multicolumn{1}{c|}{\cellcolor{blue!30}0.071} & \multicolumn{1}{c|}{\cellcolor{blue!30}0.286} & \multicolumn{1}{c|}{\cellcolor{blue!30}1.393} & \multicolumn{1}{c|}{\cellcolor{blue!30}5.645} & \multicolumn{1}{c|}{0.639} & \multicolumn{1}{c|}{0.665} & \multicolumn{1}{c|}{\cellcolor{blue!30}0.696} & \multicolumn{1}{c|}{\cellcolor{blue!30}0.917} & \multicolumn{1}{c|}{\cellcolor{blue!30}2.597} & \multicolumn{1}{c|}{\cellcolor{blue!30}10.94} & \multicolumn{1}{c|}{\cellcolor{blue!30}50.25} & \cellcolor{blue!30}237.8 \\ \cline{3-19} 
 & \multirow{-3}{*}{4} & Our PK & \multicolumn{1}{c|}{0.159} & \multicolumn{1}{c|}{0.465} & \multicolumn{1}{c|}{1.570} & \multicolumn{1}{c|}{6.136} & \multicolumn{1}{c|}{24.60} & \multicolumn{1}{c|}{44.91} & \multicolumn{1}{c|}{398.4} & \multicolumn{1}{c|}{1604} & \multicolumn{1}{c|}{\cellcolor{blue!30}0.196} & \multicolumn{1}{c|}{\cellcolor{blue!30}0.505} & \multicolumn{1}{c|}{1.643} & \multicolumn{1}{c|}{6.299} & \multicolumn{1}{c|}{25.45} & \multicolumn{1}{c|}{48.95} & \multicolumn{1}{c|}{417.5} & \multicolumn{1}{c|}{1694} \\ \cline{2-19} 

& \multirow{3}{*}{5}  & LG & \multicolumn{1}{c|}{0.107} & \multicolumn{1}{c|}{0.109} & \multicolumn{1}{c|}{0.126} & \multicolumn{1}{c|}{0.190} & \multicolumn{1}{c|}{0.541} & \multicolumn{1}{c|}{1.998} & \multicolumn{1}{c|}{7.588} & \multicolumn{1}{c|}{31.33$^*$} & \multicolumn{1}{c|}{1.939} & \multicolumn{1}{c|}{1.993} & \multicolumn{1}{c|}{2.202} & \multicolumn{1}{c|}{3.408} & \multicolumn{1}{c|}{12.54} & \multicolumn{1}{c|}{65.04} & \multicolumn{1}{c|}{289.9} & 1152$^*$ \\ \cline{3-19} 
 &  & Our SK & \multicolumn{1}{c|}{\cellcolor{blue!30}0.012} & \multicolumn{1}{c|}{\cellcolor{blue!30}0.013} & \multicolumn{1}{c|}{\cellcolor{blue!30}0.017} & \multicolumn{1}{c|}{\cellcolor{blue!30}0.030} & \multicolumn{1}{c|}{\cellcolor{blue!30}0.087} & \multicolumn{1}{c|}{\cellcolor{blue!30}0.368} & \multicolumn{1}{c|}{\cellcolor{blue!30}1.714} & \multicolumn{1}{c|}{\cellcolor{blue!30}7.003} & \multicolumn{1}{c|}{0.914} & \multicolumn{1}{c|}{0.964} & \multicolumn{1}{c|}{\cellcolor{blue!30}1.023} & \multicolumn{1}{c|}{\cellcolor{blue!30}1.558} & \multicolumn{1}{c|}{\cellcolor{blue!30}4.551} & \multicolumn{1}{c|}{\cellcolor{blue!30}18.71} & \multicolumn{1}{c|}{\cellcolor{blue!30}85.79} & \cellcolor{blue!30}373.4 \\ \cline{3-19} 
 & \multirow{-3}{*}{5} & Our PK & \multicolumn{1}{c|}{0.232} & \multicolumn{1}{c|}{0.693} & \multicolumn{1}{c|}{2.487} & \multicolumn{1}{c|}{9.858} & \multicolumn{1}{c|}{39.33} & \multicolumn{1}{c|}{158.1} & \multicolumn{1}{c|}{637.5} & \multicolumn{1}{c|}{2816} & \multicolumn{1}{c|}{\cellcolor{blue!30}0.292} & \multicolumn{1}{c|}{\cellcolor{blue!30}0.796} & \multicolumn{1}{c|}{2.547} & \multicolumn{1}{c|}{10.09} & \multicolumn{1}{c|}{40.25} & \multicolumn{1}{c|}{162.2} & \multicolumn{1}{c|}{655.0} & 2979 \\ \cline{2-19} 

& \multirow{3}{*}{7} & LG & \multicolumn{1}{c|}{0.154} & \multicolumn{1}{c|}{0.165} & \multicolumn{1}{c|}{0.174} & \multicolumn{1}{c|}{0.281} & \multicolumn{1}{c|}{0.795} & \multicolumn{1}{c|}{2.894} & \multicolumn{1}{c|}{10.92} & \multicolumn{1}{c|}{$-$} & \multicolumn{1}{c|}{3.484} & \multicolumn{1}{c|}{3.583} & \multicolumn{1}{c|}{3.921} & \multicolumn{1}{c|}{5.84} & \multicolumn{1}{c|}{23.42} & \multicolumn{1}{c|}{111.2} & \multicolumn{1}{c|}{484.5} & $-$ \\ \cline{3-19} 
 &  & Our SK & \multicolumn{1}{c|}{\cellcolor{blue!30}0.019} & \multicolumn{1}{c|}{\cellcolor{blue!30}0.021} & \multicolumn{1}{c|}{\cellcolor{blue!30}0.028} & \multicolumn{1}{c|}{\cellcolor{blue!30}0.048} & \multicolumn{1}{c|}{\cellcolor{blue!30}0.156} & \multicolumn{1}{c|}{\cellcolor{blue!30}0.607} & \multicolumn{1}{c|}{\cellcolor{blue!30}2.817} & \multicolumn{1}{c|}{$-$} & \multicolumn{1}{c|}{2.051} & \multicolumn{1}{c|}{2.079} & \multicolumn{1}{c|}{\cellcolor{blue!30}2.214} & \multicolumn{1}{c|}{\cellcolor{blue!30}3.470} & \multicolumn{1}{c|}{\cellcolor{blue!30}12.92} & \multicolumn{1}{c|}{\cellcolor{blue!30}56.50} & \multicolumn{1}{c|}{\cellcolor{blue!30}245.6} & $-$ \\ \cline{3-19} 
 & \multirow{-3}{*}{7} & Our PK & \multicolumn{1}{c|}{0.463} & \multicolumn{1}{c|}{1.365} & \multicolumn{1}{c|}{5.030} & \multicolumn{1}{c|}{19.43} & \multicolumn{1}{c|}{77.22} & \multicolumn{1}{c|}{310.0} & \multicolumn{1}{c|}{1247} & \multicolumn{1}{c|}{$-$} & \multicolumn{1}{c|}{\cellcolor{blue!30}0.550} & \multicolumn{1}{c|}{\cellcolor{blue!30}1.469} & \multicolumn{1}{c|}{5.158} & \multicolumn{1}{c|}{19.73} & \multicolumn{1}{c|}{78.59} & \multicolumn{1}{c|}{316.3} & \multicolumn{1}{c|}{1276} & $-$ \\ \cline{2-19} 

& \multirow{3}{*}{9}  & LG & \multicolumn{1}{c|}{0.222} & \multicolumn{1}{c|}{0.226} & \multicolumn{1}{c|}{0.234} & \multicolumn{1}{c|}{0.417} & \multicolumn{1}{c|}{1.216} & \multicolumn{1}{c|}{4.449} & \multicolumn{1}{c|}{17.29} & \multicolumn{1}{c|}{$-$} & \multicolumn{1}{c|}{5.501} & \multicolumn{1}{c|}{5.612} & \multicolumn{1}{c|}{6.249} & \multicolumn{1}{c|}{10.27} & \multicolumn{1}{c|}{41.69} & \multicolumn{1}{c|}{182.7} & \multicolumn{1}{c|}{793.3} & $-$ \\ \cline{3-19} 
 &  & Our SK & \multicolumn{1}{c|}{\cellcolor{blue!30}0.027} & \multicolumn{1}{c|}{\cellcolor{blue!30}0.031} & \multicolumn{1}{c|}{\cellcolor{blue!30}0.039} & \multicolumn{1}{c|}{\cellcolor{blue!30}0.075} & \multicolumn{1}{c|}{\cellcolor{blue!30}0.230} & \multicolumn{1}{c|}{\cellcolor{blue!30}0.970} & \multicolumn{1}{c|}{\cellcolor{blue!30}4.293} & \multicolumn{1}{c|}{$-$} & \multicolumn{1}{c|}{3.363} & \multicolumn{1}{c|}{3.402} & \multicolumn{1}{c|}{\cellcolor{blue!30}3.895} & \multicolumn{1}{c|}{\cellcolor{blue!30}7.161} & \multicolumn{1}{c|}{\cellcolor{blue!30}30.09} & \multicolumn{1}{c|}{\cellcolor{blue!30}126.3} & \multicolumn{1}{c|}{\cellcolor{blue!30}678.8} & $-$ \\ \cline{3-19} 
 & \multirow{-3}{*}{9} & Our PK & \multicolumn{1}{c|}{0.764} & \multicolumn{1}{c|}{2.271} & \multicolumn{1}{c|}{8.074} & \multicolumn{1}{c|}{31.64} & \multicolumn{1}{c|}{126.7} & \multicolumn{1}{c|}{507.0} & \multicolumn{1}{c|}{2039} & \multicolumn{1}{c|}{$-$} & \multicolumn{1}{c|}{\cellcolor{blue!30}0.880} & \multicolumn{1}{c|}{\cellcolor{blue!30}2.375} & \multicolumn{1}{c|}{8.238} & \multicolumn{1}{c|}{32.19} & \multicolumn{1}{c|}{128.7} & \multicolumn{1}{c|}{515.5} & \multicolumn{1}{c|}{2080} & $-$ \\ \cline{2-19}  

& \multirow{3}{*}{10} & LG & \multicolumn{1}{c|}{0.228} & \multicolumn{1}{c|}{0.243} & \multicolumn{1}{c|}{0.276} & \multicolumn{1}{c|}{0.514} & \multicolumn{1}{c|}{1.479} & \multicolumn{1}{c|}{5.467} & \multicolumn{1}{c|}{$-$} & \multicolumn{1}{c|}{$-$} & \multicolumn{1}{c|}{6.736} & \multicolumn{1}{c|}{6.846} & \multicolumn{1}{c|}{8.335} & \multicolumn{1}{c|}{13.56} & \multicolumn{1}{c|}{55.58} & \multicolumn{1}{c|}{238.3} & \multicolumn{1}{c|}{$-$} & $-$ \\ \cline{3-19} 
 &  & Our SK & \multicolumn{1}{c|}{\cellcolor{blue!30}0.031} & \multicolumn{1}{c|}{\cellcolor{blue!30}0.036} & \multicolumn{1}{c|}{\cellcolor{blue!30}0.043} & \multicolumn{1}{c|}{\cellcolor{blue!30}0.088} & \multicolumn{1}{c|}{\cellcolor{blue!30}0.286} & \multicolumn{1}{c|}{\cellcolor{blue!30}1.183} & \multicolumn{1}{c|}{$-$} & \multicolumn{1}{c|}{$-$} & \multicolumn{1}{c|}{3.965} & \multicolumn{1}{c|}{4.232} & \multicolumn{1}{c|}{\cellcolor{blue!30}4.821} & \multicolumn{1}{c|}{\cellcolor{blue!30}9.334} & \multicolumn{1}{c|}{\cellcolor{blue!30}41.97} & \multicolumn{1}{c|}{\cellcolor{blue!30}175.9} & \multicolumn{1}{c|}{$-$} & $-$ \\ \cline{3-19} 
\multirow{-18}{*}{\begin{tabular}[c]{@{}c@{}}Time.\\      (s)\\      LAN\end{tabular}} & \multirow{-3}{*}{10} & Our PK & \multicolumn{1}{c|}{0.865} & \multicolumn{1}{c|}{2.800} & \multicolumn{1}{c|}{9.944} & \multicolumn{1}{c|}{39.09} & \multicolumn{1}{c|}{155.9} & \multicolumn{1}{c|}{622.7} & \multicolumn{1}{c|}{\cellcolor{blue!30}2503} & \multicolumn{1}{c|}{$-$} & \multicolumn{1}{c|}{\cellcolor{blue!30}0.970} & \multicolumn{1}{c|}{\cellcolor{blue!30}2.912} & \multicolumn{1}{c|}{10.11} & \multicolumn{1}{c|}{39.58} & \multicolumn{1}{c|}{158.4} & \multicolumn{1}{c|}{634.6} & \multicolumn{1}{c|}{\cellcolor{blue!30}2556} & $-$ \\ \hline

\multirow{24}{*}{}  & \multirow{3}{*}{3}  & LG & \multicolumn{1}{c|}{4.502} & \multicolumn{1}{c|}{4.505} & \multicolumn{1}{c|}{4.522} & \multicolumn{1}{c|}{4.914} & \multicolumn{1}{c|}{6.272} & \multicolumn{1}{c|}{8.744} & \multicolumn{1}{c|}{\cellcolor[HTML]{FFFFFF}17.78} & \multicolumn{1}{c|}{54.13$^*$} & \multicolumn{1}{c|}{12.46} & \multicolumn{1}{c|}{13.59} & \multicolumn{1}{c|}{15.76} & \multicolumn{1}{c|}{19.52} & \multicolumn{1}{c|}{30.73} & \multicolumn{1}{c|}{78.74} & \multicolumn{1}{c|}{282.8} & 1188$^*$ \\ \cline{3-19} 
 &  & Our SK & \multicolumn{1}{c|}{\cellcolor{blue!30}2.165} & \multicolumn{1}{c|}{\cellcolor{blue!30}2.166} & \multicolumn{1}{c|}{\cellcolor{blue!30}2.332} & \multicolumn{1}{c|}{\cellcolor{blue!30}3.157} & \multicolumn{1}{c|}{\cellcolor{blue!30}3.734} & \multicolumn{1}{c|}{\cellcolor{blue!30}4.444} & \multicolumn{1}{c|}{\cellcolor{blue!30}9.705} & \multicolumn{1}{c|}{\cellcolor{blue!30}33.10} & \multicolumn{1}{c|}{8.403} & \multicolumn{1}{c|}{9.710} & \multicolumn{1}{c|}{12.62} & \multicolumn{1}{c|}{16.30} & \multicolumn{1}{c|}{25.02} & \multicolumn{1}{c|}{\cellcolor{blue!30}56.88} & \multicolumn{1}{c|}{\cellcolor{blue!30}194.7} & \cellcolor{blue!30}801.2 \\ \cline{3-19} 
 & \multirow{-3}{*}{3} & Our PK & \multicolumn{1}{c|}{4.419} & \multicolumn{1}{c|}{4.555} & \multicolumn{1}{c|}{5.553} & \multicolumn{1}{c|}{7.984} & \multicolumn{1}{c|}{18.21} & \multicolumn{1}{c|}{59.77} & \multicolumn{1}{c|}{226.3} & \multicolumn{1}{c|}{900.3} & \multicolumn{1}{c|}{\cellcolor{blue!30}5.168} & \multicolumn{1}{c|}{\cellcolor{blue!30}5.306} & \multicolumn{1}{c|}{\cellcolor{blue!30}6.472} & \multicolumn{1}{c|}{\cellcolor{blue!30}8.968} & \multicolumn{1}{c|}{\cellcolor{blue!30}19.65} & \multicolumn{1}{c|}{62.73} & \multicolumn{1}{c|}{237.3} & 946.1 \\ \cline{2-19} 

& \multirow{3}{*}{4}   & LG & \multicolumn{1}{c|}{5.696} & \multicolumn{1}{c|}{5.880} & \multicolumn{1}{c|}{6.540} & \multicolumn{1}{c|}{7.094} & \multicolumn{1}{c|}{7.323} & \multicolumn{1}{c|}{11.36} & \multicolumn{1}{c|}{23.13} & \multicolumn{1}{c|}{78.81$^*$} & \multicolumn{1}{c|}{17.94} & \multicolumn{1}{c|}{20.99} & \multicolumn{1}{c|}{27.63} & \multicolumn{1}{c|}{32.69} & \multicolumn{1}{c|}{50.59} & \multicolumn{1}{c|}{145.7} & \multicolumn{1}{c|}{554.7} & \cellcolor[HTML]{FFFFFF}2167$^*$ \\ \cline{3-19} 
 &  & Our SK & \multicolumn{1}{c|}{\cellcolor{blue!30}2.967} & \multicolumn{1}{c|}{\cellcolor{blue!30}2.969} & \multicolumn{1}{c|}{\cellcolor{blue!30}3.298} & \multicolumn{1}{c|}{\cellcolor{blue!30}3.976} & \multicolumn{1}{c|}{\cellcolor{blue!30}4.618} & \multicolumn{1}{c|}{\cellcolor{blue!30}6.507} & \multicolumn{1}{c|}{\cellcolor{blue!30}17.10} & \multicolumn{1}{c|}{\cellcolor{blue!30}59.21} & \multicolumn{1}{c|}{11.46} & \multicolumn{1}{c|}{13.93} & \multicolumn{1}{c|}{20.21} & \multicolumn{1}{c|}{26.77} & \multicolumn{1}{c|}{47.49} & \multicolumn{1}{c|}{133.2} & \multicolumn{1}{c|}{520.0} & 2141 \\ \cline{3-19} 
 & \multirow{-3}{*}{4} & Our PK & \multicolumn{1}{c|}{5.622} & \multicolumn{1}{c|}{5.899} & \multicolumn{1}{c|}{7.929} & \multicolumn{1}{c|}{12.17} & \multicolumn{1}{c|}{32.40} & \multicolumn{1}{c|}{113.8} & \multicolumn{1}{c|}{440.9} & \multicolumn{1}{c|}{1761} & \multicolumn{1}{c|}{\cellcolor{blue!30}6.773} & \multicolumn{1}{c|}{\cellcolor{blue!30}7.052} & \multicolumn{1}{c|}{\cellcolor{blue!30}9.25} & \multicolumn{1}{c|}{\cellcolor{blue!30}13.53} & \multicolumn{1}{c|}{\cellcolor{blue!30}34.27} & \multicolumn{1}{c|}{\cellcolor{blue!30}118.9} & \multicolumn{1}{c|}{\cellcolor{blue!30}461.3} & {\cellcolor{blue!30}1857} \\ \cline{2-19} 

& \multirow{3}{*}{5}  & LG & \multicolumn{1}{c|}{7.385} & \multicolumn{1}{c|}{7.708} & \multicolumn{1}{c|}{8.621} & \multicolumn{1}{c|}{9.198} & \multicolumn{1}{c|}{9.687} & \multicolumn{1}{c|}{14.62} & \multicolumn{1}{c|}{32.83} & \multicolumn{1}{c|}{126.3$^*$} & \multicolumn{1}{c|}{23.64} & \multicolumn{1}{c|}{26.82} & \multicolumn{1}{c|}{37.39} & \multicolumn{1}{c|}{48.48} & \multicolumn{1}{c|}{88.32} & \multicolumn{1}{c|}{263.3} & \multicolumn{1}{c|}{1053} & 4394$^*$ \\ \cline{3-19} 
 &  & Our SK & \multicolumn{1}{c|}{\cellcolor{blue!30}3.768} & \multicolumn{1}{c|}{\cellcolor{blue!30}3.733} & \multicolumn{1}{c|}{\cellcolor{blue!30}4.471} & \multicolumn{1}{c|}{\cellcolor{blue!30}4.800} & \multicolumn{1}{c|}{\cellcolor{blue!30}5.521} & \multicolumn{1}{c|}{\cellcolor{blue!30}8.938} & \multicolumn{1}{c|}{\cellcolor{blue!30}25.95} & \multicolumn{1}{c|}{\cellcolor{blue!30}95.40} & \multicolumn{1}{c|}{17.53} & \multicolumn{1}{c|}{21.10} & \multicolumn{1}{c|}{30.28} & \multicolumn{1}{c|}{44.30} & \multicolumn{1}{c|}{88.15} & \multicolumn{1}{c|}{278.9} & \multicolumn{1}{c|}{1119} & 4820 \\ \cline{3-19} 
 & \multirow{-3}{*}{5} & Our PK & \multicolumn{1}{c|}{6.849} & \multicolumn{1}{c|}{7.928} & \multicolumn{1}{c|}{10.50} & \multicolumn{1}{c|}{17.08} & \multicolumn{1}{c|}{49.75} & \multicolumn{1}{c|}{179.8} & \multicolumn{1}{c|}{703.3} & \multicolumn{1}{c|}{2873} & \multicolumn{1}{c|}{\cellcolor{blue!30}8.424} & \multicolumn{1}{c|}{\cellcolor{blue!30}9.483} & \multicolumn{1}{c|}{\cellcolor{blue!30}12.22} & \multicolumn{1}{c|}{\cellcolor{blue!30}18.86} & \multicolumn{1}{c|}{\cellcolor{blue!30}52.07} & \multicolumn{1}{c|}{\cellcolor{blue!30}185.4} & \multicolumn{1}{c|}{\cellcolor{blue!30}724.5} & {\cellcolor{blue!30}2980} \\ \cline{2-19} 

& \multirow{3}{*}{7} & LG & \multicolumn{1}{c|}{9.312} & \multicolumn{1}{c|}{9.833} & \multicolumn{1}{c|}{10.55} & \multicolumn{1}{c|}{11.35} & \multicolumn{1}{c|}{12.14} & \multicolumn{1}{c|}{21.29} & \multicolumn{1}{c|}{66.93} & \multicolumn{1}{c|}{$-$} & \multicolumn{1}{c|}{34.92} & \multicolumn{1}{c|}{45.69} & \multicolumn{1}{c|}{66.26} & \multicolumn{1}{c|}{94.89} & \multicolumn{1}{c|}{203.4} & \multicolumn{1}{c|}{705.8} & \multicolumn{1}{c|}{2898} & $-$ \\ \cline{3-19} 
 &  & Our SK & \multicolumn{1}{c|}{\cellcolor{blue!30}5.373} & \multicolumn{1}{c|}{\cellcolor{blue!30}5.381} & \multicolumn{1}{c|}{\cellcolor{blue!30}6.207} & \multicolumn{1}{c|}{\cellcolor{blue!30}6.644} & \multicolumn{1}{c|}{\cellcolor{blue!30}8.164} & \multicolumn{1}{c|}{\cellcolor{blue!30}17.67} & \multicolumn{1}{c|}{\cellcolor{blue!30}56.894} & \multicolumn{1}{c|}{$-$} & \multicolumn{1}{c|}{34.00} & \multicolumn{1}{c|}{44.95} & \multicolumn{1}{c|}{61.34} & \multicolumn{1}{c|}{95.03} & \multicolumn{1}{c|}{228.7} & \multicolumn{1}{c|}{817.4} & \multicolumn{1}{c|}{3506} & $-$ \\ \cline{3-19} 
 & \multirow{-3}{*}{7} & Our PK & \multicolumn{1}{c|}{9.504} & \multicolumn{1}{c|}{12.32} & \multicolumn{1}{c|}{15.14} & \multicolumn{1}{c|}{29.73} & \multicolumn{1}{c|}{92.66} & \multicolumn{1}{c|}{348.5} & \multicolumn{1}{c|}{1377.4} & \multicolumn{1}{c|}{$-$} & \multicolumn{1}{c|}{\cellcolor{blue!30}11.88} & \multicolumn{1}{c|}{\cellcolor{blue!30}14.71} & \multicolumn{1}{c|}{\cellcolor{blue!30}17.72} & \multicolumn{1}{c|}{\cellcolor{blue!30}32.35} & \multicolumn{1}{c|}{\cellcolor{blue!30}95.86} & \multicolumn{1}{c|}{\cellcolor{blue!30}356.5} & \multicolumn{1}{c|}{\cellcolor{blue!30}1409} & $-$ \\ \cline{2-19} 

& \multirow{3}{*}{9} & LG & \multicolumn{1}{c|}{11.41} & \multicolumn{1}{c|}{12.21} & \multicolumn{1}{c|}{13.34} & \multicolumn{1}{c|}{14.41} & \multicolumn{1}{c|}{15.09} & \multicolumn{1}{c|}{33.84} & \multicolumn{1}{c|}{115.5} & \multicolumn{1}{c|}{$-$} & \multicolumn{1}{c|}{56.65} & \multicolumn{1}{c|}{75.24} & \multicolumn{1}{c|}{104.1} & \multicolumn{1}{c|}{169.1} & \multicolumn{1}{c|}{406.0} & \multicolumn{1}{c|}{1503} & \multicolumn{1}{c|}{6387} & $-$ \\ \cline{3-19} 
 &  & Our SK & \multicolumn{1}{c|}{\cellcolor{blue!30}6.977} & \multicolumn{1}{c|}{\cellcolor{blue!30}7.068} & \multicolumn{1}{c|}{\cellcolor{blue!30}7.830} & \multicolumn{1}{c|}{\cellcolor{blue!30}8.387} & \multicolumn{1}{c|}{\cellcolor{blue!30}12.24} & \multicolumn{1}{c|}{\cellcolor{blue!30}29.20} & \multicolumn{1}{c|}{\cellcolor{blue!30}105.7} & \multicolumn{1}{c|}{$-$} & \multicolumn{1}{c|}{58.81} & \multicolumn{1}{c|}{84.80} & \multicolumn{1}{c|}{107.6} & \multicolumn{1}{c|}{182.0} & \multicolumn{1}{c|}{502.5} & \multicolumn{1}{c|}{1915} & \multicolumn{1}{c|}{8137} & $-$ \\ \cline{3-19} 
 & \multirow{-3}{*}{9} & Our PK & \multicolumn{1}{c|}{13.67} & \multicolumn{1}{c|}{18.84} & \multicolumn{1}{c|}{22.96} & \multicolumn{1}{c|}{45.54} & \multicolumn{1}{c|}{148.2} & \multicolumn{1}{c|}{570.9} & \multicolumn{1}{c|}{$-$} & \multicolumn{1}{c|}{$-$} & \multicolumn{1}{c|}{\cellcolor{blue!30}16.87} & \multicolumn{1}{c|}{\cellcolor{blue!30}22.04} & \multicolumn{1}{c|}{\cellcolor{blue!30}26.31} & \multicolumn{1}{c|}{\cellcolor{blue!30}48.98} & \multicolumn{1}{c|}{\cellcolor{blue!30}152.3} & \multicolumn{1}{c|}{\cellcolor{blue!30}581.5} & \multicolumn{1}{c|}{\cellcolor{blue!30}2034} & $-$ \\ \cline{2-19} 

& \multirow{3}{*}{10} & LG & \multicolumn{1}{c|}{11.77} & \multicolumn{1}{c|}{12.24} & \multicolumn{1}{c|}{15.80} & \multicolumn{1}{c|}{16.49} & \multicolumn{1}{c|}{17.48} & \multicolumn{1}{c|}{45.20} & \multicolumn{1}{c|}{$-$} & \multicolumn{1}{c|}{$-$} & \multicolumn{1}{c|}{66.19} & \multicolumn{1}{c|}{92.22} & \multicolumn{1}{c|}{125.1} & \multicolumn{1}{c|}{219.8} & \multicolumn{1}{c|}{582.1} & \multicolumn{1}{c|}{2179} & \multicolumn{1}{c|}{$-$} & $-$ \\ \cline{3-19} 
 &  & Our SK & \multicolumn{1}{c|}{\cellcolor{blue!30}7.780} & \multicolumn{1}{c|}{\cellcolor{blue!30}8.032} & \multicolumn{1}{c|}{\cellcolor{blue!30}8.635} & \multicolumn{1}{c|}{\cellcolor{blue!30}9.203} & \multicolumn{1}{c|}{\cellcolor{blue!30}14.54} & \multicolumn{1}{c|}{\cellcolor{blue!30}38.31} & \multicolumn{1}{c|}{$-$} & \multicolumn{1}{c|}{$-$} & \multicolumn{1}{c|}{71.58} & \multicolumn{1}{c|}{109.2} & \multicolumn{1}{c|}{132.7} & \multicolumn{1}{c|}{242.7} & \multicolumn{1}{c|}{684.0} & \multicolumn{1}{c|}{2687} & \multicolumn{1}{c|}{$-$} & $-$ \\ \cline{3-19} 
\multirow{-18}{*}{\begin{tabular}[c]{@{}c@{}}Time.\\      (s)\\      WAN\end{tabular}} & \multirow{-3}{*}{10} & Our PK & \multicolumn{1}{c|}{16.32} & \multicolumn{1}{c|}{23.05} & \multicolumn{1}{c|}{28.66} & \multicolumn{1}{c|}{59.79} & \multicolumn{1}{c|}{184.8} & \multicolumn{1}{c|}{708.5} & \multicolumn{1}{c|}{\cellcolor{blue!30}2792} & \multicolumn{1}{c|}{$-$} & \multicolumn{1}{c|}{\cellcolor{blue!30}19.90} & \multicolumn{1}{c|}{\cellcolor{blue!30}26.66} & \multicolumn{1}{c|}{\cellcolor{blue!30}32.44} & \multicolumn{1}{c|}{\cellcolor{blue!30}63.66} & \multicolumn{1}{c|}{\cellcolor{blue!30}189.3} & \multicolumn{1}{c|}{\cellcolor{blue!30}720.9} & \multicolumn{1}{c|}{\cellcolor{blue!30}2846} & $-$ \\ \hline \hline

\multirow{24}{*}{} & \multirow{3}{*}{3}  & LG & \multicolumn{1}{c|}{0.157} & \multicolumn{1}{c|}{0.284} & \multicolumn{1}{c|}{0.962} & \multicolumn{1}{c|}{3.662} & \multicolumn{1}{c|}{14.43} & \multicolumn{1}{c|}{57.58} & \multicolumn{1}{c|}{229.8} & \multicolumn{1}{c|}{917.0$^*$} & \multicolumn{1}{c|}{6.311} & \multicolumn{1}{c|}{7.904} & \multicolumn{1}{c|}{13.37} & \multicolumn{1}{c|}{32.58} & \multicolumn{1}{c|}{114.6} & \multicolumn{1}{c|}{474.3} & \multicolumn{1}{c|}{2052} & 8973$^*$ \\ \cline{3-19} 
 &  & Our SK & \multicolumn{1}{c|}{\cellcolor{blue!30}0.032} & \multicolumn{1}{c|}{\cellcolor{blue!30}0.111} & \multicolumn{1}{c|}{\cellcolor{blue!30}0.426} & \multicolumn{1}{c|}{\cellcolor{blue!30}1.690} & \multicolumn{1}{c|}{\cellcolor{blue!30}6.788} & \multicolumn{1}{c|}{\cellcolor{blue!30}27.87} & \multicolumn{1}{c|}{\cellcolor{blue!30}112.7} & \multicolumn{1}{c|}{\cellcolor{blue!30}455.6} & \multicolumn{1}{c|}{2.542} & \multicolumn{1}{c|}{3.582} & \multicolumn{1}{c|}{8.529} & \multicolumn{1}{c|}{30.91} & \multicolumn{1}{c|}{132.7} & \multicolumn{1}{c|}{588.8} & \multicolumn{1}{c|}{2614} & 11529 \\ \cline{3-19} 
 & \multirow{-3}{*}{3} & Our PK & \multicolumn{1}{c|}{1.815} & \multicolumn{1}{c|}{1.983} & \multicolumn{1}{c|}{2.655} & \multicolumn{1}{c|}{5.418} & \multicolumn{1}{c|}{16.36} & \multicolumn{1}{c|}{60.78} & \multicolumn{1}{c|}{239.3} & \multicolumn{1}{c|}{959.5} & \multicolumn{1}{c|}{\cellcolor{blue!30}2.084} & \multicolumn{1}{c|}{\cellcolor{blue!30}2.325} & \multicolumn{1}{c|}{\cellcolor{blue!30}3.073} & \multicolumn{1}{c|}{\cellcolor{blue!30}5.908} & \multicolumn{1}{c|}{\cellcolor{blue!30}16.92} & \multicolumn{1}{c|}{\cellcolor{blue!30}61.41} & \multicolumn{1}{c|}{\cellcolor{blue!30}240.0} & \cellcolor{blue!30}960.3 \\ \cline{2-19} 

&\multirow{3}{*}{4}   & LG & \multicolumn{1}{c|}{0.242} & \multicolumn{1}{c|}{0.449} & \multicolumn{1}{c|}{1.536} & \multicolumn{1}{c|}{5.868} & \multicolumn{1}{c|}{23.15} & \multicolumn{1}{c|}{92.38} & \multicolumn{1}{c|}{368.6} & \multicolumn{1}{c|}{1471$^*$} & \multicolumn{1}{c|}{9.591} & \multicolumn{1}{c|}{12.44} & \multicolumn{1}{c|}{23.89} & \multicolumn{1}{c|}{67.25} & \multicolumn{1}{c|}{253.7} & \multicolumn{1}{c|}{1074} & \multicolumn{1}{c|}{4674} & 20419$^*$ \\ \cline{3-19} 
 &  & Our SK & \multicolumn{1}{c|}{\cellcolor{blue!30}0.058} & \multicolumn{1}{c|}{\cellcolor{blue!30}0.204} & \multicolumn{1}{c|}{\cellcolor{blue!30}0.791} & \multicolumn{1}{c|}{\cellcolor{blue!30}3.145} & \multicolumn{1}{c|}{\cellcolor{blue!30}12.81} & \multicolumn{1}{c|}{\cellcolor{blue!30}51.69} & \multicolumn{1}{c|}{\cellcolor{blue!30}208.8} & \multicolumn{1}{c|}{\cellcolor{blue!30}843.7} & \multicolumn{1}{c|}{3.941} & \multicolumn{1}{c|}{6.385} & \multicolumn{1}{c|}{16.96} & \multicolumn{1}{c|}{66.89} & \multicolumn{1}{c|}{293.9} & \multicolumn{1}{c|}{1312} & \multicolumn{1}{c|}{5834} & 25751 \\ \cline{3-19} 
 & \multirow{-3}{*}{4} & Our PK & \multicolumn{1}{c|}{2.723} & \multicolumn{1}{c|}{2.975} & \multicolumn{1}{c|}{3.983} & \multicolumn{1}{c|}{8.126} & \multicolumn{1}{c|}{24.54} & \multicolumn{1}{c|}{101.9} & \multicolumn{1}{c|}{359.0} & \multicolumn{1}{c|}{1439} & \multicolumn{1}{c|}{\cellcolor{blue!30}3.127} & \multicolumn{1}{c|}{\cellcolor{blue!30}3.488} & \multicolumn{1}{c|}{\cellcolor{blue!30}4.610} & \multicolumn{1}{c|}{\cellcolor{blue!30}8.861} & \multicolumn{1}{c|}{\cellcolor{blue!30}25.38} & \multicolumn{1}{c|}{\cellcolor{blue!30}102.8} & \multicolumn{1}{c|}{\cellcolor{blue!30}360.1} & \multicolumn{1}{c|}{\cellcolor{blue!30}1440} \\ \cline{2-19}

&\multirow{3}{*}{5}  & LG & \multicolumn{1}{c|}{0.330} & \multicolumn{1}{c|}{0.630} & \multicolumn{1}{c|}{2.173} & \multicolumn{1}{c|}{8.325} & \multicolumn{1}{c|}{32.86} & \multicolumn{1}{c|}{131.2} & \multicolumn{1}{c|}{523.5} & \multicolumn{1}{c|}{2090$^*$} & \multicolumn{1}{c|}{12.92} & \multicolumn{1}{c|}{17.75} & \multicolumn{1}{c|}{36.48} & \multicolumn{1}{c|}{110.7} & \multicolumn{1}{c|}{435.2} & \multicolumn{1}{c|}{1868} & \multicolumn{1}{c|}{8228} & 35737$^*$ \\ \cline{3-19} 
 &  & Our SK & \multicolumn{1}{c|}{\cellcolor{blue!30}0.090} & \multicolumn{1}{c|}{\cellcolor{blue!30}0.323} & \multicolumn{1}{c|}{\cellcolor{blue!30}1.257} & \multicolumn{1}{c|}{\cellcolor{blue!30}5.007} & \multicolumn{1}{c|}{\cellcolor{blue!30}20.36} & \multicolumn{1}{c|}{\cellcolor{blue!30}82.11} & \multicolumn{1}{c|}{\cellcolor{blue!30}331.5} & \multicolumn{1}{c|}{\cellcolor{blue!30}1339} & \multicolumn{1}{c|}{5.504} & \multicolumn{1}{c|}{10.18} & \multicolumn{1}{c|}{30.44} & \multicolumn{1}{c|}{124.7} & \multicolumn{1}{c|}{548.9} & \multicolumn{1}{c|}{2445} & \multicolumn{1}{c|}{10832} & 47635 \\ \cline{3-19} 
 & \multirow{-3}{*}{5} & Our PK & \multicolumn{1}{c|}{3.630} & \multicolumn{1}{c|}{3.967} & \multicolumn{1}{c|}{5.311} & \multicolumn{1}{c|}{10.84} & \multicolumn{1}{c|}{32.72} & \multicolumn{1}{c|}{121.6} & \multicolumn{1}{c|}{478.7} & \multicolumn{1}{c|}{1919} & \multicolumn{1}{c|}{\cellcolor{blue!30}4.169} & \multicolumn{1}{c|}{\cellcolor{blue!30}4.650} & \multicolumn{1}{c|}{\cellcolor{blue!30}6.147} & \multicolumn{1}{c|}{\cellcolor{blue!30}11.82} & \multicolumn{1}{c|}{\cellcolor{blue!30}33.84} & \multicolumn{1}{c|}{\cellcolor{blue!30}122.8} & \multicolumn{1}{c|}{\cellcolor{blue!30}480.1} & {\cellcolor{blue!30}1921} \\ \cline{2-19} 

&\multirow{3}{*}{7}  & LG & \multicolumn{1}{c|}{0.519} & \multicolumn{1}{c|}{1.038} & \multicolumn{1}{c|}{3.635} & \multicolumn{1}{c|}{13.99} & \multicolumn{1}{c|}{55.29} & \multicolumn{1}{c|}{220.8} & \multicolumn{1}{c|}{881.3} & \multicolumn{1}{c|}{$-$} & \multicolumn{1}{c|}{19.92} & \multicolumn{1}{c|}{29.90} & \multicolumn{1}{c|}{41.59} & \multicolumn{1}{c|}{243.3} & \multicolumn{1}{c|}{997.9} & \multicolumn{1}{c|}{4326} & \multicolumn{1}{c|}{18924} & $-$ \\ \cline{3-19} 
 &  & Our SK & \multicolumn{1}{c|}{\cellcolor{blue!30}0.172} & \multicolumn{1}{c|}{\cellcolor{blue!30}0.634} & \multicolumn{1}{c|}{\cellcolor{blue!30}2.489} & \multicolumn{1}{c|}{\cellcolor{blue!30}10.03} & \multicolumn{1}{c|}{\cellcolor{blue!30}40.31} & \multicolumn{1}{c|}{\cellcolor{blue!30}162.5} & \multicolumn{1}{c|}{\cellcolor{blue!30}655.4} & \multicolumn{1}{c|}{$-$} & \multicolumn{1}{c|}{8.827} & \multicolumn{1}{c|}{18.80} & \multicolumn{1}{c|}{64.76} & \multicolumn{1}{c|}{275.4} & \multicolumn{1}{c|}{1226} & \multicolumn{1}{c|}{5474} & \multicolumn{1}{c|}{24275} & $-$ \\ \cline{3-19} 
 & \multirow{-3}{*}{7} & Our PK & \multicolumn{1}{c|}{5.445} & \multicolumn{1}{c|}{5.950} & \multicolumn{1}{c|}{7.966} & \multicolumn{1}{c|}{16.25} & \multicolumn{1}{c|}{49.07} & \multicolumn{1}{c|}{182.3} & \multicolumn{1}{c|}{718.0} & \multicolumn{1}{c|}{$-$} & \multicolumn{1}{c|}{\cellcolor{blue!30}6.253} & \multicolumn{1}{c|}{\cellcolor{blue!30}6.98} & \multicolumn{1}{c|}{\cellcolor{blue!30}9.220} & \multicolumn{1}{c|}{\cellcolor{blue!30}17.72} & \multicolumn{1}{c|}{\cellcolor{blue!30}50.76} & \multicolumn{1}{c|}{\cellcolor{blue!30}184.2} & \multicolumn{1}{c|}{\cellcolor{blue!30}720.1} & $-$ \\ \cline{2-19} 

&\multirow{3}{*}{9}  & LG & \multicolumn{1}{c|}{0.723} & \multicolumn{1}{c|}{1.509} & \multicolumn{1}{c|}{5.347} & \multicolumn{1}{c|}{20.65} & \multicolumn{1}{c|}{81.72} & \multicolumn{1}{c|}{326.3} & \multicolumn{1}{c|}{1303} & \multicolumn{1}{c|}{$-$} & \multicolumn{1}{c|}{28.15} & \multicolumn{1}{c|}{44.74} & \multicolumn{1}{c|}{115.0} & \multicolumn{1}{c|}{415.33} & \multicolumn{1}{c|}{1742} & \multicolumn{1}{c|}{7609} & \multicolumn{1}{c|}{33360} & $-$ \\ \cline{3-19} 
 &  & Our SK & \multicolumn{1}{c|}{\cellcolor{blue!30}0.279} & \multicolumn{1}{c|}{\cellcolor{blue!30}1.046} & \multicolumn{1}{c|}{\cellcolor{blue!30}4.122} & \multicolumn{1}{c|}{\cellcolor{blue!30}16.60} & \multicolumn{1}{c|}{66.76} & \multicolumn{1}{c|}{269.0} & \multicolumn{1}{c|}{1084} & \multicolumn{1}{c|}{$-$} & \multicolumn{1}{c|}{13.99} & \multicolumn{1}{c|}{33.04} & \multicolumn{1}{c|}{119.9} & \multicolumn{1}{c|}{516.3} & \multicolumn{1}{c|}{2295} & \multicolumn{1}{c|}{10207} & \multicolumn{1}{c|}{45079} & $-$ \\ \cline{3-19} 
 & \multirow{-3}{*}{9} & Our PK & \multicolumn{1}{c|}{7.260} & \multicolumn{1}{c|}{7.933} & \multicolumn{1}{c|}{10.62} & \multicolumn{1}{c|}{21.67} & \multicolumn{1}{c|}{\cellcolor{blue!30}65.43} & \multicolumn{1}{c|}{\cellcolor{blue!30}243.1} & \multicolumn{1}{c|}{\cellcolor{blue!30}957.3} & \multicolumn{1}{c|}{$-$} & \multicolumn{1}{c|}{\cellcolor{blue!30}8.338} & \multicolumn{1}{c|}{\cellcolor{blue!30}9.300} & \multicolumn{1}{c|}{\cellcolor{blue!30}12.29} & \multicolumn{1}{c|}{\cellcolor{blue!30}23.63} & \multicolumn{1}{c|}{\cellcolor{blue!30}67.68} & \multicolumn{1}{c|}{\cellcolor{blue!30}245.6} & \multicolumn{1}{c|}{\cellcolor{blue!30}960.1} & $-$ \\ \cline{2-19} 

&\multirow{3}{*}{10} & LG & \multicolumn{1}{c|}{0.831} & \multicolumn{1}{c|}{1.768} & \multicolumn{1}{c|}{6.296} & \multicolumn{1}{c|}{24.36} & \multicolumn{1}{c|}{96.43} & \multicolumn{1}{c|}{385.1} & \multicolumn{1}{c|}{$-$} & \multicolumn{1}{c|}{$-$} & \multicolumn{1}{c|}{32.32} & \multicolumn{1}{c|}{54.34} & \multicolumn{1}{c|}{148.7} & \multicolumn{1}{c|}{549.4} & \multicolumn{1}{c|}{2314} & \multicolumn{1}{c|}{10081} & \multicolumn{1}{c|}{$-$} & $-$ \\ \cline{3-19} 
 &  & Our SK & \multicolumn{1}{c|}{\cellcolor{blue!30}0.341} & \multicolumn{1}{c|}{\cellcolor{blue!30}1.288} & \multicolumn{1}{c|}{\cellcolor{blue!30}5.086} & \multicolumn{1}{c|}{\cellcolor{blue!30}20.48} & \multicolumn{1}{c|}{82.39} & \multicolumn{1}{c|}{332.0} & \multicolumn{1}{c|}{$-$} & \multicolumn{1}{c|}{$-$} & \multicolumn{1}{c|}{16.21} & \multicolumn{1}{c|}{40.56} & \multicolumn{1}{c|}{149.9} & \multicolumn{1}{c|}{651.6} & \multicolumn{1}{c|}{2901} & \multicolumn{1}{c|}{12908} & \multicolumn{1}{c|}{$-$} & $-$ \\ \cline{3-19} 
\multirow{-18}{*}{\begin{tabular}[c]{@{}c@{}}Comm.\\      (MB)\end{tabular}} & \multirow{-3}{*}{10} & Our PK & \multicolumn{1}{c|}{8.168} & \multicolumn{1}{c|}{8.925} & \multicolumn{1}{c|}{11.95} & \multicolumn{1}{c|}{24.38} & \multicolumn{1}{c|}{\cellcolor{blue!30}73.61} & \multicolumn{1}{c|}{\cellcolor{blue!30}273.5} & \multicolumn{1}{c|}{\cellcolor{blue!30}1077} & \multicolumn{1}{c|}{$-$} & \multicolumn{1}{c|}{\cellcolor{blue!30}9.380} & \multicolumn{1}{c|}{\cellcolor{blue!30}10.46} & \multicolumn{1}{c|}{\cellcolor{blue!30}13.83} & \multicolumn{1}{c|}{\cellcolor{blue!30}26.59} & \multicolumn{1}{c|}{\cellcolor{blue!30}76.14} & \multicolumn{1}{c|}{\cellcolor{blue!30}276.3} & \multicolumn{1}{c|}{\cellcolor{blue!30}1080} & $-$ \\ \hline

\end{tabular}}
\caption{Online and total running time / communication costs of LG and our protocols in LAN and WAN settings. $m$ is the number of parties. Our SK /PK denotes our SK-MPSU / PK-MPSU protocol. Cells with $*$ denotes trials that the original LG code will give wrong results. Cells with $-$ denotes trials that ran out of memory. The best protocol within a setting is marked in \textcolor{blue!30}{blue}.}
\label{tab:time}
\end{table*}

\begin{trivlist}
\item \textbf{Acknowledgements.} We would like to thank Jiahui Gao for the clarification of their work.
\end{trivlist}

%
\bibliographystyle{alpha}
\bibliography{MPSU}

\appendix

\section{Leakage Analysis of~\cite{GNT-eprint-2023}}\label{sec:attack}

The MPSU protocol in~\cite{GNT-eprint-2023} is claimed to be secure in the presence of arbitrary colluding participants. However, our analysis would suggest that the protocol fails to achieve this security, and also requires the non-collusion assumption as LG. First, we give a brief review of the protocol.

Apart from MKR-PKE, their protocol utilizes two new ingredients: 
1) The conditional oblivious pseudorandom function (cOPRF), an extension they develop on the OPRF, where the sender $\Sd$ additionally inputs a set $Y$. If $x \notin Y$, $\Rcv$ receives $F_k(x)$, else $\Rcv$ receives a random value sampled by $\Sd$.
2) The membership Oblivious Transfer (mOT), where $\Sd$ inputs an element $y$ and two messages $u_0,u_1$, while $\Rcv$ inputs a set $X$ and receives $u$, one of $u_0,u_1$. If $y \in X$, $u = u_0$, else $u = u_1$.

To illustrate insecurity of their protocol, we consider a three-party case where $P_1$ and $P_3$ each possess a single item $X_1 = \{x_1\}$ and $X_3 = \{x_3\}$ respectively, while $P_2$ possesses a set $X_2$. We assume that $x_1 = x_3$. According to the protocol (cf. Figure 8 in their paper), in step 3.(a), $P_1$ and $P_2$ invoke the OPRF where $P_1$ acts as $\Rcv$ inputting $x_1$ and $P_2$ acts as $\Sd$ inputting its PRF key $k_2$. $P_1$ receives the PRF value $F_{k_2}(x_1)$. Meanwhile, in step 3.(c), $P_2$ and $P_3$ invoke the cOPRF where $P_3$ acts as $\Rcv$ inputting $x_3$, and $P_2$ acts as $\Sd$ inputting its PRF key $k_2$ and the set $X_2$. $P_3$ receives the output $w$ from the cOPRF. By the definition of cOPRF functionality, if $x_3 \notin X_2$, $w = F_{k_2}(x_3)$, otherwise $w$ is a random value. 

If $P_1$ and $P_3$ collude, they can distinguish the cases where $x_3 \in X_2$ and $x_3 \notin X_2$ by comparing $P_1$'s output $F_{k_2}(x_1)$ from the OPRF and $P_3$'s output $w$ from the cOPRF for equality. To elaborate, we recall that $x_1 = x_3$, so $F_{k_2}(x_1) = F_{k_2}(x_3)$. If $F_{k_2}(x_1) = w$, it implies that $P_3$ receives $F_{k_2}(x_3)$ from the cOPRF, so the coalition learns that $x_3 \notin X_2$; On the contrary, if $F_{k_2}(x_1) \ne w$, it implies that $P_3$'s output from the cOPRF is not $F_{k_2}(x_3)$, so it is a random value, then the coalition learns that $x_3 \in X_2$. More generally, as long as $P_1$ and $P_3$ collude, they can identify whether each element $x \in X_1 \cap X_3$ belongs to $X_2$ or not, by comparing the PRF value $F_k(x)$ from the OPRF between $P_1$ and $P_2$ and the cPRF value (whose condition depends on $x \in X_2$ or not) from the cOPRF between $P_2$ and $P_3$. This acquired knowledge is information leakage in MPSU. Therefore, the non-collusion assumption is required.

\section{Public-Key Encryption}\label{sec:pke}

A public-key encryption (PKE) scheme is a tuple of PPT algorithms $(\Gen, \Enc, \Dec)$ such that:
\begin{itemize}
\item The key-generation algorithm $\Gen$ takes as input the security parameter $1^\lambda$ and outputs a pair of keys $(pk, sk) \in \mathcal{PK} \times \mathcal{SK}$.
\item The encryption algorithm $\Enc$ takes as input a public key $pk$ and a plaintext $x \in \mathcal{M}$, and outputs a ciphertext $\ct$. 
\item The decryption algorithm $\Dec$ takes as input a secret key $sk$ and a ciphertext $\ct$, and outputs a message $x$ or or an error symbol $\perp$.
\end{itemize}

\begin{trivlist}
\item \textbf{Correctness.} For any $(pk,sk)$ outputed by $\Gen(1^\lambda)$, and any $x \in \mathcal{M}$, it holds that $\Dec(sk,(\Enc(pk,x))) = x$.
\end{trivlist}

The IND-CPA security of PKE implies security for encryption of multiple messages whose definition is as follows:

\begin{definition}\label{def:mmi2}
A public-key encryption scheme $\mathcal{E} = (\Gen, \Enc, \Dec)$ has indistinguishable multiple encryptions if for all PPT adversaries $\mathcal{A}$ s.t. any tuples $(m_1, \cdots, m_q)$ and $({m}_1', \cdots, {m}_{q}')$ chosen by $\mathcal{A}$ (where $q$ is polynomial in $\lambda$):
\begin{gather*}
   \{\Enc(pk,m_1), \cdots, \Enc(pk,m_q):(pk,sk) \gets \Gen(1^\lambda)\} \overset{c}{\approx} \\
   \{\Enc(pk,{m}_1'), \cdots, \Enc(pk,{m}_{q}'):(pk,sk) \gets \Gen(1^\lambda)\}
\end{gather*}
\end{definition}


\section{Construction of MKR-PKE from ElGamal}\label{sec:elgamal}

MKR-PKE is instantiated with ElGamal encryption below:

\begin{itemize}
\item The key-generation algorithm $\Gen$ takes as input the security parameter $1^\lambda$ and generates $(\mathbb{G},g,p)$, where $\mathbb{G}$ is a cyclic group, $g$ is the generator and $q$ is the order. Outputs $sk$ and $pk = g^{sk}$. 
\item The randomized encryption algorithm $\Enc$ takes as input a public key $pk$ and a plaintext message $x \in \mathbb{G}$, samples $r \gets \mathbb{Z}_q$, and outputs $\ct = (\ct_1, \ct_2) = (g^r, x \cdot pk^r)$. 
\item The partial decryption algorithm $\ParDec$ takes as input a secret key share $sk$ and a ciphertext $\ct = (\ct_1, \ct_2)$, and outputs $\ct' = (\ct_1, \ct_2 \cdot \ct_1^{-sk})$.
\item The decryption algorithm $\Dec$ takes as input a secret key $sk$ and a ciphertext $\ct = (\ct_1, \ct_2)$, and outputs $x = \ct_2 \cdot \ct_1^{-sk}$. 
\item The rerandomization algorithm $\ReRand$ takes as input $pk$ and a ciphertext $\ct = (\ct_1, \ct_2)$, samples $r \gets \mathbb{Z}_q$, and outputs $\ct' = (\ct_1 \cdot g^{r'}, \ct_2 \cdot pk^{r'})$.
\end{itemize}

\section{Missing Security Proofs}
\subsection{The Proof of Theorem \ref{theorem:mssrot}}\label{proof:mssrot}

We consider the case that $\text{ch}_0 = 1$ and $\text{ch}_1 = m$, and $J = \{1,\cdots,m\}$. Let $P_1$ and $P_m$ input $b_1 \in \{0, 1\}$ and $b_m \in \{0, 1\}$, and $P_i$ ($1 \le i \le m$) input $\Delta_i$ respectively. We turn to proving the correctness and security of the protocol in Figure~\ref{fig:proto-mssrot} in this particular case. Note that in the different cases, the proof is essentially the same.

\begin{trivlist}
\item \textbf{Correctness.} From the description of the protocol, we have the following equations:
\begin{gather}
    r_1 = \bigoplus_{j=2}^{m}(r_{j,1}^{b_1} \oplus b_1 \cdot u_{j,1}) \oplus r_{1,m}^0 \oplus b_1 \cdot \Delta_1,  \label{eq:1} \\
    r_i = r_{i,1}^0 \oplus r_{i,m}^0 , 1 < i < m, \label{eq:2} \\
    r_m = \bigoplus_{j=1}^{m-1}(r_{j,m}^{b_m} \oplus b_m \cdot u_{j,m}) \oplus r_{m,1}^0 \oplus b_m \cdot \Delta_m, \label{eq:3} \\
    u_{j,1} = \Delta_j \oplus r_{j,1}^0 \oplus r_{j,1}^1, 
    u_{j,m} = \Delta_j \oplus r_{j,m}^0 \oplus r_{j,m}^1\label{eq:4}
\end{gather}
From the definition of Random OT functionality (Figure~\ref{fig:func-rot}), we have the following equations:
\begin{gather}
r_{j,1}^{b_1} = r_{j,1}^0 \oplus b_1 \cdot (r_{j,1}^0 \oplus r_{j,1}^1),\label{eq:5} \\
r_{j,m}^{b_m} = r_{j,m}^0 \oplus b_m \cdot (r_{j,m}^0 \oplus r_{j,m}^1),\label{eq:6} 
\end{gather}
Substitute Equation~\ref{eq:4},~\ref{eq:5},~\ref{eq:6} into Equation~\ref{eq:1},~\ref{eq:2},~\ref{eq:2} and cancel out the same terms, we obtain:
\begin{gather}
    r_1 = \bigoplus_{j=2}^{m}(r_{j,1}^{b_1} \oplus b_1 \cdot \Delta_j) \oplus r_{1,m}^0 \oplus b_1 \cdot \Delta_1,  \label{eq:7} \\
    r_m = \bigoplus_{j=1}^{m-1}(r_{j,m}^{b_m} \oplus b_m \cdot \Delta_j) \oplus r_{m,1}^0 \oplus b_m \cdot \Delta_m, \label{eq:8} 
\end{gather}
Substitute Equation~\ref{eq:2},~\ref{eq:7},~\ref{eq:8} into $r_1 \oplus (\bigoplus_{j=2}^{m-1}r_j) \oplus r_m$, we have $r_1 \oplus (\bigoplus_{j=2}^{m-1}r_j) \oplus r_m = \bigoplus_{i=1}^{m} (b_1 \oplus b_m) \cdot \Delta_i$. Then we can summarize that if $b_1 \oplus b_m = 0$, $r_1 = \bigoplus_{j=2}^m r_j$, else $r_1 = \Delta_1 \oplus (\bigoplus_{j=2}^m (r_j \oplus \Delta_j))$. This is exactly the functionality $\FuncmssROT$.

\item \textbf{Security.} We now prove the security of the protocol.

\begin{proof}
Let $\Corr$ denote the set of all corrupted parties and $\Honest$ denote the set of all honest parties. $\lvert \Corr \rvert = t$. 

Intuitively, the protocol is secure because all things the parties do are invoking $\FuncROT$ and receiving random messages. The simulator can easily simulate these outputs from $\FuncROT$ and protocol messages by generating random values, which are independent of honest parties' inputs. 

To elaborate, in the case that $P_1 \notin \Corr$ and $P_m \notin \Corr$, simulator receives all outputs $r_c$ of $P_c \in \Corr$ and needs to emulate each $P_c$'s view, including its private input $\Delta_c$, outputs $(r_{c,1}^0, r_{c,1}^1)$ and $(r_{c,m}^0, r_{c,m}^1)$ from $\FuncROT$. The simulator for corrupted $P_c$ runs the protocol honestly except that it simulates uniform outputs from $\FuncROT$ under the constraint that $r_{c,1}^0 \oplus r_{c,m}^0 = r_c$. Clearly, the joint distribution of all outputs $r_c$ of $P_c \in \Corr$, along with their view emulated by simulator, is indistinguishable from that in the real process.

In the case that $P_1 \in \Corr$ or $P_m \in \Corr$, since the protocol is symmetric with respect to the roles of $P_1$ and $P_m$, we focus on the case of corrupted $P_1$. The simulator receives all outputs $r_c$ of $P_c \in \Corr$. For $P_1$, its view consists of the choice bit $b_1$, its private input $\Delta_1$, outputs $(r_{1,m}^0, r_{1,m}^1)$, $\{r_{j,1}^{b_1}\}_{1 < j \le m}$ from $\FuncROT$ and protocol messages $\{u_{j,1}\}_{1 < j \le m}$ from $P_j$. For each $P_c(c \ne 1)$, its view consists of its private input $\Delta_c$, outputs $(r_{c,1}^0, r_{c,1}^1)$ and $(r_{c,m}^0, r_{c,m}^1)$ from $\FuncROT$. 

For $P_1$'s view, the simulator runs the protocol honestly except that it simulates uniform outputs $(r_{1,m}^0, r_{1,m}^1)$, $r_{i,1}^{b_1}$ from $\FuncROT$ and uniformly random messages $u_{i,1}$ from $P_i$ under the constraint $\bigoplus_{j=2}^{m}(r_{j,1}^{b_1} \oplus b_1 \cdot u_{j,1}) \oplus r_{1,m}^0 \oplus b_1 \cdot \Delta_1 = r_1$, where $P_i \in \Honest$. For other corrupted parties' view, it runs the protocol honestly except that it sets $r_{c,m}^0 = r_{c,1}^0 \oplus r_c$ and simulates uniform output $r_{c,m}^1$ from $\FuncROT$.

In the real execution, $P_1$ receives $u_{i,1} = \Delta_i \oplus r_{i,1}^0 \oplus r_{i,1}^1$. From the definition of ROT functionality, $r_{i,1}^0$ (when $b_1 =0$) or $r_{i,1}^1$(when $b_1 =0$) is uniform and independent of $P_1$'s view. Therefore, $u_{i,1}$ is uniformly at random from the perspective of $P_1$. Clearly, the joint distribution of all outputs $r_c$ of $P_c \in \Corr$, along with their view emulated by simulator, is indistinguishable from that in the real process.

In the case that $P_1 \in \Corr$ and $P_m \in \Corr$, the simulator receives all outputs $r_c$ of $P_c \in \Corr$. For $P_1$, its view consists of the choice bit $b_1$, its private input $\Delta_1$, outputs $(r_{1,m}^0, r_{1,m}^1)$, $\{r_{j,1}^{b_1}\}_{1 < j \le m}$ from $\FuncROT$ and protocol messages $\{u_{j,1}\}_{1 < j \le m}$ from $P_j$. For each $P_c(c \ne 1 , c \ne m)$, its view consists of its private input $\Delta_c$, outputs $(r_{c,1}^0, r_{c,1}^1)$ and $(r_{c,m}^0, r_{c,m}^1)$ from $\FuncROT$. For $P_m$, its view consists of its private input $\Delta_m$, outputs $(r_{m,1}^0, r_{m,1}^1)$, $\{r_{j,m}^{b_m}\}_{1 \le j < m}$ from $\FuncROT$ and protocol messages $\{u_{j,m}\}_{1 \le j < m}$ from $P_j$. 

For $P_1$'s view, the simulator runs the protocol honestly except that it simulates uniform outputs $r_{i,1}^{b_1}$ from $\FuncROT$ and uniformly random messages $u_{i,1}$ from $P_i$ under the constraint $\bigoplus_{j=2}^{m}(r_{j,1}^{b_1} \oplus b_1 \cdot u_{j,1}) \oplus r_{1,m}^0 \oplus b_1 \cdot \Delta_1 = r_1$, where $P_i \in \Honest$. 
For the view of $P_c(c \ne 1 , c \ne m)$, it runs the protocol honestly except that it sets $r_{c,m}^0 = r_{c,1}^0 \oplus r_c$ and simulates uniform output $r_{c,m}^1$ from $\FuncROT$.
For $P_m$'s view, it runs the protocol honestly with the following changes:
\begin{itemize}
    \item It simulates uniform outputs $r_{i,m}^{b_m}$ from $\FuncROT$ and uniform messages $u_{i,m}$ from $P_i$ under the constraint $\bigoplus_{j=1}^{m-1} (r_{j,m}^{b_m} \oplus b_m \cdot u_{j,m}) \oplus r_{m,1}^0 \oplus b_m \cdot \Delta_m = r_m$, where $P_i \in \Honest$. 
    \item It sets the output $r_{c,m}^{b_m}$ from $\FuncROT$ to be consistent with the partial view $(r_{c,m}^0, r_{c,m}^1)$ of each corrupted $P_c$ in preceding simulation, where $c \ne 1$ and $c \ne m$.
\end{itemize}
Clearly, the joint distribution of all outputs $r_c$ of $P_c \in \Corr$, along with their view emulated by simulator, is indistinguishable from that in the real process.
\end{proof}
\end{trivlist}

\subsection{The Proof of Theorem~\ref{theorem:ske}}~\label{proof:ske}

\begin{proof} 
This proof is supposed to be divided into two cases in terms of whether $P_1 \in \Corr$, since this determines whether the adversary has knowledge of the output. Nevertheless, the simulation of these two cases merely differ in the output reconstruction stage, thus we combine them together for the sake of simplicity. Specifically, the simulator receives the input $X_c$ of $P_c \in \Corr$ and the output $\bigcup_{i=1}^m X_i$ if $P_1 \in \Corr$. 

For each $P_c$, its view consists of its input $X_c$, outputs from $\FuncbssPMT$, $\FuncmssROT$, output $\vec{sh}_{c}'$ from $\FuncMS$ as its share, sampled values as shares of $\Delta$ for $\FuncmssROT$ and $m-1$ sets of shares $\{\vec{sh}_{i}'\}_{1 < i \le m}$($P_i$'s output from $\FuncMS$) from $P_i$ if $c = 1$. The simulator emulates each $P_c$'s view by running the protocol honestly with the following changes:
\begin{itemize}
\item In step 2, it simulates uniform outputs $\{e_{c,j}^b\}_{c < j \le m}$ and $\{e_{c,i}^b\}_{1 \le i < c}$ from $\FuncbssPMT$, on condition that $P_i,P_j \in \Honest$.
\item In step 3, it simulates uniform outputs $\{r_{c,ji}^b\}_{c < j \le m,  1 \le i < j}$ from $\FuncmssROT$, on condition that $\exists min(2,i) \le d \le j, P_d \in \Honest$. If $c \ne 1$, it simulates uniform $\{\Delta_{c,ji}^b\}_{c < j \le m,  1 \le i < j}$ as $P_c$'s random tapes.
\item In step 4, it simulates uniformly output $\vec{sh}_{c}'$ from $\FuncMS$.
\end{itemize}

Now we discuss the case when $P_1 \in \Corr$. In step 4 and 5, it computes $Y=\bigcup_{i=1}^m X_i \setminus X_1$ and constructs $\vec{v} \in (\{0,1\}^{l+\kappa})^{(m-1)B}$ as follows:
\begin{itemize}
    \item For $\forall x_i \in Y$, $v_i = x_i \Vert \Hash(x)$, $1 \leq i \leq \lvert Y \rvert$.
    \item For $\lvert Y \rvert <i \le (m-1)B$, samples $v_i \gets \{0,1\}^{l+\kappa}$.
\end{itemize}
Then it samples a random permutation $\pi:[(m-1)B] \to [(m-1)B]$ and computes $\vec{v'}=\pi(\vec{v})$. For $1 \le i \le m$, it samples share $\vec{sh}_{i}' \gets (\{0,1\}^{l+\kappa})^{(m-1)B}$, which satisfies $\bigoplus_{i = 1}^m \vec{sh}_{i}' = \vec{v'}$ and is consistent with the previous sampled $\vec{sh}_{c}'$ for each corrupted $P_c$. Add all $\vec{sh}_{i}'$ to $P_1$'s view and $\vec{sh}_{c'}'$ to each corrupted $P_{c'}$'s view($c' \ne 1$) as its output from $\FuncMS$, respectively.

The changes of outputs from $\FuncbssPMT$ and $\FuncmssROT$ have no impact on $P_c$'s view, for the following reasons. By the definition of $\FuncbssPMT$, each output $e_{c,j}^b$ and $e_{c,i}^b$ from $\FuncbssPMT$ is uniformly distributed as a secret-share between $P_c$ and $P_j$, or $P_i$ and $P_c$, where $P_i,P_j \in \Honest$. By the definition of $\FuncmssROT$, each output $r_{c,ji}^b$ from $\FuncmssROT$ is a secret-share of $0$ among $P_{min(2, i)}, \cdots, P_j$ if $e_{i,j}^b \oplus e_{j,i}^b = 0$, or a secret-share of $\bigoplus_{d=2}^j \Delta_{d,ji}^b$ if $e_{i,j}^b \oplus e_{j,i}^b = 1$. Therefore, even if $P_c$ colludes with others, $r_{c,ji}^b$ is still uniformly random from the perspective of adversary, since there always exists a party $P_d \in \Honest (min(2,i) \le d \le j)$ holding one share. 

It remains to demonstrate that the output $\vec{sh}_{c}'$ from $\FuncMS$($P_1 \notin \Corr$) or all outputs $\{\vec{sh}_{i}'\}_{1 \le i \le m}$ from $\FuncMS$($P_1 \in \Corr$) does not leak any other information except for the union. The former case is easier to tackle with. The output $\vec{sh}_{c}'$ is distributed as a secret-share among all parties, so it is uniformly distributed from the perspective of adversary. 

We now proceed to explain the latter case. For all $1 < j \le m$, consider an element $x \in X_j$ and $x$ is placed in the $b$th bin by $P_j$. In the real protocol, if there is no $X_i(1 \le i < j)$ s.t. $x \in X_i$, then for all $1 \le i < j$, $e_{i,j}^{y} \oplus e_{j,i}^{y} = 0$. By the $\FuncmssROT$ functionality in Figure~\ref{fig:func-mssrot}, each $r_{d,ji}^b$ is uniform in $\{0,1\}^ {l+\kappa}$ conditioned on $\bigoplus_{d=min(2,i)}^j r_{d,ji}^b = 0$. From the descriptions of the protocol, We derive that each $u_{d,j}^b$ is uniform in $\{0,1\}^ {l+\kappa}$ conditioned on $\bigoplus_{d=min(2,i)}^j u_{d,j}^b = x \Vert \Hash(x)$, namely, they are additive shares of $x \Vert \Hash(x)$ among all parties. This is exactly identical to the simulation. 

If there exists some $X_i(1 \le i < j)$ s.t. $x \in X_i$, then $e_{i,j}^b \oplus e_{j,i}^b = 1$. By the $\FuncmssROT$ functionality in Figure~\ref{fig:func-mssrot}, each $r_{d,ji}^b$ is uniform conditioned on $\bigoplus_{d=min(2,i)}^j r_{d,ji}^b = \bigoplus_{j'=2}^j \Delta_{j',ji}^b$, where each $\Delta_{j',ji}^b$ is uniformly held by $P_{j'}$ ($1 < j' \le j$). From the descriptions of the protocol, We derive that each $u_{d,j}^b$ is uniform conditioned on $\bigoplus_{d=min(2,i)}^j u_{d,j}^b = x \Vert \Hash(x) \oplus \bigoplus_{j'=2}^j \Delta_{j',ji}^b \oplus r$, where $r$ is the sum of remaining terms. Then, even if $P_1$ colludes with others, $\bigoplus_{d=min(2,i)}^j u_{d,j}^b$ is still uniformly random from the perspective of adversary, since there always exists a party $P_{j'} \in \Honest (1 < j' \le j)$ holding one uniform $\Delta_{j',ji}^b$ and independent of all honest parties' inputs. For all empty bins, $u_{d,j}^b$ is chosen uniformly random, so the corresponding $\bigoplus_{d=1}^j u_{d,j}^b$ is also uniformly at random, which is identical to the simulation. By the definition of $\FuncMS$, all parties additively share $\bigoplus_{d=1}^j u_{d,j}^b$ in a random permutation that maintains privacy against a coalition of arbitrary corrupted parties, and receive back $\{\vec{sh}_{i}'\}$, respectively. We conclude that all outputs $\{\vec{sh}_{i}'\}_{1 \le i \le m}$ from $\FuncMS$ distribute identically between the real and ideal executions.
\end{proof}

\subsection{The Proof of Theorem \ref{theorem:pke}}~\label{proof:pke}

\begin{proof}
The simulator receives the input $X_c$ of $P_c \in \Corr$ and the output $\bigcup_{i=1}^m X_i$ if $P_1 \in \Corr$. 

For each $P_c$, its view consists of its input $X_c$, outputs from $\FuncbssPMT$ and $\FuncROT$, protocol messages $\{u_{j,c,0}^b\}_{c < j \le m}, \{u_{j,c,1}^b\}_{c < j \le m}$ from $P_j$, rerandomization messages $\{{v}_{c,i}'^b\}_{1 < i < c}$ from $P_i$, $\pi_c$, permuted partial decryption messages $\vec{\ct}_{c-1}''$ from $P_{c-1}$ if $c > 1$, or $\vec{\ct}_{m}''$ from $P_{m}$ if $c = 1$. The simulator emulates each $P_c$'s view by running the protocol honestly with the following changes:

\begin{itemize}
\item In step 2, it simulates uniform outputs $\{e_{c,j}^b\}_{c < j \le m}$ and $\{e_{c,i}^b\}_{1 \le i < c}$ from $\FuncbssPMT$, on condition that $P_i,P_j \in \Honest$.
\item In step 3, it simulates uniform outputs $\{r_{c,i,0}^b\}_{1 \le i < c}$, $\{r_{c,i,1}^b\}_{1 \le i < c}$ from $\FuncROT$, and $\{r_{j,c,e_{c,j}^b}^b\}_{c < j \le m}$ from $\FuncROT$, on condition that $P_i,P_j \in \Honest$.
For $c<j \le m$, it computes $u_{j,c,e_{c,j}^b}^b = r_{j,c,e_{c,j}^b}^b \oplus \Enc(pk, \perp)$ and simulates $u_{j,c,e_{c,j}^b \oplus 1}^b$ uniformly at random, on condition that $P_j \in \Honest$.
For $1 < i < c$, it simulates ${v}_{c,i}'^b = \Enc(pk, \perp)$, on condition that $P_i \in \Honest$.
\item If $P_1 \notin \Corr$, in step 4, for $1 \le i \le (m-1)B$, it computes ${\ct}_{c-1}'^i = \Enc(pk, \perp)$, and then simulates the vector $\vec{\ct}_{c-1}'' = {\pi}(\vec{\ct}_{c-1}')$ from $P_{c-1}$, where $\pi$ is sampled uniformly random and $P_{c-1} \in \Honest$.
\end{itemize}

Now we discuss the case when $P_1 \in \Corr$. In step 4, assume $d$ is the largest number that $P_d \in \Honest$, namely, $P_{d+1}, \cdots, P_m \in \Corr$. The simulator emulates the partial decryption messages $\vec{\ct}_{d}''$ from $P_{d}$ in the view of $P_{d+1}$ as follows:
\begin{itemize}
    \item For $\forall x_i \in Y = \bigcup_{j=1}^m X_j$, ${\ct}_{d}'^i = \Enc(pk_{A}, x_i)$, $1 \leq i \leq \lvert Y \rvert$.
    \item For $\lvert Y \rvert <i \le (m-1)B$, sets ${\ct}_{d}'^i = \Enc(pk_{A}, \perp)$.
\end{itemize}
where $pk_{A_d} = pk_1 \cdot \prod_{j= d+1}^m pk_j$. Then it samples a random permutation $\pi:[(m-1)B] \to [(m-1)B]$ and computes $\vec{\ct}_{d}''=\pi(\vec{\ct}_{d}')$.

For other corrupted $P_{d'+1} \in \{P_2, \cdots, P_{d-1}\}$, if $P_{d'} \in \Honest$, it simulates each partial decryption message ${\ct}_{d'}'^i = \Enc(pk, \perp)$ for $1 \le i \le (m-1)B$, and then computes the vector $\vec{\ct}_{d'}'' = {\pi}(\vec{\ct}_{d'}')$ from $P_{d'}$, where $\pi_{d'}$ is sampled uniformly random. Append $\vec{\ct}_{d'}''$ to the view of $P_{d'+1}$.

The changes of outputs from $\FuncbssPMT$ and $\FuncROT$ have no impact on $P_c$'s view, for similar reasons in Theorem~\ref{theorem:ske}. 

Indeed, $u_{j,c,e_{c,j}^b \oplus 1}^b$ is uniform in the real process, as $r_{j,c,e_{c,j}^b \oplus 1}^b$ (which is one of $P_j$'s output from $\FuncROT$ hidden from $P_c$, and is used to mask the encrypted message in $u_{j,c,e_{c,j}^b \oplus 1}^b$) is uniform and independent of $r_{j,c,e_{c,j}^b}^b$ from $P_c$'s perspective. 

It's evident from the descriptions of the protocol and the simulation that the simulated $u_{j,c,e_{c,j}^b}^b$ is identically distributed to that in the real process, conditioned on the event $e_{c,j}^b \oplus e_{j,c}^b = 1$. The analysis in the case of $e_{c,j}^b \oplus e_{j,c}^b = 0$ can be further divided into two subcases, $c \ne 1$ and  $c = 1$. We first argue that when $c \ne 1$, $u_{j,c,e_{c,j}^b}^b$ emulated by simulator is indistinguishable from that in the real process.

In the real process, for $1 < c < j \le m, 1 \le b \le B$, if $\Elem(\mathcal{C}_j^b) \in X_j \setminus (X_2 \cup \cdots \cup X_{c-1})$, $c_j^b = \Enc(pk, \Elem(\mathcal{C}_j^b))$, $u_{j,c,e_{c,j}^b}^b = r_{j,c,e_{c,j}^b}^b \oplus \Enc(pk, \Elem(\mathcal{C}_j^b))$; else $c_j^b = \Enc(pk, \perp)$, $u_{j,c,e_{c,j}^b}^b = r_{j,c,e_{c,j}^b}^b \oplus \Enc(pk, \perp)$. In the real process, for $1 < c < j \le m, 1 \le b \le B$, $u_{j,c,e_{c,j}^b}^b = r_{j,c,e_{c,j}^b}^b \oplus \Enc(pk, \perp)$. If there exists an algorithm that distinguishes these two process, it implies the existence of an algorithm that can distinguish two lists of encrypted messages, with no knowledge of $sk$ (since $sk$ is secret-shared among $m$ parties, it is uniformly distributed for any coalition of $m-1$ parties). Consequently, this implies the existence of a adversary to break the indistinguishable multiple encryptions of $\mathcal{E}$ in Definition~\ref{def:mmi2} (where $q = (m-1)B$).  

When $c = 1$, $u_{j,e_{1,j}^b}^b$ emulated by simulator is indistinguishable from that in the real process for the similar reason as the above analysis when $c > 1$.

Next, we start demonstrating that all ${v}_{c,i}'^b = \Enc(pk, \perp)$ emulated by simulator are indistinguishable from the real ones via the sequences of hybrids:

\begin{itemize}
    \item $\mathsf{Hyb}_{0}$ The real interaction. For $1 < i < c, 1 \le b \le B$: If $\Elem(\mathcal{C}_c^b) \in X_c \setminus (X_1 \cup \cdots \cup X_i)$, ${v}_{c,i}^b = \Enc(pk, \Elem(\mathcal{C}_c^b))$; else ${v}_{c,i}^b = \Enc(pk, \perp)$. ${v}_{c,i}'^b = \ReRand(pk, {v}_{c,i}^b)$.
    \item $\mathsf{Hyb}_{1}$ For $1 < i < c, 1 \le b \le B$: If $\Elem(\mathcal{C}_c^b) \in X_c \setminus (X_1 \cup \cdots \cup X_i)$, ${v}_{c,i}'^b = \Enc(pk, \Elem(\mathcal{C}_c^b))$; else ${v}_{c,i}'^b = \Enc(pk, \perp)$. This change is indistinguishable by the rerandomizable property of $\mathcal{E}$. 
    \item $\mathsf{Hyb}_{2}$ For $1 < i < c, 1 \le b \le B$: ${v}_{c,i}'^b = \Enc(pk, \perp)$. This change is indistinguishable by the indistinguishable multiple encryptions of $\mathcal{E}$.
\end{itemize}

When $P_1 \in \Corr$, we prove that $\vec{\ct}_{d}''$ emulated by simulator is indistinguishable from that in the real process via the sequences of hybrids:

\begin{itemize}
    \item $\mathsf{Hyb}_{0}$ The real interaction. $\vec{\ct}_{1}'' = {\pi_{1}}(\vec{\ct}_{1}')$. For $2 \le j \le d, 1 \le i \le (m - 1)B$:  ${\ct}_{j}^i = \ParDec(sk_{j}, {\ct}_{j-1}''^i), {\ct}_{j}'^i = \ReRand(pk_{A_{j}}, {\ct}_{j}^i)$, $\vec{\ct}_{j}'' = {\pi_{j}}(\vec{\ct}_{j}')$.
    \item $\mathsf{Hyb}_{1}$ For $2 \le j \le d, 1 \le i \le (m - 1)B$: ${\ct}_{j}^i = \ParDec(sk_{j}, {\ct}_{j-1}^i)$. $\vec{\ct}_{d}' = \ReRand(pk_{A_{d}},\vec{\ct}_{d})$, $\vec{\ct}_{d}'' = {\pi}(\vec{\ct}_{d}')$, where $\pi = \pi_1 \circ \cdots \circ \pi_{d}$. 
    It's easy to see that $\mathsf{Hyb}_{1}$ is identical to $\mathsf{Hyb}_{0}$.
    \item $\mathsf{Hyb}_{2}$ $\vec{\ct}_{1}$ is replaced by the following:
    \begin{itemize}
    \item For $\forall x_i \in Y = \bigcup_{j=1}^m X_j$, $\ct_1^i = \Enc(pk, x_i)$, $1 \leq i \leq \lvert Y \rvert$.
    \item For $\lvert Y \rvert < i \le (m-1)B$, sets $\ct_1^i = \Enc(pk, \perp)$.
    \end{itemize}
    $\mathsf{Hyb}_{2}$ rearranges $\vec{\ct}_{1}$ and it is identical to $\mathsf{Hyb}_{1}$ as the adversary is unaware of $\pi_{d}$ s.t. the order of elements in $\vec{\ct}_{1}$ has no effect on the result of $\vec{\ct}_{d}''$.
    \item $\mathsf{Hyb}_{3}$ $\vec{\ct}_{d}$ is replaced by the following:
    \begin{itemize}
    \item For $\forall x_i \in Y = \bigcup_{j=1}^m X_j$, $\ct_d^i = \Enc(pk_{A_{d}}, x_i)$, $1 \leq i \leq \lvert Y \rvert$.
    \item For $\lvert Y \rvert < i \le (m-1)B$, sets $\ct_d^i = \Enc(pk_{A_{d}}, \perp)$.
    \end{itemize}
    The indistinguishability between $\mathsf{Hyb}_{3}$ and $\mathsf{Hyb}_{2}$ is implied by the partially decryptable property of $\mathcal{E}$.
    \item $\mathsf{Hyb}_{4}$ $\vec{\ct}_{d}'$ is replaced by the following:
    \begin{itemize}
    \item For $\forall x_i \in Y = \bigcup_{j=1}^m X_j$, ${\ct}_{d}'^i = \Enc(pk_{A_{d}}, x_i)$, $1 \leq i \leq \lvert Y \rvert$.
    \item For $\lvert Y \rvert < i \le (m-1)B$, sets ${\ct}_{d}'^i = \Enc(pk_{A_{d}}, \perp)$.
    \end{itemize}
    $\mathsf{Hyb}_{4}$ is indistinguishable to $\mathsf{Hyb}_{3}$ because of the rerandomizable property of $\mathcal{E}$.
    \item $\mathsf{Hyb}_{5}$ The only change in $\mathsf{Hyb}_{5}$ is that $\pi$ are sampled uniformly by the simulator. $\mathsf{Hyb}_{5}$ generates the same $\vec{\ct}_{d}''$ as in the simulation. Given that $\pi_{d}$ is uniform in the adversary's perspective, the same holds for $\pi$, so $\mathsf{Hyb}_{5}$ is identical to $\mathsf{Hyb}_{4}$.
\end{itemize}
 
When $P_1 \notin \Corr$, the simulator is unaware of the final union, so it  has to emulate the partial decryption messages $\vec{\ct}_{c-1}''$ as permuted $\Enc(pk, \perp)$ in the view of $P_{c}$. Compared to the above hybrid argument, we only need to add one additional hybrid after $\mathsf{Hyb}_{4}$ to replace all rerandomized partial decryption messages $\vec{\ct}_{c-1}'$ with $\Enc(pk,\perp)$. This change is indistinguishable by the indistinguishable multiple encryptions of $\mathcal{E}$, as the adversary cannot distinguish two lists of messages with no knowledge of the partial secret key $sk_{A_{c-1}} = sk_1 + sk_c + \cdots + sk_m$ (it is unaware of $sk_1$).

The same applies for the simulation of the partial decryption messages $\vec{\ct}_{d'}''$ in the view of corrupted $P_{d'+1}$ ($d' \ne d$) when $P_1 \in \Corr$. $\vec{\ct}_{d'}''$ is also emulated by permuted $\Enc(pk,\perp)$. To avoid repetition, we omit the analysis here.
\end{proof}

\section{Complexity Analyses and Comparisons}~\label{sec:analysis} 

\subsection{Complete Analysis of LG}\label{sec:analysis-LG}

The costs of each stage in LG are calculated as follows.

\begin{trivlist}
\item \textbf{mq-ssPMT.} The cost of this stage consists of three parts:
\begin{itemize}
\item SKE encryption: The computation complexity of encrypting $0, 1,\cdots, n-1$ is $O(n)$.
\item OKVS: The computation complexities of Encode and Decode algorithms are both $O(n)$. The size of OKVS is
 $1.28 \kappa n$ bits, where $\kappa$ is the size of a SKE ciphertext with the same range as before.
\item ssVODM: The ssVODM protocol requires a total of $(T + l - \log n - 1)n$ AND gates, where $T$ is the number of AND gates in the SKE decryption circuit.
\end{itemize}

For $1 \le i < j \le m$, $P_i$ and $P_j$ invoke mq-ssPMT of size $B$. Overall, each party $P_j$ engages in $(m-1)$ instances of mq-ssPMT.

\begin{itemize}
\item \textbf{Offline phase.} The computation complexity per party is $O((T + l - \log n) m n \log ((T + l - \log n) n))$. The communication complexity per party is $O(t \lambda m \log (((T + l - \log n) n) / t))$. The round complexity is $O(1)$.

\item \textbf{Online phase.} The computation complexity per party is $O((T + l - \log n) m n)$. The communication complexity per party is $O((T + l + \kappa - \log n) m n)$. The round complexity is $O(\log (l - \log n))$.
\end{itemize}

\item \textbf{Random oblivious transfers.} For $1 \le i < j \le m$, $P_i$ and $P_j$ invoke ROT extension of size $B$. Overall, each party $P_j$ engages in $(m-1)$ instances of ROT extension of size $B$.

\begin{itemize}
\item \textbf{Offline phase.} The computation complexity per party is $O(m n\log n)$. The communication complexity per party is $O(t \lambda m \log (n / t))$. The round complexity is $O(1)$.

\item \textbf{Online phase.} The computation complexity per party is $O(m n)$. The communication complexity per party is $O(m n)$. The round complexity is $O(1)$.
\end{itemize}

The costs of the remaining two steps, \textbf{multi-party secret-shared shuffle} and \textbf{output reconstruction}, are exactly the same as those in our SK-MPSU.(cf. Appendix ~\ref{sec:analysis-ske}).
\end{trivlist}

\begin{trivlist}
\item \textbf{Total costs.} 
\begin{itemize}
\item \textbf{Offline phase.} The computation complexity per party is $O((T + l - \log n) m n \log ((T + l - \log n) n) + m^2 n \log m n)$. The communication complexity per party is $O(t \lambda m \log (((T + l - \log n) n) / t) + \lambda m^2 n (\log m + \log n))$. The round complexity is $O(1)$.

\item \textbf{Online phase.} The computation complexity of $P_1$ is $O((T + l - \log n) m n + m^2 n)$. The communication complexity of $P_1$ is $O((T + l + \kappa - \log n) m n + (l + \kappa) m^2 n)$. The computation complexity of $P_j$ is $O((T + l - \log n) m n)$. The communication complexity of $P_j$ is $O((T + l + \kappa - \log n) m n)$. The round complexity is $O(\log (l - \log n) + m)$.
\end{itemize}
\end{trivlist}

\subsection{Theoretical Analysis of Batch ssPMT}~\label{sec:analysis-bsspmt}

\subsubsection{Complete Analysis}

The costs of each stage in $\ProtobssPMT$ (Figure~\ref{fig:proto-bssPMT}) are calculated as follows.

\begin{trivlist}
\item \textbf{Batch OPPRF.} The cost of batch OPPRF mainly consists of two parts: 
\begin{itemize}
    \item Batch OPRF: There are several options for instantiating batch OPRF functionality~\cite{KKRT-CCS-2016,BuiC23}. We opt for subfield vector oblivious linear evaluation (subfield-VOLE)~\cite{BCGIKRS19,BCGIKS19,RRT23} to instantiate batch OPRF using the approach in~\cite{BuiC23}. 
    
    For $B = O(n)$ instances of OPRF, we require performing a subfield-VOLE of size $B$ in the offline phase, and sending $B$ derandomization messages of length $l$ in the online phase.
    We resort to \textit{Dual LPN with Fixed Weight and Regular Noise}~\cite{BCGIKRS19,BCGIKS19} to improve efficiency of subfield-VOLE, and instantiate the code family with Expand-Convolute codes~\cite{RRT23}, which enables near-linear time syndrome computation. The computation complexity is $O(n\log n)$. The computation complexity is $O(t\lambda \log n / t)$, where $t$ is the fixed noise weight\footnote{For instance, $t \approx 128$.}. The round complexity is $O(1)$.

    We denote the output length of OPPRF as $\gamma$. The lower bound of $\gamma$ is relevant to the total number of the batch OPPRF invocations. In our SK-MPSU and PK-MPSU, for $1 \le i < j \le m$, $P_i$ and $P_j$ invoke batch OPPRF. Overall, there are $1 + 2 + \cdots + (m-1) = (m^2 - m) / 2$ invocations of batch OPPRF. Considering all these invocations, we set $\gamma \ge \sigma + \log((m^2 - m) / 2) + \log_2 B$, so that the probability of any $t_i \ne s_i$ occurring if $x \notin X_i$, which is $((m^2 - m) / 2)) B \cdot 2^{-\gamma}$, is less than or equal to $2^{-\sigma}$.

    \item OKVS: The size of key-value pairs encoded into OKVS is $\lvert K_1 \rvert + \cdots + \lvert K_B \rvert$. Taking into account the invocation of batch ssPMT in our two MPSU protocols in advance, this size is $3n$. We use the OKVS construction of~\cite{RR-CCS-2022}, and the computation complexities of $\Encode$ and $\Decode$ algorithms are both $O(n)$. As we employ their $w = 3$ scheme with a cluster size of $2^{14}$, the size of OKVS is $1.28 \gamma \cdot (3n)$ bits.
\end{itemize}


\begin{itemize}
\item \textbf{Offline phase.} The computation complexity per party is $O(n\log n)$. The communication complexity per party is $O(t \lambda \log (n / t))$. The round complexity is $O(1)$.

\item \textbf{Online phase.} The computation complexity per party is $O(n)$. The communication complexity per party is $O(\gamma n)$. The round complexity is $O(1)$.
\end{itemize}

\item \textbf{Secret-shared private equality tests.} 
Like \cite{PSTY-EUROCRYPT-2019}, we instantiate ssPEQT (Figure~\ref{fig:func-sspeqt}) using the generic MPC techniques. The circuit of $\FuncssPEQT$ is composed of $\gamma-1$ AND gates in GMW~\cite{GMW-STOC-1987}, where the inputs are already in the form of secret-shaing. Executing $\gamma-1$ AND gates in sequence would incur $\gamma-1$ rounds. To reduce the round complexity, we leverage a divide-and-conquer strategy and recursively organize the AND gates within a binary tree structure, where each layer requires one round. This ensures that the number of rounds is directly related to the depth of the tree (i.e., $O(\log{\gamma})$). To sum up, each party invoke $B$ instances of ssPEQT, which amounts to $(\gamma-1) B$ AND gates and takes $O(\log{\gamma})$ rounds.

We use silent OT extension~\cite{BCGIKRS19} to generate Beaver triples in offline phase, then each AND gate only requires 4 bits communication and $O(1)$ computation in the online phase. As a result, an invocation of ssPEQT requires $4 \gamma$ bits communication and $O(1)$ computation in the online phase\footnote{When calculating computational complexity, one evaluation on PRG or one hash operation is usually considered as $O(1)$ operation. However, here, the computational unit is bitwise XOR operation, and it's evident that performing $O(\gamma)$ bitwise XOR operations is much faster than executing $O(1)$ PRG evaluation or hash operation. Therefore, we count the online computation complexity of ssPEQT as $O(1)$}.

\begin{itemize}
\item \textbf{Offline phase.} The computation complexity per party is $O(\gamma n\log n)$. The communication complexity per party is $O(t \lambda \log (\gamma n / t))$. The round complexity is $O(1)$.

\item \textbf{Online phase.} The computation complexity per party is $O(n)$. The communication complexity per party is $O(\gamma n)$. The round complexity is $O(\log \gamma)$.
\end{itemize}
\end{trivlist}

\subsubsection{Comparison with mq-ssPMT}\label{compare:lg}

Liu and Gao adapt the SKE-based mq-RPMT in~\cite{ZCLZL-USENIX-2023} to construct mq-ssPMT. The most expensive part of their construction is secret-shared oblivious decryption-then-matching (ssVODM)~\cite{ZCLZL-USENIX-2023}, which is to implement a decryption circuit and a comparison circuit by the GMW protocol~\cite{GMW-STOC-1987}. In total, the ssVODM requires $(T + l -\log n -1)n$ AND gates, where $l$ is the bit length of set elements, and $T$ is the number of AND gates in the SKE decryption circuit and is set to be considerably large ($\approx 600$) according to their paper.

The costs of batch ssPMT of size $B = 1.27n$\footnote{We use stash-less Cuckoo hashing \cite{PSTY-EUROCRYPT-2019} with 3 hash functions, where $B = 1.27n$.} consist of the costs of batch OPPRF of size $1.27n$ and the costs of $1.27n$ instances of ssPEQT, where batch OPPRF can be instantiated by extremely fast specialized protocol and ssPEQT is implemented by the GMW protocol. 
Moreover, the state-of-the-art batch OPPRF construction \cite{CGS22,RS-EUROCRYPT-2021,RR-CCS-2022} can achieve linear computation and communication with respect to $n$ . In ssPEQT, the parties engage in $1.27 (\gamma-1) n$ AND gates, where $\gamma$ is the output length of OPPRF. In the typical setting where $n \le 2^{24}, l \le 128, \gamma \le 64$, we have $(T + l -\log n -1) n \gg 1.27 (\gamma-1) n$, which means that the number of AND gates desired involved in batch ssPMT is far smaller than mq-ssPMT. Therefore, the construction built on batch ssPMT greatly reduces the dependency on general 2PC and significantly decreases both computation and communication complexity by a considerable factor, compared to mq-ssPMT.

\subsection{Theoretical Analysis of Our SK-MPSU}~\label{sec:analysis-ske}

\subsubsection{Complete Analysis}

The costs of each stage in $\ProtoMPSUTwo$ (Figure~\ref{fig:proto-mpsu2}) are calculated as follows.

\begin{trivlist}

\item \textbf{Batch ssPMT.} To achieve linear communication of this stage, we use stash-less Cuckoo hashing \cite{PSTY-EUROCRYPT-2019}. To render the failure probability (failure is defined as the event where an item cannot be stored in the table and must be stored in the stash) less than $2^{-40}$, we set $B = 1.27n = O(n)$ for 3-hash Cuckoo hashing. The cost of batch ssPMT follows the complexity analysis in Section~\ref{bssPMT}. For $1 \le i < j \le m$, $P_i$ and $P_j$ invoke batch ssPMT. Overall, each party $P_j$ engages in $m-1$ invocations of batch ssPMT, acting as $\Rcv$ in the first $j-1$ invocations, and acting as $\Sd$ in the last $m-j$ invocations.

\begin{itemize}
\item \textbf{Offline phase.} The computation complexity per party is $O(\gamma m n\log n)$. The communication complexity per party is $O(t \lambda m \log (\gamma n / t))$. The round complexity is $O(1)$.

\item \textbf{Online phase.} The computation complexity per party is $O(m n)$. The communication complexity per party is $O(\gamma m n)$. The round complexity is $O(\log \gamma)$.
\end{itemize}

\item \textbf{Multi-party secret-shared random oblivious transfers.} 
Each invocation of mss-ROT involves pairwise executions of two-party OT, where each OT execution consists of two parts: In the offline phase, the parties invoke random-choice-bit ROT, then $\Sd$ sends one messages of length $l + \kappa$ ($\kappa = \sigma + \log (m-1) +\log n$) to $\Rcv$; In the online phase, $\Rcv$ sends a 1-bit derandomization message to transform ROT into a chosen-input version. If $P_i$ and $P_j$ hold choice bits, then $P_i$ and $P_j$ invoke OT with the parties in $\{P_2, \cdots, P_j\}$ except itself. 

For $1 \le i < j \le m$, $P_{min(2,i)}, \cdots, P_j$ engage in $B$ instances of mss-ROT, where $P_i$ and $P_j$ hold the two choice bits. Considering the overall invocations of mss-ROT: $P_1$ invokes $[\sum_{j = 2}^m (j-1)]B = \frac{m^2B -mB}{2}$ instances of two-party OT; For $1 < i \le m$, each party $P_i$ invokes $[(i - 1) + \sum_{j = i+1}^m 3(j-2)]B = \frac{3m^2B - 3i^2B -9mB +11iB -2B}{2}$ instances of two-party OT.

\begin{itemize}
\item \textbf{Offline phase.} The computation complexity per party is $O(m^2 n\log n)$. The communication complexity per party is $O(t \lambda m^2 \log (n / t))$. The round complexity is $O(1)$.

\item \textbf{Online phase.} The computation complexity per party is $O(m^2 n)$. The communication complexity per party is $O(m^2 n)$. The round complexity is $O(1)$.
\end{itemize}

\item \textbf{Multi-party secret-shared shuffle.} We use the construction in~\cite{EB22}. In the offline
phase, each pair of parties runs a Share Translation protocol \cite{CGP-ASIACRYPT-2020} of size $(m-1)B$ and length $l+\kappa$.

\begin{itemize}
\item \textbf{Offline phase.} The computation complexity per party is $O(m^2 n \log (mn))$. The communication complexity per party is $O(\lambda m^2 n \log (mn))$. The round complexity is $O(1)$.

\item \textbf{Online phase.} The computation complexity of $P_1$ is $O(m^2 n)$. The communication complexity of $P_1$ is $O((l+\kappa) m^2 n)$. The computation complexity of $P_j$ is $O(m n)$. The communication complexity of $P_j$ is $O((l+\kappa) m n)$. The round complexity is $O(m)$.
\end{itemize}

\item \textbf{Output reconstruction.} For $1 < j \le m$, $P_j$ sends $\vec{sh}_{j}'  \in (\{0,1\}^{l+\kappa})^{(m-1)B}$ to $P_1$. $P_1$ reconstructs $(m-1)B$ secrets, each having $m$ shares.

\begin{itemize}
\item \textbf{Online phase.} The computation complexity of $P_1$ is $O(m^2 n)$. The communication complexity of $P_1$ is $O((l+\kappa) m^2 n)$. The communication complexity of $P_j$ is $O((l+\kappa) m n)$. The round complexity is $O(1)$.
\end{itemize}

\item \textbf{Total costs.} 
\begin{itemize}
\item \textbf{Offline phase.} The offline computation complexity per party is $O(\gamma m n \log n + m^2 n (\log m + \log n))$. The offline communication complexity per party is $O(t \lambda m \log (\gamma n / t) + t \lambda m^2 \log (n / t) + \lambda m^2  n (\log m + \log n))$. The offline round complexity is $O(1)$.

\item \textbf{Online phase.} The online computation complexity per party is $O(\gamma m n + m^2 n)$. The online communication complexity of $P_1$ is $O(\gamma m n + (l + \kappa) m^2 n)$. The online communication complexity of $P_j$ is $O(\gamma m n + m^2 n + (l + \kappa) m n)$. The online round complexity is $O(\log \gamma + m)$.
\end{itemize}
\end{trivlist}

\subsubsection{Comparison with LG}\label{sec:compare-lg}

We present a comparison of the theoretical computation and communication complexity for each party in both offline and online phases between LG and our SK-MPSU.

\begin{table*}
\centering
\resizebox{\textwidth}{!}{%
  \begin{tabular}{|c|c|c|c|c|c|c|c|c|}
        \hline
        & \multicolumn{4}{c|}{\textbf{Computation}} & \multicolumn{4}{c|}{\textbf{Communication}} \\
        \cline{2-9}
        & \multicolumn{2}{c|}{\textbf{Offline}} & \multicolumn{2}{c|}{\textbf{Online}} & \multicolumn{2}{c|}{\textbf{Offline}} & \multicolumn{2}{c|}{\textbf{Online}} \\
        \cline{4-5} \cline{8-9}
        & \multicolumn{2}{c|}{} & Leader  & Client &  \multicolumn{2}{c|}{} & Leader  & Client \\
        \hline
      ~\cite{LG-ASIACRYPT-2023} & \multicolumn{2}{c|}{$(T + l + m) m n \log n$} & $(T + l + m) m n$ & $(T + l) m n$ & \multicolumn{2}{c|}{$\lambda m^2 n \log n$} & $(T + l) m n + l m^2 n$ & $(T + l) m n$ \\
        \hline
        Ours& \multicolumn{2}{c|}{$(\gamma + m) m n \log n$} & \multicolumn{2}{c|}{$m^2 n$} & \multicolumn{2}{c|}{$\lambda m^2 n \log n$} & $\gamma m n + l m^2 n$ & $(\gamma + l+ m) m n$ \\
        \hline
  \end{tabular}}
  \caption{Asymptotic communication (bits) and computation costs of LG and our SK-MPSU in the offline and online phases. $n$ is the set size. $m$ is the number of participants. $\lambda$ and $\sigma$ are computational and statistical security parameter respectively and $\lambda = 128$, $\sigma = 40$. $T$ is the number of AND gate in an SKE decryption circuit, $T \approx 600$. $l$ is the bit length of input elements. $\gamma$ is the output length of OPPRF. $t$ is the noise weight in dual LPN, $t \approx 128$. }\label{tab:compare-ske}
\end{table*}


\begin{table*}
\resizebox{\textwidth}{!}{%
  \begin{tabular}{|c|c|c|c|c|}
        \hline
        & \multicolumn{2}{c|}{\textbf{Computation}} & \multicolumn{2}{c|}{\textbf{Communication}} \\
        \cline{2-5}
        & \textbf{Offline} & \textbf{Online} & \textbf{Offline} & \textbf{Online} \\
        \hline
      ~\cite{GNT-eprint-2023} & $\gamma m n \log n (\log n / \log \log n)$ & $m n (\log n / \log \log n)$ & $t \lambda m \log n (\log n / \log \log n)$ & $(\gamma + \lambda) m n (\log n / \log \log n)$ \\
        \hline
        Ours & $\gamma m n \log n$& $m n$ & $t \lambda m \log n$ & $(\gamma + \lambda) m n$ \\
        \hline
  \end{tabular}}
  \caption{Asymptotic communication (bits) and computation costs of~\cite{GNT-eprint-2023} and our PK-MPSU in the offline and online phases. In the offline phase, the computation is composed of symmetric-key operations; In the online phase, the computation is composed of public-key operations since we ignore symmetric-key operations. $n$ is the set size. $m$ is the number of participants. $\lambda$ and $\sigma$ are computational and statistical security parameter respectively and $\lambda = 128$, $\sigma = 40$. $l$ is the bit length of input elements. $\gamma$ is the output length of OPPRF. $t$ is the noise weight in dual LPN, $t \approx 128$. }\label{tab:compare-pke}
\end{table*}

\subsection{Theoretical Analysis of Our PK-MPSU}~\label{sec:analysis-pke} 

\subsubsection{Complete Analysis}

The costs of each stage in $\ProtoMPSUFour$ (Figure~\ref{fig:proto-mpsu4}) are calculated as follows.

\begin{trivlist}

\item \textbf{Batch ssPMT} The cost of this stage is the same as that in $\ProtoMPSUTwo$ (cf. Appendix~\ref{sec:analysis-pke}).

\begin{itemize}
\item \textbf{Offline phase.} The computation complexity per party is $O(\gamma m n\log n)$. The communication complexity per party is $O(t \lambda m \log (\gamma n / t))$. The round complexity is $O(1)$.

\item \textbf{Online phase.} The computation complexity per party is $O(m n)$. The communication complexity per party is $O(\gamma m n)$. The round complexity is $O(\log \gamma)$.
\end{itemize}

\item \textbf{Random oblivious transfers and messages rerandomization.} We use EC ElGamal encryption to instantiate the MKR-PKE scheme. So each of encryption, partial decryption and rerandomization takes one point scalar operation. The length of ciphertext is $4 \lambda$. 

For $1 \le i < j \le m$, $P_i$ and $P_j$ invoke $B$ instances of silent ROT correlations with random inputs during the offline phase. In the online phase, $P_i$ sends 1-bit derandomization message to transform each ROT into a chosen-input version. For each OT correlation, each $P_j$ executes two encryptions and sends two $4 \lambda$-bit messages to $P_i$. $P_i$ executes one rerandomization. If $i \ne 1$, $P_i$ sends one ciphertext to $P_j$ and $P_j$ executes one rerandomization as well.

\begin{itemize}
\item \textbf{Offline phase.} The computation complexity per party is $O(m n\log n)$. The communication complexity per party is $O(t \lambda m \log (n / t))$. The round complexity is $O(1)$.

\item \textbf{Online phase.} The computation complexity per party is $O(m n)$ public-key operations. The communication complexity per party is $O(\lambda m n)$. The round complexity is $O(1)$.
\end{itemize}

\item \textbf{Messages decryptions and shufflings.} Each party shuffles $(m-1)B$ ciphertexts and executes $(m-1)B$ partial decryptions before sending them to the next party.

\begin{itemize}
\item \textbf{Online phase.} The computation complexity per party is $O(m n)$ public-key operations. The communication complexity per party is $O(\lambda m n)$. The round complexity is $O(m)$.
\end{itemize}

\item \textbf{Total costs.} 
\begin{itemize}
\item \textbf{Offline phase.} The computation complexity per party is $O(\gamma m n \log n)$. The communication complexity per party is $O(t \lambda m \log (\gamma n / t)$. The round complexity is $O(1)$.

\item \textbf{Online phase.} The computation complexity per party is $O(m n)$ symmetric-key operations and $O(m n)$ public-key operations. The communication complexity per party is $O((\gamma + \lambda) m n)$. The round complexity is $O(\log \gamma + m)$.
\end{itemize}
\end{trivlist}

\subsubsection{Comparison with~\cite{GNT-eprint-2023}}

In Table~\ref{tab:compare-pke}, we present a comparison of the theoretical computation and communication complexity for each party in both offline and online phases between~\cite{GNT-eprint-2023} and our PK-MPSU.

We conclude that the complexity of our PK-MPSU surpasses theirs in all respects, including the computation and communication complexity of the leader and clients in the offline and online phases, as depicted in the Table~\ref{tab:compare-pke}.


\section{Multi-Party Private-ID}\label{sec:mpid}

It is natural to generalize the two-party private-ID~\cite{BKMSTV-ePrint-2020} to multi-party setting. Suppose there are $m$ parties, each possessing a set of $n$ elements. The multi-party private-ID functionality assigns a unique random identifier to each element across all input sets, ensuring that identical elements in different sets obtain the same identifiers. Each party receives identifiers associated with its own input elements, as well as identifiers associated with the union of all parties' input sets. With multi-party private-ID, the parties can sort their private sets based on a global set of identifiers and perform desired private computations item by item, ensuring alignment of identical elements across their sets. The formal definition of the multi-party private-ID is depicted in Figure~\ref{fig:func-mpid}. 
We build a concrete multi-party private-ID protocol based on the DDH assumption (described in Figure~\ref{fig:proto-mpid}) 
by extending the ``distributed OPRF+PSU" paradigm~\cite{CZZDL-PKC-2024} to multi-party setting.

\begin{figure}[!hbtp]
\begin{framed}
\begin{minipage}[center]{\textwidth}
\begin{trivlist}
\item \textbf{Parameters.} $m$ parties $P_1, \cdots P_m$. Size $n$ of input sets. The bit length $l$ of set elements. The range $D$ of identifiers. 
\item \textbf{Functionality.} On input $X_i = \{x_i^1,\cdots, x_i^n\} \subseteq \{0,1\}^l$ from $P_i$,
\begin{itemize}
    \item For every element $x \in \bigcup_{i=1}^m X_i$, choose a random identifier $R(x) \gets D$.
    \item Define $R^* = \{R(x) \vert x \in \bigcup_{i=1}^m X_i\}$ and $R_i = \{R(x) \vert x \in X_i\}$ for $i \in [m]$.
    \item Give output $(R^*, R_i)$ to $P_i$.
\end{itemize}
\end{trivlist}
\end{minipage}
\end{framed}
\caption{Multi-Party Private ID Functionality $\FuncMPID$}\label{fig:func-mpid}
\end{figure}

\begin{figure}[!hbtp]
\begin{framed}
\begin{minipage}[center]{\textwidth}
\begin{trivlist}
\item \textbf{Parameters.} $m$ parties $P_1, \cdots P_m$. Size $n$ of input sets. The bit length $l$ of set elements. A cyclic group $\mathbb{G}$, where $g$ is the generator and $q$ is the order. The identifie range $D = \mathbb{G}$. Hash function $\Hash(x): \{0,1\}^l \to \mathbb{G}$. 
\item \textbf{Inputs.} Each $P_i$ has input $X_i = \{x_i^1,\cdots, x_i^n\} \subseteq \{0,1\}^l$.
\item \textbf{Protocol.}
\begin{enumerate}
\item For $1 \le i \le m$, $P_i$ samples $a_i \gets \mathbb{Z}_q$ and $k_i \gets \mathbb{Z}_q$, then sends $\{\Hash(x_i^1)^{a_i},\cdots, \Hash(x_i^n)^{a_i}\}$ to $P_{(i+1) \mod m}$. For $1 \le j < m$, $P_{(i+j) \mod m}$ receives $\{y_i^1, \cdots, y_i^n\}$ from $P_{(i+j -1) \mod m}$. $P_{(i+j) \mod m}$ computes $\{(y_i^1)^{k_{(i+j) \mod m}}, \cdots, (y_i^n)^{k_{(i+j) \mod m}}\}$ and sends to $P_{(i+j+1) \mod m}$.
\item For $1 \le i \le m$, $P_i$ receives $\Hash(x_i^j)^{a_i k_{i+1} \cdots k_{m} k_1 \cdots k_{i-1}}$, $j \in [n]$, and computes $(\Hash(x_i^j)^{a_i k_{i+1} \cdots k_{m} k_1 \cdots k_{i-1}})^{-a_i k_i} = \Hash(x_i^j)^{k_{1} \cdots k_{m}}$. We denote $\{\Hash(x_i^1)^{k_{1} \cdots k_{m}}, \Hash(x_i^n)^{k_{1} \cdots k_{m}}\}$ as $R_i = \{r_i^1, \cdots, r_i^n\}$, where each $r_i^j \in \mathbb{G}$.
\item The parties invoke $\FuncMPSU$ where $P_i$ inputs $R_i = \{r_i^1, \cdots, r_i^n\}$. $P_1$ receives the union $R^* = \bigcup_{i=1}^m Y_i$, and sends it to other parties.
\item Each party $P_i$ outputs $(R^*, R_i)$.
\end{enumerate}
\end{trivlist}
\end{minipage}
\end{framed}
\caption{DH-based Multi-Party Private ID $\ProtoMPID$}\label{fig:proto-mpid}
\end{figure}

\begin{theorem}\label{theorem:mpid}
Protocol $\ProtoMPID$ securely implements $\FuncMPID$ against any semi-honest adversary corrupting $t < m$ parties in the $\FuncMPSU$-hybrid model, assuming the DDH assumption and $\Hash$ is a random oracle.
\end{theorem} 

\begin{trivlist}
\item \textbf{Correctness.} The first two steps essentially realize a multi-party distributed OPRF protocol, where each party $P_i$ inputs a set $\{x_i^1,\cdots, x_i^n\}$ and receives its own PRF key $k_i$ along with the PRF values computed on its input set using all parties' keys $k_1,\cdots, k_m$, denoted as $\{\PRF_{k_1, \cdots,k_m}(x_i^1),\cdots, \PRF_{k_1, \cdots,k_m}(x_i^n)\}$. In Figure~\ref{fig:proto-mpid}, each party $P_i$ first masks the hash value of each element by raising it to the power of $a_i$. Then these masked values are then cycled among the parties, with each party raising them to the power of its key. Finally, $P_i$ unmasks the values and raises them to the power of its won key $k_i$, resulting in the distributed PRF values $\PRF_{k_1, \cdots,k_m}(x) = \Hash(x)^{k_1 \cdots k_m}$. Even if $m-1$ parties collude, there still exists one exponent private to the adversary, ensuring that the PRF values remain pseudorandom according to the DDH assumption.

In the step 3, the parties invoke an MPSU protocol on their distributed PRF values, so than they can securely receive all PRF values on the union of input sets.

\item \textbf{Security.} Assuming $\Hash$ is a random oracle, the security of protocol follows immediately from the DDH assumption and the security of our PK-MPSU.

Modern DH-based cryptosystems often use elliptic curves for the underlying cyclic group, due to their compact size and better efficiency at a given level of security. In this context, the first two steps of the protocol have relatively low communication costs, meanwhile, the element space of the MPU protocol in step 3 is EC points, which aligns perfectly with the element space of our PK-MPSU. This alignment allows us to instantiate the MPSU functionality using our PK-MPSU, leveraging its lowest communication costs. As a result, the entire multi-party private-ID protocol achieves extremely low communication costs, making it well-suited for deployment in bandwidth-restricted environments.

Another benefit of instantiating the MPSU functionality with our PK-MPSU is to achieve linear computation and communication complexity. Since the multi-party distributed OPRF phase has linear computation and communication, by integrating our PK-MPSU, the entire multi-party private-ID protocol can maintain the linear complexities throughout.
\end{trivlist}

\end{document}